\documentclass[12pt]{article}
\usepackage[pass]{geometry}

\usepackage{graphicx} 
\usepackage{subfigure} 
\usepackage{lscape}	
\usepackage{pst-plot, pstricks} 
\usepackage{fancybox,amssymb,color} 
\usepackage[english]{babel}
\usepackage{setspace} 
\usepackage{epstopdf}
\usepackage[capposition=top]{floatrow} 
\usepackage{float}
\usepackage{enumitem}
\usepackage{longtable}
\usepackage{tabularx,calc}
\usepackage{multirow}
\usepackage{booktabs}
\usepackage{acro}
\usepackage{setspace}
\usepackage{lscape}
\usepackage{mathtools}
\usepackage{rotating}
\usepackage{xcolor}
 \usepackage{subcaption}

\usepackage[authoryear]{natbib}
\usepackage{caption}

\usepackage{amsmath}
\usepackage{amsthm} 
\usepackage{amssymb}
\usepackage{amsbsy}
\usepackage{bm}

\DeclareMathOperator*{\argmin}{arg\,min}

\usepackage{changes}
\definechangesauthor[name={Sascha}, color=green]{Sascha}

\newtheorem{proposition}{Proposition}

  \title{Uncertain Short-Run Restrictions and Statistically Identified Structural Vector Autoregressions}
  \author{Sascha  A. Keweloh  \\
    TU Dortmund University }

 \date{\today}

\begin{document}
\clearpage\maketitle
\thispagestyle{empty}

\begin{abstract}
\noindent  This study proposes a combination of a statistical identification approach with potentially invalid short-run zero restrictions. The estimator shrinks towards imposed restrictions and stops shrinkage when the data provide evidence against a restriction.
Simulation results demonstrate how incorporating valid restrictions through the shrinkage approach enhances the accuracy of the statistically identified estimator and how the impact of invalid restrictions decreases with the sample size. The estimator is applied to analyze the interaction between the stock and oil market. The results indicate that incorporating stock market data into the analysis is crucial, as it enables the identification of information shocks, which are shown to be important drivers of the oil price.

\end{abstract}

\noindent%
\vfill

\newpage

\section{Introduction}
\label{sec: Introduction} 
Traditional approaches to identifying structural vector autoregressions (SVAR) typically involve imposing economically motivated restrictions, often restricting how structural shocks affect the variables in the SVAR simultaneously.
More recently, alternative approaches imposing structure on the stochastic properties of the shocks, such as time-varying volatility or non-Gaussian and independent shocks, have been used for identification.
Although statistical identification methods do not rely on economically motivated restrictions for identification, prior economic knowledge is still required to label the shocks.
Put differently, some form of prior economic knowledge beyond the stochastic properties of the shocks remains necessary, even if it is not required for identification. 

Consequently, the question is how do we utilize our prior economic knowledge in a statistically identified SVAR? 
The question relates to the critique of the "all-or-nothing approach" w.r.t. prior economic knowledge raised by \cite{baumeister2019structural}. Traditional methods often treat prior knowledge as indisputable truth, enforcing restrictions without the ability to update them, while simultaneously ignoring other prior knowledge entirely.  In this context, estimators relying on statistical identification approaches, which disregard any available restrictions, represent the extreme end of the "nothing approach."

This study proposes an approach to incorporate potentially invalid short-run restrictions on impulse responses into the estimation of a statistically identified SVAR. The estimator relies on non-Gaussian and (mean) independent shocks for identification and adds penalizes deviations from imposed zero restrictions on the simultaneous impulse responses. 
The study goes beyond merely proposing a test for overidentifying restrictions; instead, it advocates a non-dogmatic approach to incorporate restrictions.  
In contrast to traditional estimation approaches that treat restrictions as binding constraints, the proposed shrinkage estimator can stop shrinkage towards restrictions if the data present evidence against them, offering a non-dogmatic approach to incorporate restrictions. This approach seeks to enhance the efficiency of the statistically identified estimator through valid restrictions while mitigating the impact of invalid restrictions when the data provide evidence against them.

In this study, short-run zero restrictions are incorporated using a ridge penalty with adaptive weights (see, e.g., \cite{zou2006adaptive}).  
The adaptive weights induce an important feature: It becomes cheap to deviate from invalid restrictions and costly to deviate from valid restrictions. As a result, the weights determine the importance of a given restriction and, consequently, the degree of shrinkage toward it in a data-driven manner.
Therefore, in contrast to traditional estimators, which dogmatically rely on restrictions included as binding constraints, the ridge penalty offers a non-dogmatic alternative where the data determine the degree of shrinkage towards imposed restrictions.
This approach is only possible when restrictions are not required for identification. Therefore, a separate identification approach is required to determine the weights by providing evidence in favor or against the imposed restrictions.

Identification in this study relies on non-Gaussian and (mean) independent shocks. 
The assumption of independent shocks often faces criticism, with the common objection that shocks driven by the same volatility process are not independent, as discussed in \cite{montiel2022svar}. 
In response to this critique, recent developments in the non-Gaussian SVAR literature yield identification results under more relaxed assumptions regarding the (in)dependencies of shocks; see \cite{guay2021identification}, \cite{mesters2022non}, \cite{anttonen2023bayesian}, or \cite{lewis2023identification} for a comprehensive overview.
This study adds to the literature by providing an identification result that allows for a two-stage identification approach based on the non-Gaussianity of the shocks. For skewed shocks, identification requires mean independent shocks and allows for a common volatility process. For shocks with zero-skewness but non-zero excess kurtosis, identification requires their independence.

The non-Gaussian estimator considered in this study aims at achieving robustness by relying as little as possible on structure imposed on the stochastic properties of the shocks. Specifically, the estimator only minimizes second- to fourth-order  moment conditions implied by mean independent shocks, thus circumventing the need to impose a specific distribution on the shocks and enabling the identification of shocks driven by a common volatility process. Although this approach of imposing minimal structure on the stochastic properties of the shocks enhances robustness, it comes at the cost of efficiency.

The motivation of the proposed ridge estimator is to combine the statistical identification approach with short-run restrictions, leveraging prior economic knowledge on the simultaneous impulse responses, to enhance the efficiency of the estimator. Monte Carlo simulations show how economically motivated restrictions and a statistical identification approach complement each other; Valid restrictions improve the accuracy of the statistically identified estimator, and the impact of invalid restrictions decreases with evidence of the statistically identified estimator against them.

Bayesian approaches offer a natural way to incorporate economic knowledge using the prior distribution of the parameters.  Moreover,  Bayesian SVARs identified by independence and non-Gaussianity allow for updating economically motivated priors, see  \cite{lanne2020identification}, \cite{anttonen2021statistically},  \cite{braun2021importance}, and \cite{keweloh2023estimating}.
Nevertheless, the incorporation of prior economic knowledge, specifically the imposition of economically motivated zero restrictions, has deep roots in the frequentist SVAR literature as well. However,  the frequentist approach to including restrictions often adopts a dogmatic stance, lacking the ability to gather and utilize evidence against a given restriction.
This study introduces a non-dogmatic approach to include restrictions in the frequentist SVAR estimation framework, allowing for a more flexible and data-driven treatment of restrictions.

The application analyzes the interaction of the oil and stock market.
\cite{kilian2009impact} propose  recursive   restrictions  to identify and estimate the effects of different oil and stock market shocks. The proposed restrictions are widely used to analyze the impact of oil market shocks on the stock market, see, e.g., \cite{apergis2009structural}, \cite{abhyankar2013oil}, \cite{kang2013oil},  \cite{sim2015oil}, \cite{ahmadi2016global}, \cite{lambertides2017effects}, \cite{mokni2020time},  \cite{arampatzidis2021oil}, \cite{kwon2022impacts}, or \cite{arampatzidis2023identification}. However, the impact and importance of stock market information shocks on the oil price is usually not analyzed. The application in this study fills this gap. I present evidence that oil and stock prices cannot be ordered recursively. By allowing both variables to interact simultaneously, the study reveals that information shocks originating from the stock market contain crucial information on oil prices, which explains approximately $25$ \% of the fluctuations in oil prices.

The remainder of the paper is organized as follows:
Section \ref{sec: Overview: SVAR} contains a brief overview on SVAR models. 
Section \ref{sec: nG identification}  derives the non-Gaussian identification and estimation approach.
Section \ref{sec: Ridge SVAR-GMM  estimator} introduces the ridge estimator to incorporate potentially invalid restrictions.
Section \ref{sec: Finite Sample Performance}  uses  simulations to illustrate the ability of the estimator to exploit correctly and discard falsely imposed restrictions.
Section \ref{sec: Application}  applies the  estimator to an oil and stock market SVAR.
Section \ref{sec: Conclusion} concludes.

\section{Overview: SVAR}
\label{sec: Overview: SVAR}  
Consider an SVAR with $n$ variables
\begin{align}
\label{eq: SVAR}
y_t =  \nu +  A_1 y_{t-1} + ... + A_p y_{t-p}  + u_t 
\quad \text{and} \quad
u_t=   B_0  \varepsilon_{t} ,
\end{align}
with  $B_0 \in \mathbb{B} := \{B \in \mathbb{R}^{n \times n} | det(B)\neq 0 \}$ and $A_0:= B_0^{-1}$ and 
 $n$-dimensional vectors of  time series  $y_t=[y_{1t} ,...,y_{nt} ]'$,   reduced form shocks $u_t=[u_{1t} ,...,u_{nt} ]'$, and  structural shocks $\varepsilon_t=[\varepsilon_{1t},...,\varepsilon_{nt}]'$ with mean zero and unit variance.
The parameter matrices $A_1,...,A_p$ and the intercept term can be consistently estimated to obtain the reduced form shocks. To simplify, I treat the reduced form shocks as observable random variables and focus on identifying and estimating the simultaneous interaction $u_t =   B_0  \varepsilon_{t}$.\footnote{
	In practice, an SVAR can be estimated using a two-step approach where the VAR is estimated in the first step and the simultaneous interaction is estimated in the second step. Simulations analyzing the performance of the two-step approach can be found in Appendix \ref{appendix: sec: Finite sample performance} and show little differences compared to the simulations in the main text.
}

Define the  innovations 
\begin{align}
\label{eq: define unmixed innovations}
e(B)_t := B^{-1} u_t, 
\end{align}
equal to the innovations obtained by unmixing the reduced form shocks with a matrix $B\in \mathbb{B}$.
For $B=B_0$, the innovations are equal to the structural shocks.
Identification of the SVAR comes down to formulating a set of equations that guarantee the equivalence between innovations and structural shocks.

Typically, SVAR models are identified based on the assumption of uncorrelated structural shocks. 
Therefore, the matrix $B$ should generate uncorrelated innovations with unit variance, which yields $(n+1)n/2$ moment conditions. However, the matrix $B$ has $n^2$ coefficients. Consequently, infinitely many matrices $B \in \mathbb{B}$ generate uncorrelated innovations with unit variance, meaning that the assumption of uncorrelated structural shocks is not sufficient to identify the SVAR. 

Traditional identification methods solve the identification problem by imposing structure on the interaction of the variables or impact of the shocks (e.g. short-run restrictions in \cite{sims1980macroeconomics}, long-run restrictions in \cite{blanchard1989dynamic}, sign restrictions in \cite{uhlig2005effects}, or proxy variables in \cite{mertens2013dynamic}). The structure probably most frequently imposed are short-run restrictions, meaning restrictions on coefficients of the $B$ matrix to reduce the number of free coefficients to $(n+1)n/2$ such that the remaining unrestricted coefficients are identified by the $(n+1)n/2$ moment conditions implied by uncorrelated shocks with unit variance.   
Note that identification requires at least $(n-1)n/2$ restrictions, and incorrect restrictions lead to inconsistent estimates. 
Additionally, with  $(n-1)n/2$ restrictions, the SVAR is just identified. Therefore, even when the sample size goes to infinity, we are unable to detect incorrect restrictions.

More recently, identification approaches based on additional structure imposed on the stochastic properties of the structural shocks have been put forward in the literature. These approaches use properties such as time-varying volatility (see, e.g.,  \cite{rigobon2003identification}, \cite{lanne2010structural},   \cite{lutkepohl2017structural}, \cite{lewis2021identifying}, or \cite{bertsche2022identification}) or the non-Gaussianity and independence of the shocks (see, e.g., \cite{matteson2017independent}, \cite{herwartz2016macroeconomic},  \cite{gourieroux2017statistical}, \cite{lanne2017identification}, \cite{maxand2020identification}, \cite{lanne2021gmm}, \cite{keweloh2020generalized},   \cite{guay2021identification}, \cite{mesters2022non}, \cite{lanne2022identifying}, \cite{herwartz2023point},   \cite{drautzburg2023refining}, or \cite{fiorentini2023discrete}) to ensure identification.

\section{Non-Gaussian SVAR}
\label{sec: nG identification}
This section first provides an intuition of how non-Gaussian and independent shocks allow to solve the identification problem and discusses different degrees of (in)dependence assumptions, emphasizing their economic significance.
Subsequently, the section derives explicit conditions under which third- and fourth-order moment conditions derived from the assumption of mutually mean independent shocks identify the SVAR up to labeling of the shocks. Moreover, I propose an approach to label the shocks based on a first-step estimator. 
Finally,   the last subsection  introduces the non-Gaussian moment based  estimator used in the remainder of the study.

\subsection{Intuition}
\begin{figure}[h!] 
	\centering
	\caption{Illustrating the Role of Coskewness in Identifying Skewed Shocks}
	\includegraphics[width=0.8\textwidth]{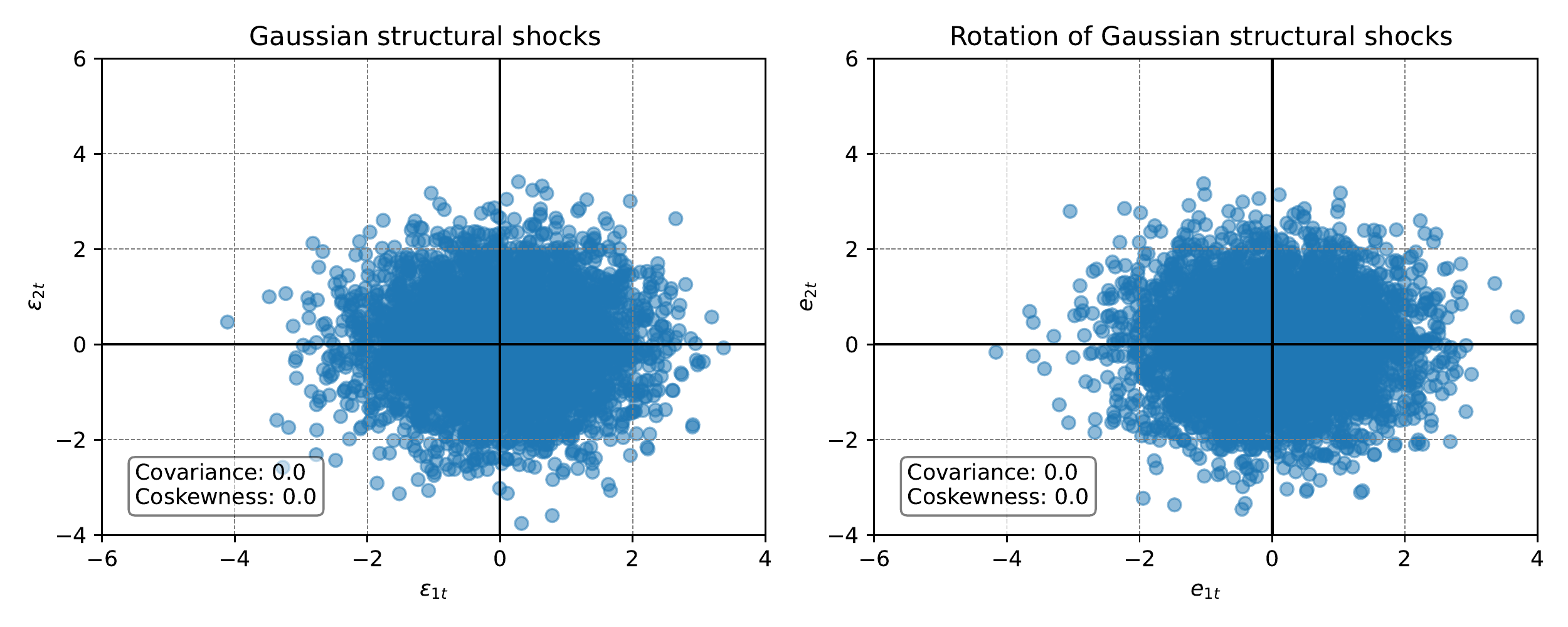}
	\includegraphics[width=0.8\textwidth]{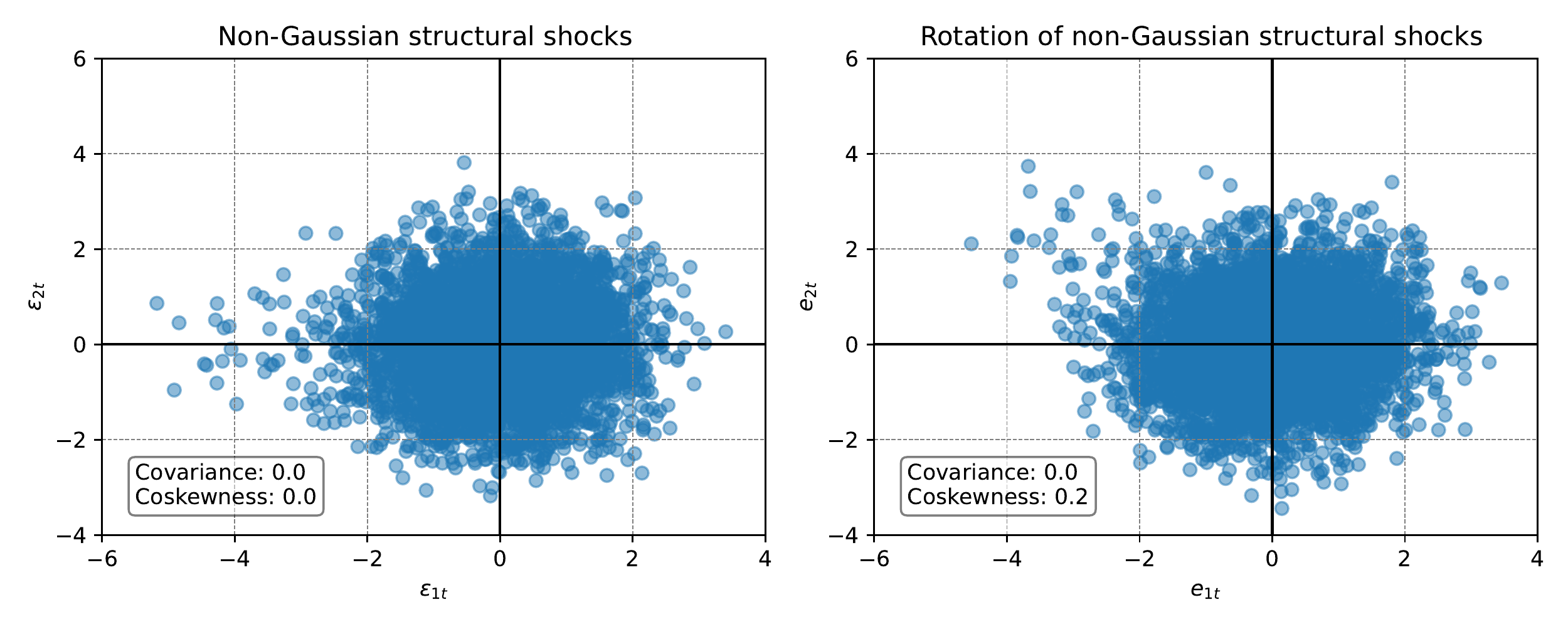}
	\label{fig: Illustration} 
	\begin{minipage}{1\textwidth} %
		{   \footnotesize  
			\textit{Note:} 
			The left side displays independent structural shocks $\varepsilon_{1t}$ and $\varepsilon_{2t}$. The right side exhibits a rotation $\begin{bmatrix}
				e_{1t}\\ e_{2t} 
			\end{bmatrix} = \begin{bmatrix}
			cos(\pi/5) & 	sin(\pi/5)\\ 	
				-sin(\pi/5) & 	cos(\pi/5)
			\end{bmatrix}  \begin{bmatrix}
				\epsilon_{1t}\\ \epsilon_{2t}.
			\end{bmatrix}
			 $.
			In the upper row, the shocks   are independently   drawn from a standard normal distribution.
			In the lower row, the first shock, $\varepsilon_{1t}$, is drawn from a mixture of normal distributions, resulting in a left-skewed distribution. Each plot presents the covariance $E[\varepsilon_{1t}  \varepsilon_{2t}]$ and $E[e_{1t} e_{2t}]$, as well as the coskewness $E[\varepsilon_{1t}^2 \varepsilon_{2t}]$ and $E[e_{1t}^2 e_{2t}]$ respectively.
			\par}
	\end{minipage}
\end{figure}

Assumptions on the mutual (in)dependence of the structural shocks can be used to derive higher-order moment conditions and identify the SVAR. For example, the coskewness $E[\epsilon_{1t}^2 \epsilon_{2t}]$ of two independent shocks is zero. 
Figure \ref{fig: Illustration}  illustrates how the coskewness   can be used to identify skewed shocks.
The left side shows plots of independent structural shocks $\varepsilon_{1t}$ and $\varepsilon_{2t}$, while the right side shows a rotation $e_{1t}$ and $e_{2t}$ of the shocks.
In the upper row, the shocks $\varepsilon_{1t}$ and $\varepsilon_{2t}$ are independently drawn from a standard normal distribution. Any rotation of the shocks again leads to uncorrelated and independent innovations. Specifically, the covariance and coskewness are equal to zero for any rotation of the shocks. 
In the lower row, the first shock $\varepsilon_{1t}$ is drawn from a mixture of normal distributions, that is, the shock is drawn from a standard normal distribution with a probability of $99$\% and with a probability of $1$\% the shock is drawn from a normal distribution with mean four and variance one, which leads to a skewed distribution of the shock $\varepsilon_{1t}$. Rotating the skewed shocks leads to uncorrelated but dependent shocks. In particular, for the rotation depicted in the bottom right, the coskewness is positive, indicating that high absolute values of $e_{1t}$ are correlated with positive values of $e_{2t}$. Consequently, knowing the value of the first shock, $e_{1t}$, conveys information about the other shock, $e_{2t}$, although both shocks are uncorrelated.  By utilizing the fact that the structural shocks are independent, the coskewness allows to immediately detect that the bottom right panel only shows a rotation of the structural shocks.

The key difference between non-Gaussian identification approaches and traditional restriction based  approaches is that non-Gaussian approaches impose and utilize more structure on the dependency of the structural shocks. Traditional approaches typically only utilize the assumption of uncorrelated structural shocks, whereas non-Gaussian approaches rely on stronger assumptions, i.e., on the assumption of independent shocks.  
However, non-Gaussian identification does not necessarily require the assumption of fully independent shocks, but instead can work with weaker assumptions on the dependencies of the shocks, see \cite{lanne2021gmm}, \cite{guay2021identification}, \cite{mesters2022non}, or \cite{anttonen2023bayesian}. For instance, the illustration in Figure \ref{fig: Illustration} uses only a coskewness condition.

This raises the question of what constitutes an appropriate assumption regarding the dependencies among structural shocks.
The assumption of independent shocks is often criticized for being overly restrictive, as it does not account for the possibility that shocks are influenced by a common volatility process, see  \cite{montiel2022svar}.
In contrast, one could also argue that the assumption of uncorrelated shocks may be too weak.  Shocks can be uncorrelated, but still be highly dependent in a manner that may not be suitable for structural shocks. 
For instance, consider the lower right panel in Figure \ref{fig: Illustration}, which shows uncorrelated but evidently dependent shocks. In this scenario, if the first shock represents a demand shock and the second a supply shock, knowing the value of the demand shock would immediately provide information about the mean of the supply shock, despite the fact that both shocks are not correlated. Alternatively, consider an even more extreme example with $\varepsilon_1 \sim \mathcal{N}(0,1)$ and $\varepsilon_2 = \varepsilon_1^2-1$. Both random variables are uncorrelated but clearly dependent in a way that appears implausible for structural shocks.

In this study, I advocate for the assumption of mean independent shocks, i.e. $E[\varepsilon_{it} |  \varepsilon_{-it} ]=0$ for $i=1,...,n$ and $ \varepsilon_{-it}:=[\varepsilon_{1t},...,\varepsilon_{(i-1)t},\varepsilon_{(i+1)t},...,\varepsilon_{nt}]$. This assumption is stronger than mere uncorrelated shocks, yet more lenient than assuming full independence. 
Imposing the condition of mean independent shocks excludes dependency structures where one shock provides information about the mean of another shock, while still allowing for dependency structures like a common volatility process.

Indeed, it is crucial to recognize that while traditional identification approaches may primarily utilize the assumption of uncorrelated shocks, their applications implicitly rest on the assumption of mean independent shocks for interpretation. This implicit assumption is vital when making causal statements about the expected responses of variables to shocks. 
To illustrate this, consider a bivariate SVAR without lags such that the first variable is equal to $y_{1t} = b_{11} \varepsilon_{1t} + b_{12} \varepsilon_{2t}$. The simultaneous impulse response of $y_{1t}$ to shocks $\varepsilon_{1t}$ is $b_{11}$ and is typically interpreted as the expected response of $y_{1t}$ to shocks $\varepsilon_{1t}$, that is, $b_{11} = E\left[ y_{1t } | \varepsilon_{1t} =1\right]$.
Crucially, the assumption of uncorrelated shocks is not sufficient to guarantee that this equality holds. This is because $ E\left[  y_{1,t } | \varepsilon_{1t} =1\right] =  b_{11} E\left[   \varepsilon_{1t} | \varepsilon_{1t}=1 \right] + b_{12} E\left[  \varepsilon_{2t} | \varepsilon_{1t}=1 \right] $ and $ E\left[  \varepsilon_{2t} | \varepsilon_{1t}=1 \right] =0$ does not follow solely from the assumption of uncorrelated shocks.\footnote{
 One can also argue that impulse responses represents the thought experiment $b_{11} = E\left[ y_{1t } | \varepsilon_{1t}=1,\varepsilon_{2t}=0  \right]$. Although this is mathematically correct, it is not clear whether the thought experiment makes economically any sense for uncorrolated but dependent shocks. For example, the two shocks $\varepsilon_{1t} \sim N(0,1)$ and $\varepsilon_{2t}=\varepsilon_{1t}^3-3\varepsilon_{1t}$ are uncorrelated; however, the combination $\varepsilon_{1t}=1$ and $\varepsilon_{2t}=0 $ cannot even occur.
}
Instead, it requires mean independent shocks.
Therefore, the assumption of mean independent shocks is used implicitly in any SVAR application that makes causal statements about the expected response of the variables to the shocks and thus, the assumption of mean independent shocks can also be used to identify the SVAR.

\subsection{Identification}

 For $i,j,k,l \in \{1,...,n\}$,  mutually mean independent shocks with mean zero and unit variance imply	
 (co-)variance conditions 
\begin{align}
	E [e(B)_{it}^2 ] -1 &= 0, & \\  
	E [e(B)_{it}  e(B)_{jt}   ] &= 0, & \text{for } i&<j 
\end{align}
coskewness conditions
\begin{align}
	\label{eq: proof cos1}
	E [e(B)_{it}^2 e(B)_{jt} ] &= 0, & \text{for } i&\neq j \\
	\label{eq: proof cos2}
	E [e(B)_{it}  e(B)_{jt} e(B)_{kt} ] &= 0, & \text{for } i&<j<k    	
\end{align}
and cokurtosis conditions
\begin{align} 
	\label{eq: proof cokurtosis}
	E [e(B)_{it}^3  e(B)_{jt} ] &= 0, & \text{for } i&\neq j \\
	E [e(B)_{it}^2  e(B)_{jt} e(B)_{kt}]  &= 0, & \text{for } i&\neq j, i \neq k, j<k \\
	\label{eq: proof cokurtosis 3}
	E [e(B)_{it}   e(B)_{jt} e(B)_{kt} e(B)_{lt} ]&=0, & \text{for } i&< j<k<l. 
\end{align} 

In general, the moment conditions implied by   mean independent shocks  can be written as
\begin{align}
	\label{eq: f}
	E[f_m(B,u_t)] = 0 \quad &\text{with} \\
	 f_M(B,u_t):=\prod_{i=1}^{n} e(B)_{i,t}^{m_i}- c(m)   &\text{, } \quad c(m) := 
	 \begin{cases}
	 	0,& \text{if } 1 \in\{m_1,....m_n\}\\
	 	1,              & \text{else}
	 \end{cases}, 
\end{align}
with variance and 
covariance conditions for 	$M \in \mathbf{2}:=  \{M=[  m_1,....m_n] \in  \{0,1,2\}^n |
\sum_{i=1}^{n} m_i = 2  
\}$,
coskewness conditions for 	$M \in \mathbf{3}:= \{M=[  m_1,....m_n] \in   \{0,1,2\}^n |  \sum_{i=1}^{n} m_i = 3  \}$, and
cokurtosis conditions for 	$M \in \mathbf{4}:= \{  M=
[  m_1,....m_n] \in  \{0,1,2,3\}^n |
\sum_{i=1}^{n} m_i = 4  , \exists 1 \in [ m_1,....m_n] 
\}$. For a given moment condition $E[f_{M}(B,u_t)]=0$, the indices $m_i$ simply denote the power of each innovation in the moment condition, i.e. for the moment condition $E[f_{M}(B,u_t)]=	E [e(B)_{1t}^2 e(B)_{2t} ]  = 0$ the indices are $m_1=2$, $m_2=1$, and $m_{3}=...=m_{n}=0$.

The following proposition establishes  conditions under which the second- to fourth-order moment conditions implied by mutually mean independent shocks identify the SVAR.
\begin{proposition}
	\label{proposition: 1}
	Partition the SVAR $u_t = B_0 \varepsilon_t$ into three groups of shocks $\varepsilon_t = [\tilde{\varepsilon}_{1t}, \tilde{\varepsilon}_{2t}  , \tilde{\varepsilon}_{3t} ]'$ where
    $\tilde{\varepsilon}_{1t}$ contains all skewed shocks with arbitrary excess kurtosis,
	   $\tilde{\varepsilon}_{2t}$    contains all shocks with zero skewness and non-zero  excess kurtosis, 
		and   $\tilde{\varepsilon}_{3t}$  contains all shocks with zero skewness and zero excess kurtosis.
	Assume that all shocks are mutually mean independent, i.e., $E[\varepsilon_{it} |  \varepsilon_{-it} ]=0$ for $i=1,...,n$.
		
	Let $0 = E[f(B,u_t)] := [ E[f_{M_1}(B,u_t)],...,E[f_{M_K}(B,u_t)]]' $ contain all second- to fourth-order moment conditions implied by mutually mean independent shocks with unit variance.
		  
	\begin{enumerate}
		\item 	The skewed shocks   $\tilde{\varepsilon}_{1t}$  and the corresponding columns of $B_0$ are identified up to sign and permutation.  
		
		\item The shocks with excess kurtosis $\tilde{\varepsilon}_{2t}$    are identified up to sign and permutation together with the corresponding columns of $B_0$ if the shocks contained in $\tilde{\varepsilon}_{2t}$ are mutually independent.  
		
		\item The remaining shocks $\tilde{\varepsilon}_{3t}$ are identified up to an orthogonal rotation. If the set contains only one shock, the shock and the corresponding column of $B_0$ are identified.
	\end{enumerate}	
\end{proposition}
\begin{proof}
	The proof can be found in the Appendix.
	The proof of the first statement generalizes Theorem 5.3 in \cite{mesters2022non} to multiple Gaussian shocks.
	The proof of the second statement follows the proof of Theorem 5.10 in \cite{mesters2022non}, however, I replaced the genericity condition used in  \cite{mesters2022non} by the assumption of independent shocks and generalize the statement to multiple Gaussian shocks.
	The proof of the third statement is trivial.
\end{proof}
The proposition shows that mutually mean independent shocks with non-zero skewness are identified by the moment conditions. This allows to identify skewed shocks even if they are driven by the same volatility process. 
Moreover, if shocks exhibit zero skewness,  the moment conditions still identify the shocks with non-zero excess kurtosis if these shocks are mutually independent.\footnote{
	 Technically, the first statement  only requires that the shocks satisfy all coskewness conditions implied by mean independent shocks and the second statement only requires that the  shocks satisfy all cokurtosis conditions implied by independent shocks. Therefore, the second statement does not necessarily require independent shocks, however, it requires that all cokurtosis conditions resulting from independent shocks hold. However, the conditions do not follow from mean independent shocks, and finding an economically plausible process other than independent shocks that yields shocks satisfying all such conditions is not straightforward.
}
The first statement is a  generalization of moment-based identification results in the literature  (specifically for third moments in \cite{bonhomme2009consistent}   and for arbitrary moments in  \cite{mesters2022non}) to the case with multiple Gaussian shocks, similar to the partial identification results in \cite{maxand2020identification} and \cite{guay2021identification}.
The second statement is related to the identification result in \cite{lanne2021gmm} based on asymmetric fourth-order moment conditions. However, in contrast to the identification result in \cite{lanne2021gmm} which only provides a local identification result, Proposition \ref{proposition: 1} provides a global identification result up to sign and permutation.
The second statement assumes that the shocks are independent and, therefore, satisfy the symmetric cokurtosis conditions $E[\varepsilon_{it}^2 \varepsilon_{jt}^2-1]=0$ for $i\neq j$. However, these symmetric conditions are not contained in the moment conditions $E[f(B,u_t)]$. The contribution of the second statement is to show that for independent shocks with sufficient excess kurtosis, the moment conditions $E[f(B,u_t)]$  guarantee global identification.\footnote{
	Note that   while the identification result in \cite{lanne2021gmm} only uses asymmetric fourth-order moment conditions, it still requires the assumption that all cokurtosis conditions (not just the asymmetric ones) implied by mutually independent shocks hold. This assumption is used in the proof of the proposition in \cite{lanne2021gmm}. Therefore, the identification result in \cite{lanne2021gmm} cannot be used to identify an SVAR with shocks affected by the same volatility process.
} Exuding the symmetric cokurtosis moment conditions is important to guarantee identification based on mean independent and skewed shock in the first statement. Specifically, if  the moment conditions $E[f(B,u_t)]$  would contain the  symmetric cokurtosis moment conditions, the first statement would not hold.

\subsection{Labeling}
\label{subsec: Labeling}
Proposition \ref{proposition: 1} establishes  identification up to sign and permutation, e.g. for any sign-permutation matrix $P$ the models  $u_t = B_0  \epsilon_t$ and $u_t = \tilde{B} \tilde{\epsilon}_t$ with $\tilde{B}= B P^{-1}$ and $\tilde{\epsilon}_t= P \epsilon_t$ have the same dependency structure. 
Without additional guidance from the researcher, the shocks do not possess explicit structural labels.
However, imposing  restrictions on the impact of a structural shock of interest, as discussed in the next section, requires to label the shocks a priori.

One approach to address the indeterminacy of sign-permutations and label the shocks is to restrict the set of admissible $B$ matrices to a set containing a single representative of each sign-permutation class. This can be achieved, for instance, by constraining the set to:
\begin{align}
	\nonumber
	\bar{\mathbb{B}} := \{B \in \mathbb{B}|& \text{ } |C_{kk}|>|C_{kl}| 
	\text{ for } l=k+1,...,n 
	\text{ and } C_{kk} >0 \text{ for } k=1,...,n  
	\text{ and } C:=  B   
	\} ,
\end{align}
where for almost all $B \in \mathbb{B}$ there exists a unique sign-permutation matrix $P$ such that $B P^{-1} \in \bar{\mathbb{B}}$ and $ B \tilde{P}^{-1} \notin \bar{\mathbb{B}}$ for all sign-permutation matrices $\tilde{P} \neq P$, compare \cite{lanne2017identification}.
Intuitively, the constrained set $\bar{\mathbb{B}}$ imposes the restriction that   shocks $\varepsilon_{it}$ have a positive impact on $u_{it}$ and the simultaneous impact of shock $\varepsilon_{it}$ on $u_{it}$ is greater in absolute terms than the   impact of all following shocks on $u_{it}$.

Constraining the set of admissible  matrices to $\bar{\mathbb{B}}$ allows to a priori label the shocks. For example, suppose that the first variable measures government spending, and thus in $\bar{\mathbb{B}}$ the first shock always has the largest simultaneous impact on government spending and could be labeled as the government spending shock.\footnote{
	Sign restrictions are an alternative approach to restrict and a priori label the shocks. Sign restrictions restrict the set of admissible $B$ matrices to a smaller set containing unique sign-permutation representatives, and each matrix in the constrained set corresponds to a shock labeled a priori based on the sign restrictions. However, labeling based on sign restrictions is not ideal for zero restrictions which are located at the boundary of the constrained set.
} 
However, relying on $\bar{\mathbb{B}}$ to label the shocks can be problematic if  $B_0$  is located at the boundary of the set, i.e., if in the example above government spending is equally driven by government spending and output shocks.

\begin{figure}[h!] 
	\centering
	\caption{Illustration of labeling  using $\bar{\mathbb{B}} $ }
	\includegraphics[width=0.8\textwidth]{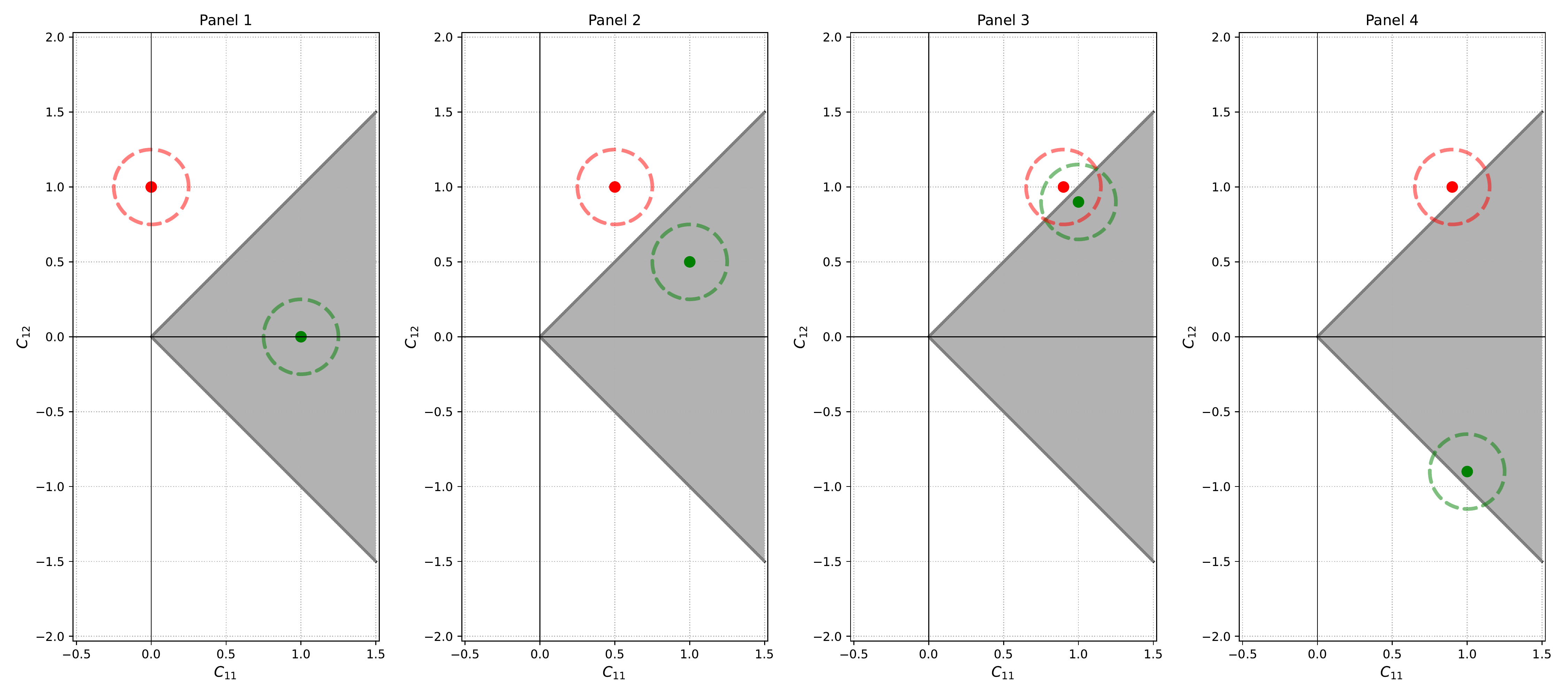} 
	\includegraphics[width=0.8\textwidth]{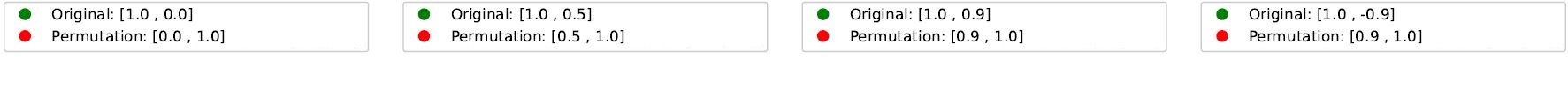} 
	\label{fig: visualization_set} 
	\begin{minipage}{1\textwidth} %
		{   \footnotesize  
			\textit{Note:} 
			The panels  display  the elements $[c_{11}, c_{12}]$ in the first row of various $B_0$ matrices.
			The green dots represent the first row of the following four matrices: 
			$B_0 = \begin{bmatrix}
				1 & 0 \\ 0& 1
			\end{bmatrix} $ in panel $1$, $ 
			B_0 = \begin{bmatrix}
				1 & 0.5 \\ 0& 1
			\end{bmatrix}  $ in panel $2$, $ 
			B_0 = \begin{bmatrix}
				1 & 0.9\\ 0& 1
			\end{bmatrix} $ in panel $3$, and $ 
			B_0 = \begin{bmatrix}
				1 & -0.9 \\ 0& 1
			\end{bmatrix}   $ in panel $4$. 
			The red dots display the first row's elements of the sign permutation of the $B_0$ matrices, which keeps the diagonal elements positive.  
			The circles around each point illustrate a set of  matrices close to the corresponding matrix.
			The shaded area indicates the set of  matrices contained in $\bar{\mathbb{B}} $.
			\par}
	\end{minipage}
\end{figure}   

Figure \ref{fig: visualization_set} illustrates potential labeling problems using $\bar{\mathbb{B}}$. To simplify, I consider bivariate models where labeling based on $\bar{\mathbb{B}}$ only relies on the two elements in the first row of a given $B_0$ matrix and thus can be easily visualized. Each panel shows a different $B_0$ matrix and plots the elements in the first row that represent the matrix as a green dot. In the bivariate SVAR, there is only one permutation of $B_0$ with positive diagonal elements, represented by the red dot that represents the elements in the first row of the permutation.  
The dotted epsilon balls around points illustrate a set of similar matrices.
The shaded area displays the set $\bar{\mathbb{B}}$ which always contains one of both permutations.  

The first and second panels display examples where $B_0$ is located in the inner area of $\bar{\mathbb{B}}$ and, therefore, the set $\bar{\mathbb{B}}$ also contains similar matrices in the epsilon balls.
In contrast, the third and fourth panels display examples where $B_0$ is located at the boundary, resulting in some similar matrices in the epsilon ball not being contained within the set.
For example, consider the third panel where 
\begin{align}
	\nonumber
	B_0 = \begin{bmatrix}
		1 & -0.9  \\
		0  &  1
	\end{bmatrix} \in \bar{\mathbb{B}} 
	\quad
	\text{ and }
	u_t = B_0 \begin{bmatrix}
		\epsilon_{1t}\\ \epsilon_{2t}.
	\end{bmatrix}
\end{align}
Now,  consider an estimator $\hat{B}$ and its sign permutation $\tilde{B}$ with
\begin{align}
	\nonumber
	\hat{B}= \begin{bmatrix}
		1 & -1.01  \\
		0.01  &  1
	\end{bmatrix}
	\quad \text{ and } \quad
	\tilde{B}= \begin{bmatrix}
		1.01  & 1  \\
		-1  &  0.01 
	\end{bmatrix}.
\end{align}
Clearly, $\hat{B}$ corresponds to the same order of shocks as $B_0$ and is contained in the epsilon ball around $B_0$, while $\tilde{B}$ corresponds to the reverse order with a sign flip and is contained in the epsilon ball around the permutation of $B_0$. However, even though $B_0 \in \bar{\mathbb{B}}$ the estimator $\hat{B}$ that is close to $B_0$ is not contained in $\bar{\mathbb{B}}$, while the estimator $\tilde{B}$ corresponding to the reverse order is included. This illustrates how using $\bar{\mathbb{B}}$ to label shocks can yield misleading results when $B_0$ is located at the boundary of the set.

\begin{figure}[h!] 
	\centering
	\caption{Illustration of labeling  using  $\bar{\mathbb{B}}_{\bar{B}}$}
	\includegraphics[width=0.8\textwidth]{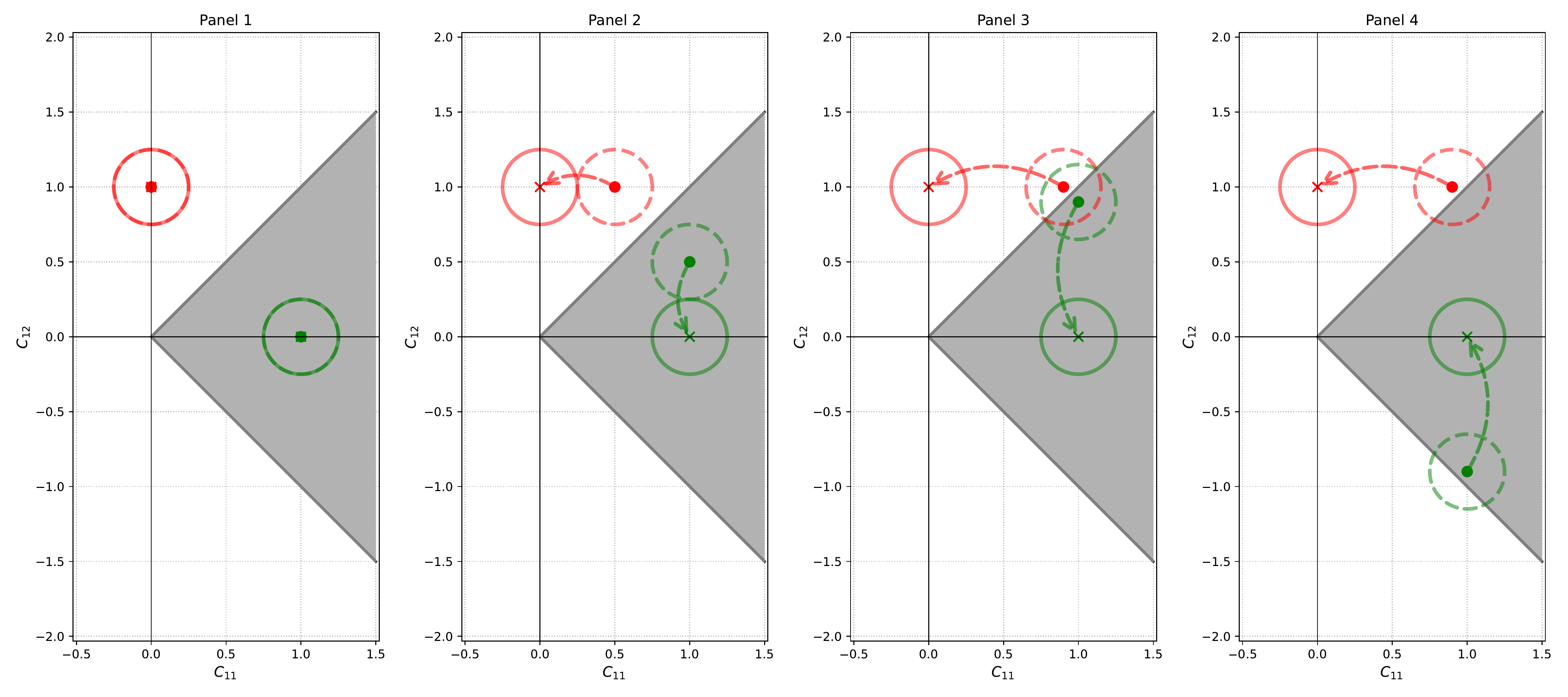} 
	\includegraphics[width=0.8\textwidth]{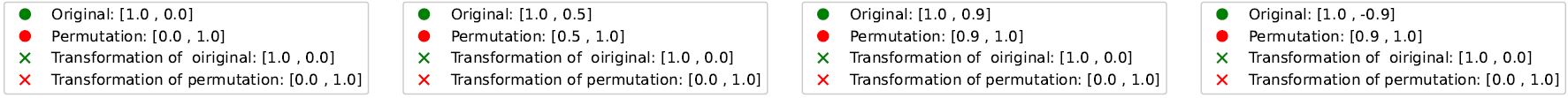} 
	\label{fig: visualization_set 2} 
	\begin{minipage}{1\textwidth} %
		{   \footnotesize  
			\textit{Note:} 
						The panels  display  the elements $[c_{11}, c_{12}]$ in the first row of various $B_0$ matrices.
			The green dots represent the first row of the following four matrices: 
			$B_0 = \begin{bmatrix}
				1 & 0 \\ 0& 1
			\end{bmatrix} $ in panel $1$, $ 
			B_0 = \begin{bmatrix}
				1 & 0.5 \\ 0& 1
			\end{bmatrix}  $ in panel $2$, $ 
			B_0 = \begin{bmatrix}
				1 & 0.9\\ 0& 1
			\end{bmatrix} $ in panel $3$, and $ 
			B_0 = \begin{bmatrix}
				1 & -0.9 \\ 0& 1
			\end{bmatrix}   $ in panel $4$. 
			The red dots display the elements of the first row of the sign permutation of the $B_0$ matrices, which keeps the diagonal elements positive.  
				Additionally, the figure shows the elements in the first row of the transformed  matrices using $C:= \bar{B}^{-1} B$  with $\bar{B} =  B_0$  and $B$ is equal to $B_0$ or equal to the sign permutation of $B_0$, represented by  green and red crosses.  
			The circles around each point illustrate a set of  matrices close to the corresponding   matrix.
			The shaded area indicates the set of  matrices contained in $\bar{\mathbb{B}}_{\bar{B}}$. 
			 
			\par}
	\end{minipage}
\end{figure}   
 
To avoid this, I propose to use an initial labeled estimator $\bar{B} $ of $B_0$ as a transformation to ensure that $B_0$ is located in the inner area of the set. 
Define the generalized set 
\begin{align}
	\nonumber
	\bar{\mathbb{B}}_{\bar{B}} := \{B \in \mathbb{B}|& \text{ } |C_{kk}|>|C_{kl}| 
	\text{ for } l=k+1,...,n 
	\text{ and } C_{kk} >0 \text{ for } k=1,...,n 
	\text{ and } C:= \bar{B}^{-1} B   
	\}.
\end{align} 
For $\bar{B}=I$ the set $\bar{\mathbb{B}}_{\bar{B}=I} $ is equal to $\bar{\mathbb{B} }$.
The key idea is to use an initial labeled estimator  $\bar{B}$ as a transformation that re-centers each $B$ matrices such that $B_0$  is located at the inner of the set  $\bar{\mathbb{B}}_{\bar{B}}$. 
Figure \ref{fig: visualization_set 2} visualizes how the transformation moves the green points and circles representing the correct permutations and their neighborhood to the center of the set, while the red points and circles, representing the incorrect permutation and their neighborhood, are pushed away from the set. This ensures that the neighborhood of the correct permutation is contained in the set, and thus all solutions in the neighborhood receive the same labels.
Importantly, the initial estimator $\bar{B} $ used for the transformation does not need to be equal to $B_0$. Figure \ref{fig: visualization_set 4}  in the appendix shows that using a $\bar{B} $ in proximity of $B_0$ is sufficient to move $  B_0  $ to the inner area of the set such that all matrices in the neighborhood still receive the same labels.

\subsection{Estimation}

Using the second- to fourth-order moment conditions, $ E[f(B,u_t)]=0$, implied by mutually mean independent shocks, the SVAR-GMM estimator can be written as
\begin{align}
	\label{eq: gmm}
	\hat{B}_T    := \argmin \limits_{B \in 	 \bar{\mathbb{B}}_{\bar{B}}  } \text{ }
	g_T(B)'
	W
	g_T(B) , 
\end{align}
with  $g_T(B)= \frac{1}{T}\sum_{t=1}^{T} f(B,u_t)$   and  a suitable weighting matrix   $W$.
Consistency and asymptotic normality of the estimator follow from standard assumptions and the identification result in Proposition \ref{proposition: 1}, see \cite{hall2005generalized}.

The asymptotically efficient weighting matrix $W =S^{-1}$ with the long-run covariance matrix $S := \underset{T \rightarrow \infty }{lim} E \left[T g_T(B_0) g_T(B_0)' \right]$  leads to the lowest possible asymptotic variance of the estimator, see \cite{hall2005generalized}. 
However, in small samples and with higher-order moment conditions, the efficient weighting matrix is difficult to estimate and the asymptotically efficient SVAR-GMM estimator exhibits a scaling bias towards innovations with a variance smaller than the normalizing unit variance assumption, \cite{keweloh2023structural}.

To address both problems, \cite{keweloh2023structural} proposes the two-step SVAR continuous scale updating estimator (SVAR-CSUE). The SVAR-CSUE assumes serially and mutually independent shocks to estimate the efficient weighting matrix and incorporates a continuously updated scaling term into the weighting matrix to eliminate the scaling bias. The two-step SVAR-CSUE is defined as follows:
\begin{align}
	\label{eq: csue}
	\hat{B}_T    := \argmin \limits_{B \in 	 \bar{\mathbb{B}}_{\bar{B}}  } \text{ }
	g_T(B)'
W(B)
	g_T(B),
\end{align} 
with $W(B)=	\left(  \hat{D}(B) \hat{W}  \hat{D}(B)  \right)$ and the continuously updated scaling term $
\hat{D}(B) := diag \left( \prod_{i=1}^{n}  \hat{d}(B)_i^{ m_{1,i}}   ,...,   \prod_{i=1}^{n} \hat{d}(B)_i^{ m_{K,i}}  \right)
$
where
$  \hat{d}(B)_i := \frac{1}{ \sqrt{1/T \sum_{t=1}^{T}e(B)_{it}^2 }} $.
The parameters $ m_{j,i}$ for $j=1,...,K$ and $i=1,...,n$ in the scaling term $\hat{D}(B) $ are equal to the power of the $i$-th innovation in the $j$-th moment condition and   $\hat{d}(B)_i$ is equal to the inverse of the 
standard deviation of the $i$-th innovation. Consequently, the scaling term increases the weight of a given moment condition if $B$ leads to innovations with a variance smaller than one, which eliminates the scaling bias towards innovations with a variance smaller than the normalizing unit variance assumption, see \cite{keweloh2023structural}.

In the first step, the weighting matrix $\hat{W}$ of the two-step SVAR-CSUE  is equal to the identity matrix, and in the second step, it is equal to the inverse of the estimated long-run covariance matrix leveraging the assumption of serially and mutually independent shocks as proposed in \cite{keweloh2023structural}.

Importantly, the assumption of serially and mutually independent shocks is only used to estimate the weighting matrix, which affects efficiency of the estimation. However, identification and consistency are ensured by the weaker assumption of mutually mean independent and sufficiently skewed shocks with Proposition \ref{proposition: 1}. 
Therefore, consistency of the SVAR-GMM and SVAR-CSUE require to impose only little structure on the stochastic properties of the shocks.
Intuitively, the precision of the estimation tends to increase with the structure imposed on the SVAR.
For example, a correctly specified maximum likelihood estimator would likely yield reduced bias, lower mean squared error, and narrower confidence intervals. However, the gain in precision of such an estimator must be weighed against the potential pitfall of misspecifying the distribution and dependence patterns of the shocks.\footnote{The simulation in summarized in Table \ref{Table: Finite sample performance - Bias MSE - MCRGMMLagsAndHet - ML} in the appendix illustrates how a non-Gaussian maximum likelihood estimator which misspecifies the common volatility process of the shocks is biased whereas the proposed moment-based estimator remains unbiased.}
Rather than imposing more structure on the stochastic properties of the shocks, the subsequent section shows how the researcher can complement the SVAR-CSUE by incorporating economically grounded short-run zero restrictions using a shrinkage approach to increase the estimator's precision.

\section{Incorporating potentially invalid restrictions}
\label{sec: Ridge SVAR-GMM  estimator}   
This section proposes an approach to incorporate potentially invalid short-run zero restrictions using a ridge penalty with adaptive weights into the non-Gaussian SVAR estimation.
Unlike conventional SVAR methods that rely on restrictions for identification, the proposed methodology leverages restrictions as a means of incorporating the researchers' a priori economic knowledge to increase the precision and efficiency of the estimation in comparison to an estimator relying only on the non-Gaussianity of the shocks.
Moreover, combining a statistical identification approach with restrictions using a shrinkage mechanism enables the data to provide evidence against a given restriction and to stop shrinkage towards invalid restrictions.

The following notation is used to denote short-run zero restrictions on $B_0$. 
Let $\mathcal{R} $ be the set of all pairs $(i,j) \in \{1,...,n\}^2 $  corresponding to elements $B_{ij}$ of $B_0$ restricted to zero. For example, imposing a recursive order implies that all elements in the upper-triangular of $B_0$ are equal to zero, which corresponds to $ \mathcal{R}=\{(i,j) \in \{1,...,n\}^2 | j>i \}$.\footnote{
	Proxy variables  can also be   implemented using  short-run restrictions and the ridge estimator proposed in this section can be applied to proxy variables in augmented proxy SVAR models.  Moreover, the ridge approach can be applied to restrictions in an $A$-type SVAR model with $A_0 u_t = \varepsilon_t$. Lastly, the approach can easily be extended to non-zero restrictions.  
}

Define the Ridge SVAR-CSUE as
 \begin{align}
 	\label{eq: rcsue}
 	\hat{B}_T    := \argmin \limits_{B \in 	\bar{\mathbb{B}}_{\bar{B}} } \text{ }
 	g_T(B)'
 W(B)
 g_T(B) + \lambda \sum_{(i,j)\in \mathcal{R}}^{ }  
   v_{ij} B_{ij}^2  ,   
 \end{align} 
with a tuning parameter $\lambda \geq 0$  and
adaptive weights $v_{ij}$, compare \cite{zou2006adaptive} for adaptive Lasso and \cite{dai2018broken} for adaptive Ridge estimators.

Before discussing the construction of the weights and tuning parameter, it is worth highlighting the relation of the proposed estimator to existing approaches.
Traditional restriction based estimators use restrictions as binding constraints, while estimators based on stochastic properties, like non-Gaussianity, typically ignore any potentially available restrictions.
The proposed shrinkage estimator  nests both approaches as special cases.
 Specifically, if the tuning parameter and the weights are manually set to converge to infinity, all restrictions become binding constraints, resembling a traditional restriction based estimator. Conversely, if the tuning parameter and weights are set to zero, deviating from restrictions induces no penalty, reducing the estimator to one based solely on non-Gaussianity. In contrast, this study's proposal involves using the data to determine the tuning parameter and weights. Consequently, the cost of deviating from restrictions is determined by empirical evidence rather than by relying on a dogmatic approach that assigns either zero or infinity to the tuning parameter and weights.

The adaptive weights play a crucial role in determining the cost associated with deviating from a given restriction. The proposal of the study is to allow these weights and consequently the cost of deviation to be guided by the data.
Let $\hat{B}$ be an initial consistent estimator of $B_0$ obtained by the SVAR-CSUE without penalties. The proposed adaptive weights are defined as
\begin{align}
	\label{eq: adaptive weights}
	v_{ij} :=  \frac{1}{ \hat{B}_{ij}^2}  \left( 1 + \frac{1}{nG_j^2 + min_{(2)}(nG_1,...,nG_n)^2}\right),
\end{align}
where the first term, $  1/\hat{B}_{ij}^2$, is driven by the distance of the first step estimator to the zero restriction,
and the second term, $\left( 1 + \frac{1}{nG_j^2 + min_{(2)}(nG_1,...,nG_n)^2}\right)$, adjusts the weights depending on the Gaussianity of the shocks corresponding to the first step estimator.
The term  $nG_j$ quantifies  the non-Gaussianity of the first step estimator's $j$-th shock, with   $nG_j := S_j^2/6 + (K_j-3)^2/24$ where $S_j$ and $K_j$ denote the skewness and kurtosis of the $j$-th shock, and $min_{(2)}(nG_1,...,nG_n)$ selects the second smallest non-Gaussianity measure of all shocks.

With the adaptive weights, the cost of deviating from a given restriction is determined by the data through the first-step estimator. 
Consider the scenario where the shocks are sufficiently non-Gaussian, allowing the first-step estimator to consistently estimate all elements of $B_0$. If a restriction is correct, the consistency of the first-step estimator ensures that the first term of the adaptive weight for that restriction converges to infinity. Consequently, it becomes costly to deviate from correct restrictions, and the estimator shrinks towards those restrictions in line with the data. However, if a restriction is not correct, the first-step estimator will not converge to the restriction, and the first term of the adaptive weight for an incorrect restriction will not diverge to infinity. As a result, the estimator can deviate from incorrect restrictions.

The introduction of the second term, which adjusts the weights based on the Gaussianity of the shocks, is designed to ensure proper weights when multiple shocks are Gaussian. In such cases, the first-step estimator can only identify and consistently estimate the impact of non-Gaussian shocks. The remaining Gaussian shocks are only identified up to a rotation of the Gaussian shocks. Therefore, if there are more than two Gaussian shocks, the first-step estimator cannot provide evidence against restrictions on the impact of the Gaussian shocks. In this case, the second term of the adaptive weights leads to an increase in the weights of the restrictions corresponding to Gaussian shocks. This ensures that only those restrictions where the data actually provide evidence against them receive low weights.

Specifically, the Gaussianity correction is constructed such that if two or more (or even all) shocks are Gaussian, the Gaussianity measure of restrictions corresponding to elements in the columns of the Gaussian shocks converges to zero and the weights of these restrictions go to infinity.
However, for the restrictions corresponding to non-Gaussian shocks, the non-Gaussianity measure remains finite and scales the weights based on the degree of non-Gaussianity.

The tuning parameter $\lambda$ scales the weights of all restrictions. The tuning parameter is determined by cross-validation with two folds and multiple repetitions.  
First, I define an increasing sequence of possible $\lambda$ values.
Larger values lead to stronger weighting of the restrictions, which in turn induces more shrinkage. Essentially, higher values of $\lambda$  express greater confidence in the validity of the restrictions imposed.
For each $\lambda$, I estimate $\hat{B}_T  $ using training fold data and calculate the loss in the let-out fold at $\hat{B}_T$.\footnote{
	Solving the optimization problems in the cross-validation can be challenging and may require a good starting value. Appendix \ref{appendix: sec: CV} describes the details of the cross-validation that involves a rule to derive reasonable starting values for each $\lambda$ and an additional regularization term.
} 
The subsequent task is to select the tuning parameter based on the losses from the cross-validation.
The selection process aims to  select the largest feasible $\lambda$ value supported by the cross-validation results.
To this end, I calculate the median, $60$\% quantile, and $40$\% quantile loss across repetitions for each $\lambda$ value. 
Subsequently, I calculate the smallest value $\lambda$  for which the median loss of all subsequent $\lambda$ values is smaller than $60$\% quantile loss of all preceding $\lambda$ values and the smallest $\lambda$ value for which the $40$\% quantile loss of all subsequent $\lambda$ values is smaller than the median loss of all preceding $\lambda$ values. The tuning parameter is set to the minimum of both values.
This approach employs a systematic ascent of the tuning parameter, translating to increased shrinkage, until a pronounced surge in losses becomes apparent, indicating a transition point where additional shrinkage is unwarranted.

\section{Finite sample performance}
\label{sec: Finite Sample Performance} 
The following Monte Carlo study illustrates the benefits of correctly imposed restrictions via the penalty term and sheds light on the  estimator's ability to distinguish between correct and incorrect restrictions. The simulations show that correctly imposed restrictions via the penalty term lead to an increase in precision, i.e. a reduced small sample bias and reduced MSE, and an increase in efficiency, i.e. narrower confidence bands. Moreover, the simulations show how the impact of false restrictions decreases with increasing sample size.

I simulate an SVAR with four variables  
\begin{align}
\label{eq: MC} 
\begin{bmatrix}
u_{1t} \\
u_{2t} \\
u_{3t} \\
u_{4t} \\
\end{bmatrix} =
\begin{bmatrix}
10 & 0  & 0 & 0 \\
5 & 10 & 0 & 0\\
5 & 5 & 10 & 5\\
5 & 5 & 5 & 10
\end{bmatrix}
\begin{bmatrix}
\varepsilon_{1t} \\
\varepsilon_{2t} \\
\varepsilon_{3t} \\
\varepsilon_{4t} \\
\end{bmatrix}  ,
\end{align}  
where the structural shocks are independently and identically drawn from the  two-component mixture
$
\epsilon_{it} \sim  0.79\; \mathcal{N}(-0.2,0.7^2) + 0.21 \;  \mathcal{N}(0.75,1.5^2),
$
where $\mathcal{N}(\mu, \sigma^2)$ indicates a normal distribution with mean $\mu$ and standard deviation $\sigma$. The shocks have skewness $0.9$ and excess kurtosis $2.4$.

The simulation compares three estimators.
The first estimator, denoted by CSUE, is the two-step SVAR-CSUE estimator and does not use restrictions. 
The second estimator, denoted by RCSUE($\mathcal{R}_1$), is the Ridge SVAR-CSUE with a penalty on the correct zero restrictions $ \mathcal{R}_1=\{(i,j) \in \{1,...,n\}^2 | j>i \text{ and } i \leq 2 \}$.
The third estimator, denoted by RCSUE($\mathcal{R}_2$), is the Ridge SVAR-CSUE with a penalty on the restrictions $ \mathcal{R}_2=\{(i,j) \in \{1,...,n\}^2 | j>i \}$ which impose a recursive structure and thus contain one incorrect restriction.  
The adaptive weights of the ridge estimators are calculated based on the unrestricted estimator.
The tuning parameter $\lambda$ of the ridge estimators is chosen using a repeated cross-validation with two folds, $10$ repetitions, and a sequence of $40$ potential $\lambda$ values. 
The Appendix contains multiple additional simulations, including simulations with Gaussian shocks,  a VAR with lags and shocks with a common volatility process, $A$-type restrictions, and  augmented proxy VAR restrictions.

\begin{table}
	\caption{Finite sample performance - average and mean squared error}
	\label{Table: Finite sample performance - Bias MSE}
	\centering
	\renewcommand{\arraystretch}{1}
	\begin{tabular}{c@{\hspace{2pt}} |@{\hspace{2pt}}c@{\hspace{2pt}} c@{\hspace{2pt}} c@{\hspace{2pt}}}
		& $CSUE$ & $RCSUE(\mathcal{R}_1)$ & $RCSUE(\mathcal{R}_2)$ \\
		\hline
 \rotatebox[origin=c]{90}{$T = 250$ } &$\begin{bmatrix}\underset{(0.59)}{9.75} & \underset{(1.37)}{\hphantom{-}0.04} & \underset{(1.35)}{\hphantom{-}0.02} & \underset{(1.45)}{\hphantom{-}0.01} \\ \underset{(1.49)}{4.87} & \underset{(0.85)}{\hphantom{-}9.74} & \underset{(1.67)}{-0.01} & \underset{(1.81)}{\hphantom{-}0.08} \\ \underset{(2.08)}{4.87} & \underset{(2.11)}{\hphantom{-}4.87} & \underset{(1.33)}{\hphantom{-}9.75} & \underset{(2.17)}{\hphantom{-}4.86} \\ \underset{(2.18)}{4.88} & \underset{(2.22)}{\hphantom{-}4.84} & \underset{(2.0)}{\hphantom{-}4.91} & \underset{(1.51)}{\hphantom{-}9.72} \\ \end{bmatrix}$ &$\begin{bmatrix}\underset{(0.4)}{9.9} & \textcolor{red}{\underset{(0.05)}{\hphantom{-}0.01}} & \textcolor{red}{\underset{(0.06)}{\hphantom{-}0.01}} & \textcolor{red}{\underset{(0.06)}{\hphantom{-}0.01}} \\ \underset{(0.49)}{4.95} & \underset{(0.44)}{\hphantom{-}9.88} & \textcolor{red}{\underset{(0.06)}{-0.0}} & \textcolor{red}{\underset{(0.08)}{0.0}} \\ \underset{(0.69)}{4.95} & \underset{(0.61)}{\hphantom{-}4.94} & \underset{(0.67)}{\hphantom{-}9.84} & \underset{(1.34)}{\hphantom{-}4.89} \\ \underset{(0.68)}{4.96} & \underset{(0.63)}{\hphantom{-}4.93} & \underset{(1.31)}{\hphantom{-}4.94} & \underset{(0.75)}{\hphantom{-}9.8} \\ \end{bmatrix}$ &$\begin{bmatrix}\underset{(0.42)}{9.87} & \textcolor{red}{\underset{(0.1)}{0.0}} & \textcolor{red}{\underset{(0.11)}{-0.0}} & \textcolor{red}{\underset{(0.15)}{-0.04}} \\ \underset{(0.54)}{4.93} & \underset{(0.47)}{\hphantom{-}9.84} & \textcolor{red}{\underset{(0.14)}{0.0}} & \textcolor{red}{\underset{(0.22)}{-0.08}} \\ \underset{(0.81)}{4.94} & \underset{(0.76)}{\hphantom{-}4.94} & \underset{(0.72)}{\hphantom{-}10.35} & \textcolor{red}{\underset{(10.85)}{\hphantom{-}2.3}} \\ \underset{(0.83)}{4.96} & \underset{(0.8)}{\hphantom{-}4.96} & \underset{(4.4)}{\hphantom{-}6.51} & \underset{(5.44)}{\hphantom{-}8.09} \\ \end{bmatrix}$ \\ [40pt]\rotatebox[origin=c]{90}{$T = 500$ } &$\begin{bmatrix}\underset{(0.24)}{9.87} & \underset{(0.66)}{-0.02} & \underset{(0.62)}{\hphantom{-}0.02} & \underset{(0.62)}{\hphantom{-}0.01} \\ \underset{(0.67)}{4.97} & \underset{(0.38)}{\hphantom{-}9.87} & \underset{(0.8)}{\hphantom{-}0.04} & \underset{(0.84)}{\hphantom{-}0.02} \\ \underset{(0.95)}{4.95} & \underset{(1.05)}{\hphantom{-}4.92} & \underset{(0.65)}{\hphantom{-}9.91} & \underset{(1.02)}{\hphantom{-}4.93} \\ \underset{(0.98)}{4.94} & \underset{(1.09)}{\hphantom{-}4.93} & \underset{(1.04)}{\hphantom{-}4.98} & \underset{(0.67)}{\hphantom{-}9.87} \\ \end{bmatrix}$ &$\begin{bmatrix}\underset{(0.19)}{9.94} & \textcolor{red}{\underset{(0.02)}{0.0}} & \textcolor{red}{\underset{(0.02)}{-0.0}} & \textcolor{red}{\underset{(0.02)}{0.0}} \\ \underset{(0.21)}{4.98} & \underset{(0.2)}{\hphantom{-}9.95} & \textcolor{red}{\underset{(0.03)}{0.0}} & \textcolor{red}{\underset{(0.03)}{0.0}} \\ \underset{(0.26)}{4.98} & \underset{(0.29)}{\hphantom{-}4.99} & \underset{(0.31)}{\hphantom{-}9.93} & \underset{(0.59)}{\hphantom{-}4.95} \\ \underset{(0.26)}{4.97} & \underset{(0.29)}{\hphantom{-}5.0} & \underset{(0.59)}{\hphantom{-}4.97} & \underset{(0.32)}{\hphantom{-}9.9} \\ \end{bmatrix}$ &$\begin{bmatrix}\underset{(0.2)}{9.92} & \textcolor{red}{\underset{(0.05)}{-0.01}} & \textcolor{red}{\underset{(0.05)}{-0.0}} & \textcolor{red}{\underset{(0.06)}{-0.02}} \\ \underset{(0.25)}{4.97} & \underset{(0.22)}{\hphantom{-}9.93} & \textcolor{red}{\underset{(0.06)}{0.0}} & \textcolor{red}{\underset{(0.09)}{-0.05}} \\ \underset{(0.32)}{4.98} & \underset{(0.35)}{\hphantom{-}4.99} & \underset{(0.42)}{\hphantom{-}10.29} & \textcolor{red}{\underset{(5.27)}{\hphantom{-}3.27}} \\ \underset{(0.34)}{4.97} & \underset{(0.37)}{\hphantom{-}5.01} & \underset{(2.31)}{\hphantom{-}6.02} & \underset{(2.43)}{\hphantom{-}8.82} \\ \end{bmatrix}$ \\ [40pt]\rotatebox[origin=c]{90}{$T = 1000$ } &$\begin{bmatrix}\underset{(0.11)}{9.96} & \underset{(0.29)}{-0.02} & \underset{(0.28)}{-0.01} & \underset{(0.28)}{\hphantom{-}0.02} \\ \underset{(0.3)}{4.99} & \underset{(0.17)}{\hphantom{-}9.94} & \underset{(0.37)}{-0.01} & \underset{(0.38)}{\hphantom{-}0.03} \\ \underset{(0.45)}{4.99} & \underset{(0.46)}{\hphantom{-}4.96} & \underset{(0.29)}{\hphantom{-}9.95} & \underset{(0.48)}{\hphantom{-}4.98} \\ \underset{(0.46)}{4.97} & \underset{(0.46)}{\hphantom{-}4.95} & \underset{(0.45)}{\hphantom{-}4.98} & \underset{(0.3)}{\hphantom{-}9.96} \\ \end{bmatrix}$ &$\begin{bmatrix}\underset{(0.09)}{9.98} & \textcolor{red}{\underset{(0.0)}{-0.0}} & \textcolor{red}{\underset{(0.0)}{-0.0}} & \textcolor{red}{\underset{(0.01)}{0.0}} \\ \underset{(0.09)}{5.0} & \underset{(0.09)}{\hphantom{-}9.97} & \textcolor{red}{\underset{(0.01)}{-0.0}} & \textcolor{red}{\underset{(0.01)}{0.0}} \\ \underset{(0.12)}{5.0} & \underset{(0.12)}{\hphantom{-}4.99} & \underset{(0.13)}{\hphantom{-}9.97} & \underset{(0.25)}{\hphantom{-}4.97} \\ \underset{(0.13)}{5.0} & \underset{(0.12)}{\hphantom{-}4.99} & \underset{(0.24)}{\hphantom{-}4.99} & \underset{(0.14)}{\hphantom{-}9.96} \\ \end{bmatrix}$ &$\begin{bmatrix}\underset{(0.1)}{9.98} & \textcolor{red}{\underset{(0.02)}{-0.0}} & \textcolor{red}{\underset{(0.01)}{-0.0}} & \textcolor{red}{\underset{(0.02)}{-0.01}} \\ \underset{(0.11)}{4.99} & \underset{(0.11)}{\hphantom{-}9.97} & \textcolor{red}{\underset{(0.02)}{-0.0}} & \textcolor{red}{\underset{(0.02)}{-0.02}} \\ \underset{(0.16)}{4.99} & \underset{(0.16)}{\hphantom{-}4.99} & \underset{(0.2)}{\hphantom{-}10.2} & \textcolor{red}{\underset{(1.84)}{\hphantom{-}4.0}} \\ \underset{(0.17)}{5.0} & \underset{(0.16)}{\hphantom{-}4.99} & \underset{(0.87)}{\hphantom{-}5.61} & \underset{(0.82)}{\hphantom{-}9.35} \\ \end{bmatrix}$ \\ [40pt]

	\end{tabular}
		\begin{minipage}{1\textwidth} %
		{   \footnotesize  
			\textit{Note:} 
			Monte Carlo simulation with $M = 2000$ replications for the SVAR in Equation (\ref{eq: MC}). The table
			shows the average, $1/M 
			\sum_{m=1}^{M}
			\hat{b}^{m}_{ij} $ and in parentheses the mean squared error $1/M 
			\sum_{m=1}^{M}
			(\hat{b}^{m}_{ij} - b_ij)^2 $
			of
			each estimated element  $\hat{b}^{m}_{ij}$ in simulation $m$. Penalized elements are highlighted in red.
			\par}
	\end{minipage}
\end{table}

Table \ref{Table: Finite sample performance - Bias MSE} shows the average and MSE of each estimated element and Table \ref{Table: Finite sample performance - Coverage} displays the coverage and average length of bootstrap confidence intervals, see Appendix \ref{appendix: sec: Bootstrap} for details on the construction of bootstrap confidence intervals.

First, the results show how imposing correct restrictions using the ridge estimator leads to an increase of the estimator's precision, meaning a smaller bias and reduced MSE, and to an increase of efficiency, meaning a reduction in the length of the confidence bands,  compared to the unpenalized estimator. 
 The most notable improvements are observed in penalized elements, where the MSE and width of confidence bands  of the RCSUE($\mathcal{R}_1$) estimator are substantially smaller compared to the unpenalized CSUE. Notably, the penalty also improves the performance of the unpenalized elements of the RCSUE($\mathcal{R}_1$), with MSE approximately three times smaller and the width of confidence bands reduced by half compared to the CSUE.

These findings underscore the ability of the ridge estimator to identify and shrink towards correct restrictions. Moreover, they highlight the utility of restrictions beyond their traditional role in ensuring identification. While CSUE relies solely on statistical properties and imposes minimal structural constraints on the SVAR, resulting in volatile estimates with large uncertainties, the ridge estimator leverages economically motivated restrictions to enhance precision and efficiency.

Secondly, the results shed light on the impact of incorrect restrictions.
The $\mathcal{R}_2$ penalty contains a false restriction, which shrinks the $b_{34}$ element to zero, contrary to its true value of five.
This invalid restriction induces an increase in bias, MSE, and distorted coverage of the confidence bands for the $b_{34}$ element. At the same time, the correct restrictions lead to a performance increase of the correctly penalized elements and result in a mostly positive impact on the unpenalized elements of the estimator.
One crucial distinction from traditional approaches, where restrictions are treated as binding constraints, is that the ridge penalty mitigates the adverse effects of incorrect restrictions, especially as the sample size increases. In small samples, the statistical identification approach may not provide robust evidence against invalid restrictions, causing the estimator to shrink towards them. However, with more data, the approach can detect that shrinking towards the incorrect restriction leads to dependent shocks, resulting in smaller tuning parameters determined by cross-validation. This reduces the impact of incorrect restriction with increasing sample size. Further details on the selected tuning parameters are provided in the Appendix.

\begin{table}
	\caption{Finite sample performance - coverage and confidence band width}
	\label{Table: Finite sample performance - Coverage}
	\centering
	\renewcommand{\arraystretch}{1.2}
	\begin{tabular}{c@{\hspace{2pt}} |@{\hspace{2pt}}c@{\hspace{2pt}} c@{\hspace{2pt}} c@{\hspace{2pt}}}
		& $CSUE$ & $RCSUE(\mathcal{R}_1)$ & $RCSUE(\mathcal{R}_2)$ \\
		\hline
		\rotatebox[origin=c]{90}{$T = 250$} &
		$\begin{bmatrix}\underset{(1.53)}{66.0} & \underset{(2.47)}{\hphantom{-}70.0} & \underset{(2.45)}{\hphantom{-}69.0} & \underset{(2.5)}{\hphantom{-}70.0} \\ \underset{(2.54)}{70.0} & \underset{(1.88)}{\hphantom{-}65.0} & \underset{(2.8)}{\hphantom{-}71.0} & \underset{(2.84)}{\hphantom{-}71.0} \\ \underset{(3.05)}{69.0} & \underset{(3.11)}{\hphantom{-}72.0} & \underset{(2.49)}{\hphantom{-}70.0} & \underset{(3.14)}{\hphantom{-}71.0} \\ \underset{(3.1)}{69.0} & \underset{(3.12)}{\hphantom{-}71.0} & \underset{(3.14)}{\hphantom{-}73.0} & \underset{(2.5)}{\hphantom{-}68.0} \\ \end{bmatrix}$

		&
$\begin{bmatrix}\underset{(1.21)}{66.0} & \textcolor{red}{\underset{(0.07)}{\hphantom{-}69.0}} & \textcolor{red}{\underset{(0.08)}{\hphantom{-}68.0}} & \textcolor{red}{\underset{(0.08)}{\hphantom{-}70.0}} \\ \underset{(1.48)}{72.0} & \underset{(1.21)}{\hphantom{-}64.0} & \textcolor{red}{\underset{(0.08)}{\hphantom{-}72.0}} & \textcolor{red}{\underset{(0.1)}{\hphantom{-}72.0}} \\ \underset{(1.76)}{71.0} & \underset{(1.67)}{\hphantom{-}71.0} & \underset{(1.77)}{\hphantom{-}69.0} & \underset{(2.54)}{\hphantom{-}69.0} \\ \underset{(1.77)}{71.0} & \underset{(1.69)}{\hphantom{-}71.0} & \underset{(2.56)}{\hphantom{-}70.0} & \underset{(1.75)}{\hphantom{-}67.0} \\ \end{bmatrix}$

		&
$\begin{bmatrix}\underset{(1.22)}{64.0} & \textcolor{red}{\underset{(0.21)}{\hphantom{-}72.0}} & \textcolor{red}{\underset{(0.24)}{\hphantom{-}69.0}} & \textcolor{red}{\underset{(0.24)}{\hphantom{-}67.0}} \\ \underset{(1.5)}{71.0} & \underset{(1.23)}{\hphantom{-}63.0} & \textcolor{red}{\underset{(0.26)}{\hphantom{-}73.0}} & \textcolor{red}{\underset{(0.29)}{\hphantom{-}62.0}} \\ \underset{(1.8)}{68.0} & \underset{(1.75)}{\hphantom{-}68.0} & \underset{(1.41)}{\hphantom{-}57.0} & \textcolor{red}{\underset{(1.37)}{\hphantom{-}18.0}} \\ \underset{(1.82)}{67.0} & \underset{(1.77)}{\hphantom{-}68.0} & \underset{(1.9)}{\hphantom{-}31.0} & \underset{(1.32)}{\hphantom{-}19.0} \\ \end{bmatrix}$

		\\
		[40pt]
		\rotatebox[origin=c]{90}{$T = 500$} &
$\begin{bmatrix}\underset{(1.0)}{66.0} & \underset{(1.73)}{\hphantom{-}70.0} & \underset{(1.74)}{\hphantom{-}72.0} & \underset{(1.74)}{\hphantom{-}70.0} \\ \underset{(1.78)}{70.0} & \underset{(1.27)}{\hphantom{-}69.0} & \underset{(2.01)}{\hphantom{-}72.0} & \underset{(2.02)}{\hphantom{-}69.0} \\ \underset{(2.2)}{72.0} & \underset{(2.24)}{\hphantom{-}72.0} & \underset{(1.73)}{\hphantom{-}70.0} & \underset{(2.24)}{\hphantom{-}72.0} \\ \underset{(2.19)}{71.0} & \underset{(2.24)}{\hphantom{-}70.0} & \underset{(2.22)}{\hphantom{-}70.0} & \underset{(1.74)}{\hphantom{-}69.0} \\ \end{bmatrix}$

		&
$\begin{bmatrix}\underset{(0.86)}{68.0} & \textcolor{red}{\underset{(0.04)}{\hphantom{-}68.0}} & \textcolor{red}{\underset{(0.04)}{\hphantom{-}69.0}} & \textcolor{red}{\underset{(0.04)}{\hphantom{-}67.0}} \\ \underset{(1.0)}{73.0} & \underset{(0.87)}{\hphantom{-}68.0} & \textcolor{red}{\underset{(0.05)}{\hphantom{-}70.0}} & \textcolor{red}{\underset{(0.05)}{\hphantom{-}70.0}} \\ \underset{(1.18)}{76.0} & \underset{(1.14)}{\hphantom{-}73.0} & \underset{(1.19)}{\hphantom{-}71.0} & \underset{(1.76)}{\hphantom{-}72.0} \\ \underset{(1.18)}{75.0} & \underset{(1.14)}{\hphantom{-}74.0} & \underset{(1.75)}{\hphantom{-}73.0} & \underset{(1.21)}{\hphantom{-}70.0} \\ \end{bmatrix}$

		&
$\begin{bmatrix}\underset{(0.87)}{67.0} & \textcolor{red}{\underset{(0.15)}{\hphantom{-}72.0}} & \textcolor{red}{\underset{(0.15)}{\hphantom{-}72.0}} & \textcolor{red}{\underset{(0.17)}{\hphantom{-}66.0}} \\ \underset{(1.03)}{70.0} & \underset{(0.88)}{\hphantom{-}66.0} & \textcolor{red}{\underset{(0.18)}{\hphantom{-}73.0}} & \textcolor{red}{\underset{(0.19)}{\hphantom{-}57.0}} \\ \underset{(1.24)}{74.0} & \underset{(1.2)}{\hphantom{-}70.0} & \underset{(1.04)}{\hphantom{-}56.0} & \textcolor{red}{\underset{(1.39)}{\hphantom{-}25.0}} \\ \underset{(1.24)}{73.0} & \underset{(1.21)}{\hphantom{-}70.0} & \underset{(1.51)}{\hphantom{-}39.0} & \underset{(1.09)}{\hphantom{-}25.0} \\ \end{bmatrix}$

		\\
		[40pt]
		\rotatebox[origin=c]{90}{$T = 1000$} &
$\begin{bmatrix}\underset{(0.67)}{68.0} & \underset{(1.18)}{\hphantom{-}70.0} & \underset{(1.18)}{\hphantom{-}71.0} & \underset{(1.18)}{\hphantom{-}70.0} \\ \underset{(1.21)}{72.0} & \underset{(0.85)}{\hphantom{-}67.0} & \underset{(1.35)}{\hphantom{-}70.0} & \underset{(1.35)}{\hphantom{-}72.0} \\ \underset{(1.49)}{70.0} & \underset{(1.5)}{\hphantom{-}70.0} & \underset{(1.17)}{\hphantom{-}70.0} & \underset{(1.51)}{\hphantom{-}70.0} \\ \underset{(1.5)}{72.0} & \underset{(1.51)}{\hphantom{-}72.0} & \underset{(1.5)}{\hphantom{-}72.0} & \underset{(1.17)}{\hphantom{-}68.0} \\ \end{bmatrix}$

		&
$\begin{bmatrix}\underset{(0.6)}{68.0} & \textcolor{red}{\underset{(0.02)}{\hphantom{-}66.0}} & \textcolor{red}{\underset{(0.02)}{\hphantom{-}67.0}} & \textcolor{red}{\underset{(0.02)}{\hphantom{-}67.0}} \\ \underset{(0.67)}{73.0} & \underset{(0.61)}{\hphantom{-}66.0} & \textcolor{red}{\underset{(0.02)}{\hphantom{-}66.0}} & \textcolor{red}{\underset{(0.02)}{\hphantom{-}68.0}} \\ \underset{(0.78)}{73.0} & \underset{(0.75)}{\hphantom{-}72.0} & \underset{(0.8)}{\hphantom{-}72.0} & \underset{(1.15)}{\hphantom{-}73.0} \\ \underset{(0.78)}{71.0} & \underset{(0.75)}{\hphantom{-}72.0} & \underset{(1.14)}{\hphantom{-}73.0} & \underset{(0.8)}{\hphantom{-}70.0} \\ \end{bmatrix}$

		&
$\begin{bmatrix}\underset{(0.61)}{67.0} & \textcolor{red}{\underset{(0.08)}{\hphantom{-}71.0}} & \textcolor{red}{\underset{(0.08)}{\hphantom{-}72.0}} & \textcolor{red}{\underset{(0.09)}{\hphantom{-}69.0}} \\ \underset{(0.7)}{71.0} & \underset{(0.62)}{\hphantom{-}65.0} & \textcolor{red}{\underset{(0.11)}{\hphantom{-}71.0}} & \textcolor{red}{\underset{(0.1)}{\hphantom{-}57.0}} \\ \underset{(0.83)}{70.0} & \underset{(0.81)}{\hphantom{-}69.0} & \underset{(0.75)}{\hphantom{-}62.0} & \textcolor{red}{\underset{(1.07)}{\hphantom{-}28.0}} \\ \underset{(0.83)}{70.0} & \underset{(0.8)}{\hphantom{-}68.0} & \underset{(1.09)}{\hphantom{-}44.0} & \underset{(0.8)}{\hphantom{-}32.0} \\ \end{bmatrix}$

		\\
	\end{tabular}
			\begin{minipage}{1\textwidth} %
		{   \footnotesize  
			\textit{Note:} 
			Monte Carlo simulation with $M = 2000$ replications for the SVAR in Equation (\ref{eq: MC}).
				 The table shows the coverage of $68$\% bootstrap confidence bands  and in parentheses the average width of the bands for 
		each estimated element. Penalized elements are highlighted in red.
			\par}
	\end{minipage}
\end{table}

The ridge estimator, by design, cannot outright disregard an invalid restriction; instead, evidence against the restriction only decreases the degree of shrinkage towards it. The ability to completely dismiss restrictions can be implemented using an additional step to select restrictions.
Table \ref{Table: Finite sample performance - select} shows the performance of the RCSUE($\mathcal{R}_2$) using an additional restriction selection step. Initially, the estimator estimates the RCSUE($\mathcal{R}_2$), then identifies all restrictions with estimated elements below an absolute threshold of $0.5$, and subsequently repeats the RCSUE estimation process using only the selected restrictions.
The results demonstrate a notable enhancement in the estimator's performance, attributable to its ability to entirely dismiss restrictions inconsistent with the data.

\begin{table}
	\caption{Finite sample performance using an additional restriction selection step }
	\label{Table: Finite sample performance - select}
	\centering
	\renewcommand{\arraystretch}{1}
	\begin{tabular}{c@{\hspace{2pt}} |@{\hspace{2pt}}c@{\hspace{2pt}} c@{\hspace{2pt}} c@{\hspace{2pt}}}
		&$T = 250$ & $T = 500$& $T = 1000$ \\
		\hline
		\rotatebox[origin=c]{90}{Bias and MSE} 
		&
	$\begin{bmatrix}\underset{(0.4)}{9.84} & \textcolor{red}{\underset{(0.31)}{0.0}} & \textcolor{red}{\underset{(0.21)}{\hphantom{-}0.03}} & \textcolor{red}{\underset{(0.44)}{0.0}} \\ \underset{(0.84)}{4.84} & \underset{(0.47)}{\hphantom{-}9.87} & \textcolor{red}{\underset{(0.52)}{-0.05}} & \textcolor{red}{\underset{(0.68)}{\hphantom{-}0.01}} \\ \underset{(1.02)}{4.97} & \underset{(1.19)}{\hphantom{-}4.99} & \underset{(1.13)}{\hphantom{-}10.07} & \textcolor{red}{\underset{(8.5)}{\hphantom{-}3.52}} \\ \underset{(1.07)}{4.9} & \underset{(1.3)}{\hphantom{-}4.9} & \underset{(4.22)}{\hphantom{-}5.81} & \underset{(4.21)}{\hphantom{-}8.95} \\ \end{bmatrix}$

		 &
		$\begin{bmatrix}\underset{(0.21)}{9.96} & \textcolor{red}{\underset{(0.16)}{\hphantom{-}0.06}} & \textcolor{red}{\underset{(0.08)}{-0.02}} & \textcolor{red}{\underset{(0.19)}{\hphantom{-}0.01}} \\ \underset{(0.3)}{4.98} & \underset{(0.21)}{\hphantom{-}9.95} & \textcolor{red}{\underset{(0.23)}{-0.0}} & \textcolor{red}{\underset{(0.19)}{-0.02}} \\ \underset{(0.44)}{5.01} & \underset{(0.51)}{\hphantom{-}5.03} & \underset{(0.44)}{\hphantom{-}9.92} & \textcolor{red}{\underset{(2.48)}{\hphantom{-}4.71}} \\ \underset{(0.54)}{5.0} & \underset{(0.5)}{\hphantom{-}5.0} & \underset{(1.42)}{\hphantom{-}5.1} & \underset{(1.06)}{\hphantom{-}9.78} \\ \end{bmatrix}$

		 &
		$\begin{bmatrix}\underset{(0.08)}{9.96} & \textcolor{red}{\underset{(0.03)}{-0.01}} & \textcolor{red}{\underset{(0.01)}{\hphantom{-}0.01}} & \textcolor{red}{\underset{(0.0)}{0.0}} \\ \underset{(0.1)}{5.01} & \underset{(0.1)}{\hphantom{-}9.96} & \textcolor{red}{\underset{(0.02)}{\hphantom{-}0.02}} & \textcolor{red}{\underset{(0.05)}{\hphantom{-}0.02}} \\ \underset{(0.14)}{4.96} & \underset{(0.14)}{\hphantom{-}4.95} & \underset{(0.13)}{\hphantom{-}9.99} & \textcolor{red}{\underset{(0.36)}{\hphantom{-}4.92}} \\ \underset{(0.11)}{4.95} & \underset{(0.15)}{\hphantom{-}4.95} & \underset{(0.28)}{\hphantom{-}5.06} & \underset{(0.2)}{\hphantom{-}9.97} \\ \end{bmatrix}$

		 \\ [40pt]\rotatebox[origin=c]{90}{\parbox{3.7cm}{  Width and coverage  \\ of confidence bands }}
		 
		 &
		 $\begin{bmatrix}\underset{(1.23)}{71.0} & \textcolor{red}{\underset{(0.19)}{\hphantom{-}64.0}} & \textcolor{red}{\underset{(0.12)}{\hphantom{-}74.0}} & \textcolor{red}{\underset{(0.22)}{\hphantom{-}70.0}} \\ \underset{(1.55)}{66.0} & \underset{(1.29)}{\hphantom{-}64.0} & \textcolor{red}{\underset{(0.2)}{\hphantom{-}73.0}} & \textcolor{red}{\underset{(0.32)}{\hphantom{-}61.0}} \\ \underset{(1.85)}{68.0} & \underset{(1.79)}{\hphantom{-}64.0} & \underset{(1.6)}{\hphantom{-}57.0} & \textcolor{red}{\underset{(1.69)}{\hphantom{-}50.0}} \\ \underset{(1.86)}{65.0} & \underset{(1.81)}{\hphantom{-}66.0} & \underset{(2.12)}{\hphantom{-}46.0} & \underset{(1.48)}{\hphantom{-}50.0} \\ \end{bmatrix}$

		 &
		 $\begin{bmatrix}\underset{(0.88)}{72.0} & \textcolor{red}{\underset{(0.14)}{\hphantom{-}64.0}} & \textcolor{red}{\underset{(0.08)}{\hphantom{-}71.0}} & \textcolor{red}{\underset{(0.12)}{\hphantom{-}68.0}} \\ \underset{(1.08)}{74.0} & \underset{(0.9)}{\hphantom{-}71.0} & \textcolor{red}{\underset{(0.14)}{\hphantom{-}70.0}} & \textcolor{red}{\underset{(0.13)}{\hphantom{-}64.0}} \\ \underset{(1.26)}{74.0} & \underset{(1.23)}{\hphantom{-}68.0} & \underset{(1.21)}{\hphantom{-}62.0} & \textcolor{red}{\underset{(1.55)}{\hphantom{-}64.0}} \\ \underset{(1.29)}{68.0} & \underset{(1.22)}{\hphantom{-}69.0} & \underset{(1.69)}{\hphantom{-}57.0} & \underset{(1.15)}{\hphantom{-}63.0} \\ \end{bmatrix}$

		 &
		 $\begin{bmatrix}\underset{(0.6)}{72.0} & \textcolor{red}{\underset{(0.03)}{\hphantom{-}62.0}} & \textcolor{red}{\underset{(0.02)}{\hphantom{-}61.0}} & \textcolor{red}{\underset{(0.01)}{\hphantom{-}64.0}} \\ \underset{(0.68)}{69.0} & \underset{(0.62)}{\hphantom{-}68.0} & \textcolor{red}{\underset{(0.02)}{\hphantom{-}69.0}} & \textcolor{red}{\underset{(0.03)}{\hphantom{-}69.0}} \\ \underset{(0.78)}{70.0} & \underset{(0.77)}{\hphantom{-}70.0} & \underset{(0.8)}{\hphantom{-}75.0} & \textcolor{red}{\underset{(1.13)}{\hphantom{-}71.0}} \\ \underset{(0.77)}{73.0} & \underset{(0.77)}{\hphantom{-}74.0} & \underset{(1.16)}{\hphantom{-}75.0} & \underset{(0.8)}{\hphantom{-}70.0} \\ \end{bmatrix}$

		 \\   [40pt]

	\end{tabular}
	
				\begin{minipage}{1\textwidth} %
		{   \footnotesize  
			\textit{Note:} 
			Monte Carlo simulation with $M = 2000$ replications for the SVAR in Equation (\ref{eq: MC}).
			The estimator considered in the simulation uses a two-step procedure to unselect invalid restrictions. In the first step, the depicted estimator   is equal to the RCSUE($\mathcal{R}_2$). In the second step, all restrictions with estimated elements below a threshold of $0.5$ in absolute value are selected and used to repeat the RCSUE estimation process with the selected restrictions.
			The table
			shows the average, $1/M 
			\sum_{m=1}^{M}
			\hat{b}^{m}_{ij} $ and in parentheses the mean squared error $1/M 
			\sum_{m=1}^{M}
			(\hat{b}^{m}_{ij} - b_ij)^2 $  for 
			each estimated element, as well as the coverage  and in parentheses the average width of   $68$\% bootstrap confidence bands  for 
			each estimated element. Penalized elements are highlighted in red.
			\par}
	\end{minipage}
\end{table}

The simulation results provide several insights for practical applications. Firstly, if credible, economically grounded restrictions are available, incorporating them into the estimation process is recommended, as they can significantly enhance the performance of the statistically identified estimator.
Second, the simulations highlight that the estimator's ability to detect and disregard invalid restrictions depends on the sample size. Therefore, in situations where specific restrictions are dubious, especially in applications with limited sample sizes, it may be more prudent to refrain from imposing such restrictions. Alternatively, an additional restriction selection step could be used to refine the set of restrictions based on their consistency with the data.

\section{Application - The oil and stock market interaction}
\label{sec: Application} 
This section analyzes the effects of different approaches to incorporate   short-run restrictions on the oil and stock market interaction. 
The traditional recursive SVAR approach suggests that the stock market does not provide additional information on the price of oil. 
In contrast, the ridge estimator reveals that the stock market and oil prices cannot be ordered recusively and that information shocks influencing the stock market play are important drivers of oil price fluctuations.

\subsection{Specification and estimators}
The  SVAR  uses  monthly data from  January $1974$ to August $2023$ with 
\begin{align} 
\label{eq: application svar}   
\begin{bmatrix}
q_t
\\
y_t 
\\
p_t
\\
s_t 
\end{bmatrix}
=
\alpha  
+
\sum_{i=1}^{12}
A_i
\begin{bmatrix}
q_{t-i}
\\
y_{t-i} 
\\
p_{t-i}
\\
s_{t-i} 
\end{bmatrix}
+
\begin{bmatrix}
b_{11} & b_{12}  &  b_{13}  & b_{14} \\
b_{21} & b_{22} & b_{23}   & b_{24}   \\
b_{31}  & b_{32} & b_{33}    &  b_{34}    \\ 
b_{41} & b_{42} & b_{43}   & b_{44} 
\end{bmatrix}
\begin{bmatrix}
\varepsilon_{S,t} 
\\
\varepsilon_{Y,t}  
\\
\varepsilon_{D,t} 
\\
\varepsilon_{SM,t} 
\end{bmatrix},
\end{align} 
where $q_t$ is $100$ times the log   of world crude oil production, $y_t$ is $100$ times the log   of  global industrial production, $p_t$ is  $100$ times the log   of the real oil price,   and $s_t$ is $100$ times the log of a monthly U.S. stock price index. The data sources can be found in Appendix \ref{appendix: sec: Application}.

\cite{kilian2009impact} estimate a similar oil and stock market SVAR and propose to identify four shocks using a recursive order.\footnote{
 	The model analyzed  in \cite{kilian2009impact} uses a slightly different specification. Specifically, the authors use an economic activity index based on shipping costs. However, as noted by \cite{baumeister2022energy}, the shipping index may not always be a reliable indicator of changes in global economic activity. Therefore,   I follow the approach taken by \cite{baumeister2019structural} and   use a conventional measure of economic activity based on industrial production.
 
}
In the recursive SVAR, oil supply shocks $\varepsilon_{S,t} $ can simultaneously affect all variables, economic activity shocks $\varepsilon_{Y,t} $ cannot simultaneously affect oil supply, oil-specific demand shocks $\varepsilon_{D,t} $ cannot simultaneously affect oil supply and economic activity, and stock market information shocks $\varepsilon_{SM,t} $ cannot simultaneously affect oil supply, economic activity, and the oil price. 

Recursive restrictions have two major limitations. 
First, they imply that oil supply cannot respond simultaneously to demand shocks.
Secondly, the reduced form price shocks that cannot be explained by supply and economic activity shocks are, by construction, identified as oil-specific demand shocks. However, if the oil price responds immediately to information shocks affecting stock prices, these information shocks would end up in the oil-specific demand shock of the recursive model. 
The former issue regarding the response of oil supply to non-supply shocks received a lot of attention in the literature, see, e.g. \cite{kilian2012agnostic}, \cite{kilian2014role}, \cite{baumeister2019structural}, \cite{caldara2019oil}, and \cite{braun2021importance}, while the latter issue on the impact of stock market information shocks on the oil price received little attention.

In contrast to the recursive estimator,  the proposed ridge estimator does not use restrictions to ensure identification. As a result, it does not require to impose the two questionable assumptions. 
Instead, the ridge estimator employs the following short-run restrictions:
\begin{enumerate}
\item $b_{12}=b_{14}=0$: Oil supply can respond to oil demand and supply shocks.
\item $b_{21}=b_{23}=b_{24}=0$: Industrial production can respond  to economic activity shocks.\footnote{
	The restriction $b_{21}=0$, which is not utilized in the recursive model, is based on the assumption that economic activity exhibits sluggish behavior and does not concurrently respond to oil supply shocks. Thus, it relies on the same rationale underpinning the zero response of economic activity to oil-specific demand and stock market information shocks employed in the recursive SVAR.
}
\end{enumerate}

Labeling of the ridge estimator is determined by the solution of the recursive SVAR. Specifically, I estimate the recursive model using the Cholesky decomposition and  use the resulting estimated simultaneous interaction to construct a set of unique-sign permutation representatives centered at the recursive solution, see Section \ref{subsec: Labeling}. This set restricts admissible $B$ matrices and determines the labeling: within the set and in line with the recursive labeling, the first shock represents an oil supply shock, the second shock is an economic activity shock, the third shock is an oil-specific demand shock, and the last shock is the stock market information shock. 
Furthermore, the tuning parameter $\lambda$ required for the ridge estimator is determined similarly to the previous section, using repeated cross-validation with two folds and $50$ repetitions. 
Lastly, the non-Gaussianity measured by the skewness, excess kurtosis, and Jarque-Bera test of the reduced form and estimated structural form shocks are shown in the appendix. The results indicate that three out of four shocks are left skewed with heavy tails, which is sufficient to ensure identification based on Proposition \ref{proposition: 1}.

\subsection{Empirical results}

\begin{figure}[h!] 
	\centering
	\caption{Impulse responses. Red: Recursive estimator. Blue: Ridge estimator.} 
	\includegraphics[width=0.80\textwidth]{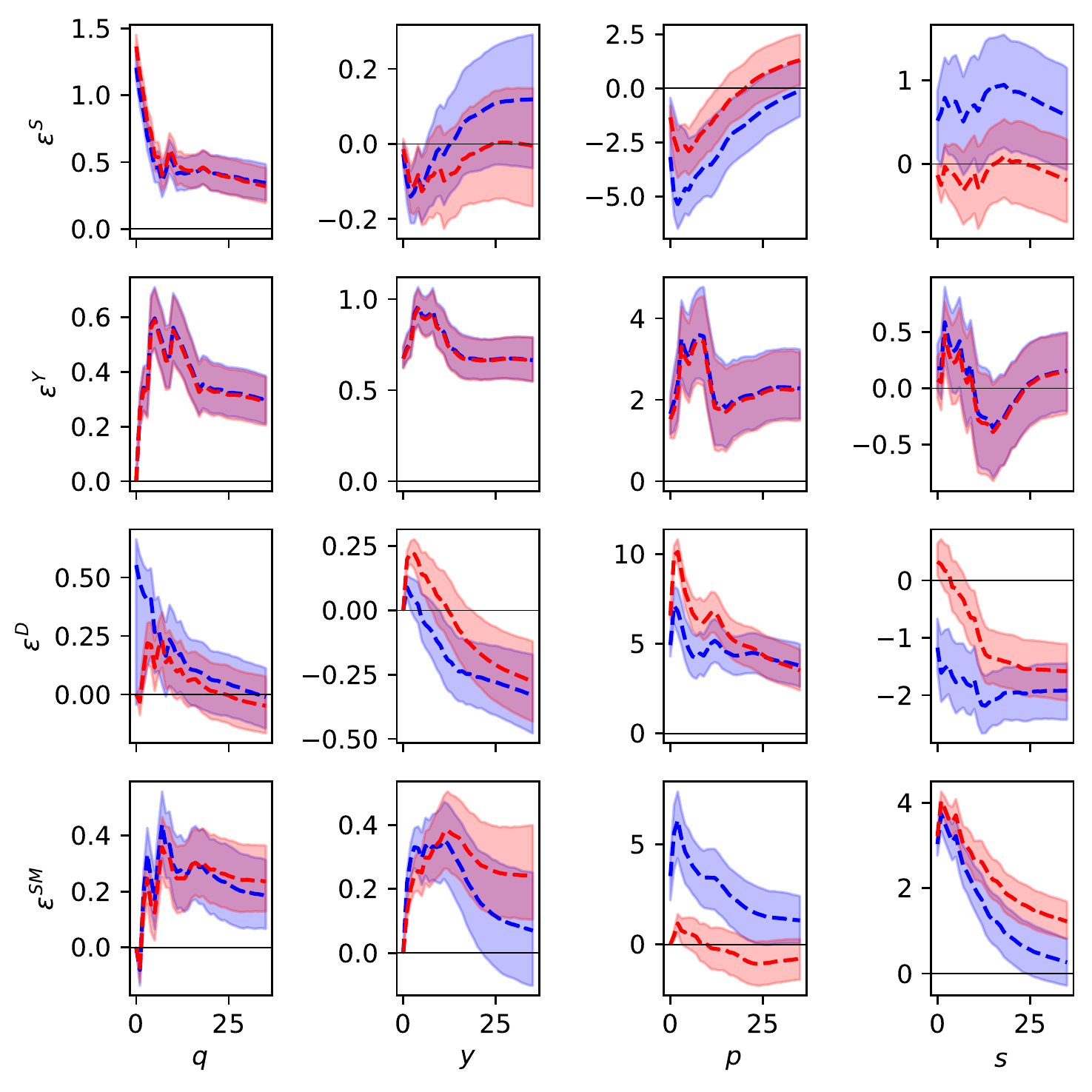}
	\label{fig: IRF} 
	\begin{minipage}{1\textwidth} %
		{   \footnotesize  
			\textit{Note:} Impulse responses to one standard deviation shocks with $68$\% bootstrap confidence bands.
			\par}
	\end{minipage}
\end{figure}

Figure \ref{fig: IRF} displays the impulse responses generated by the recursive estimator and the ridge estimator. 
Although both estimators arrive at similar conclusions on the effects of economic activity shocks, there are notable differences in their findings regarding responses to oil supply, oil demand, and stock market information shocks.

To begin, both estimators find an immediate increase in oil supply and a decrease in oil price in response to the oil supply shock.
However, the recursive estimator suggests a smaller response of the oil price and no significant reactions in economic activity and stock prices.
In contrast, the ridge estimator shows a larger initial oil price response, a more positive (though not statistically significant) long-run response of economic activity, and a positive and significant response of stock prices after the supply shock.

Secondly, both estimators find a positive response of the oil price to oil demand shocks.
In the recursive model, the simultaneous response of oil production to demand shocks is zero by construction. The ridge estimator indicates  a positive short-run response of oil production to demand shocks, however, the immediate response is not significant.
Furthermore, both estimators find a negative long-term impact on economic activity and stock prices in response to oil demand shocks. However, the recursive model suggests a significant positive response in economic activity in the medium term and a positive reaction in stock prices in the short term.  In contrast, the ridge estimator suggests an earlier negative response of economic activity and an immediate negative response of stock prices to the oil price increase caused by oil demand shocks.
 
Third, both estimators show that the stock market information shock is followed by a subsequent expansion of economic activity and oil production, along with an immediate positive response of stock prices.
Nevertheless, the response of the oil price to the stock market information shock differs substantially between both estimators. In the recursive model,  the initial response of the oil price to the information shocks is restricted to zero, and the model implies that the  shock has no noteworthy impact on oil prices in the medium and long term. In contrast, the ridge estimator does not impose such constraints. In fact, the data provide evidence against the zero restriction and suggest a significant positive response of the oil price to the stock market information shock.

Imposing a restriction that confines the oil price response to the stock market information shock to zero has significant implications for shocks characterized as oil-specific demand and stock market information shocks within the recursive model.
In the recursive model, a shock that simultaneously affects the  oil price residual unexplained by supply and economic activity shocks is by construction identified as an oil-specific demand shock.  
Consequently, information shocks about future economic activity and, henceforth, future oil demand,  which immediately affect the oil and stock market in the same direction, become subsumed within the category of oil-specific demand shocks in the recursive model.
Similarly, oil-specific demand shocks, i.e. those not originating from economic activity shocks that immediately impact both the oil and stock markets in opposing directions, also end up within the oil-specific demand shock of the recursive model.
Therefore, the recursive model  identifies the oil-specific demand shock as a mixture of oil demand and information shocks. Given that both shocks have opposing impacts on the stock price, this mixture of shocks results in an immediate stock market response that nearly offsets, leading to the conclusion that the stock market exhibits minimal immediate responsiveness to the oil-specific demand shock.
Consequently, the stock market information shock is also a mixture of oil-specific demand and information shocks in the recursive model, which leads to the conclusion that information shocks driving the stock market have almost no effect on the oil price.

Table \ref{table: A} displays the estimated simultaneous interaction transformed to an $A$-type SVAR comparable to \cite{baumeister2019structural}, \cite{caldara2019oil}, or \cite{braun2021importance} with
\begin{align}
	\label{eq: oil supply}
			q_t 
	&=  
	  \alpha_{qy}	y_t 
	+ \alpha_{qp} 	p_t 
	+ \alpha_{qs}    s_t   
	+\varepsilon^S_t
	\\
	\label{eq: econ act}
	y_t 
	&=  \rho_{yq}  q_t 
	+  \rho_{yp}   p_t 
	+  \rho_{ys}    s_t 
	+\varepsilon^Y_t 
	\\
	\label{eq: oil demand}
	q_t 
	&=
	   \beta_{qy}  y_t 
	+ \beta_{qp}    p_t 
  	+\beta_{qs}  s_t
	+\varepsilon^D_t
	\\
	\label{eq: stock market}
	s_t 
	&=    \gamma_{sq}   q_t 
+  \gamma_{sy}  y_t 
	  +  \gamma_{sp}  p_t 
	+\varepsilon^{SM}_t,  
\end{align} 
where Equation (\ref{eq: oil supply}) models oil supply, Equation (\ref{eq: econ act}) models economic activity, Equation (\ref{eq: oil demand}) models oil demand, and Equation (\ref{eq: stock market}) models the stock market.
The oil supply elasticity is equal to zero in the recursive model by construction, whereas the ridge estimator indicates an elasticity of $0.088$, close to the location of the prior used in \cite{baumeister2019structural}.
The oil demand elasticity in the recursive model is close to minus one, whereas the ridge estimator yields a demand elasticity of $-0.325$, which is close to the median posterior demand elasticity in \cite{baumeister2019structural}.
Turning to the income elasticity of oil demand, the ridge estimator suggests a value of approximately $0.7$, which is equal to the location of the corresponding prior in \cite{baumeister2019structural}, while the recursive estimator indicates a significantly higher value.
Moreover, the recursive estimator finds that the oil demand response to the stock market and the stock market response to economic activity do not differ significantly from zero, whereas the ridge estimator indicates a positive response of oil demand to the stock market, a positive stock market response to economic activity.
Lastly, the recursive estimator suggests a positive effect of the oil price on the stock market, whereas the ridge estimator suggests the opposite. 
\begin{table}[h] 
	\centering
	\caption{Summary of simultaneous interaction } 
	\begin{tabular}{c c|  c c   }
	& & Recursive estimator & Ridge estimator \\ 
	 \hline
	 
	$\alpha_{qp}$ &Oil supply elasticity  & $\underset{(-0.0 / -0.0)}{-0.0} $ & $\underset{(-0.01 / 0.11)}{0.088}$ 
	 \\
	
	$\beta_{qy}$	 &Income elasticity of oil demand  & $ \underset{(1.5 / 4.06)}{2.353}$ & $\underset{(0.18 / 1.19)}{0.702}$ 
	\\
	
	$\beta_{qp}$	 &Oil demand elasticity  & $\underset{(-2.0 / -0.6)}{-1.039}$ & $\underset{(-0.42 / -0.23)}{-0.325} $ 
	\\
	
	$\beta_{qs}$	&Effect of $s_t$ on oil demand  & $ \underset{(-0.0 / -0.0)}{-0.0}$ & $ \underset{(0.04 / 0.54)}{0.367}$ 
	\\
	
	$\gamma_{sy}$	& Effect of $y_t$ on stock prices & $\underset{(-0.27 / 0.29)}{0.0}$ & $\underset{(0.39 / 1.29)}{0.81}$ 
	\\
	
$\gamma_{sp}$	&	Effect of $p_t$ on stock prices  & $ \underset{(0.01 / 0.1)}{0.05}$ & $\underset{(-0.34 / -0.12)}{-0.222}$ 
	\\
	\end{tabular}     
	\label{table: A}
	\begin{minipage}{1\textwidth} %
		{   \footnotesize  
			\textit{Note:} The table shows the estimated  simultaneous interaction transformed to an $A$-type SVAR  together with $68$\% bootstrap confidence bands.
			\par}
	\end{minipage}
\end{table}

Table \ref{table: fevd} provides insights into the effect of recursiveness restrictions on the forecast error variance decomposition.
In the recursive model,  oil-specific demand shocks are the primary driver of the oil price, explaining more than $80$\% of the variation.
Conversely, the ridge estimator unveils a less one-sided picture, indicating that oil-specific demand shocks, although still the primary influence, explain a reduced share of only $36$\% of the variance. The remaining variance in oil prices is attributed to oil supply and stock market information shocks, both contributing approximately $25$\% to the overall variation. 
Moreover,  disentangling oil-specific demand and stock market information shocks based on their interdependence, rather than relying on  restrictions, also has a substantial impact on  the variation of stock prices explained by oil-specific demand shock.  In the recursive model, oil-specific demand only explains $2$\% of the variation in stock prices, while the ridge estimator finds a more pronounced influence of oil-specific demand shocks.

\begin{table}[h] 
	\centering
	\caption{One year ahead forecast error variance decomposition } 
	\begin{tabular}{ c|  c c c  c c   c|  c  c c c       }
		\multicolumn{5}{c}{Recursive estimator} 
		&  $\quad$&
		\multicolumn{5}{c}{Ridge estimator}  
		\\ 
		 &  $\varepsilon_{S}$ & $\varepsilon_{Y}$  & $\varepsilon_{D}$   & $\varepsilon_{SM}$   
		&  $\quad$&
		 &  $\varepsilon_{S}$ & $\varepsilon_{Y}$  & $\varepsilon_{D}$   & $\varepsilon_{SM}$   
		\\ \cline{1-5} \cline{7-11}

		$q$&  $\underset{0.6/0.74}{0.69}$ & $\underset{0.15/0.28}{0.22}$ & $\underset{0.01/0.06}{0.03}$ & $\underset{0.03/0.11}{0.06}$
		&&
		$q$&  $\underset{0.47/0.68}{0.56}$ & $\underset{0.16/0.28}{0.23}$ & $\underset{0.03/0.19}{0.13}$ & $\underset{0.05/0.14}{0.09}$ 
		\\ 
		
		$y$&  $\underset{0.0/0.04}{0.01}$ & $\underset{0.78/0.89}{0.86}$ & $\underset{0.02/0.05}{0.02}$ & $\underset{0.06/0.16}{0.1}$
		&&
		$y$& $\underset{0.0/0.04}{0.01}$ & $\underset{0.78/0.89}{0.87}$ & $\underset{0.01/0.04}{0.01}$ & $\underset{0.06/0.17}{0.11}$
		\\ 
		
		$p$&  $\underset{0.03/0.15}{0.08}$ & $\underset{0.05/0.16}{0.11}$ & $\underset{0.7/0.87}{0.81}$ & $\underset{0.0/0.02}{0.0}$
		&&
		$p$& $\underset{0.03/0.36}{0.27}$ & $\underset{0.05/0.17}{0.12}$ & $\underset{0.26/0.57}{0.36}$ & $\underset{0.12/0.39}{0.25}$
		\\ 
		
		$s$&  $\underset{0.0/0.03}{0.0}$ & $\underset{0.01/0.03}{0.0}$ & $\underset{0.01/0.06}{0.02}$ & $\underset{0.89/0.96}{0.98}$
		&&
		$s$& $\underset{0.01/0.08}{0.04}$ & $\underset{0.01/0.04}{0.01}$ & $\underset{0.12/0.38}{0.24}$ & $\underset{0.53/0.84}{0.71}$

	\end{tabular}     
	\label{table: fevd}
	\begin{minipage}{1\textwidth} %
		{   \footnotesize  
			\textit{Note:} The table shows the estimated contribution of each shock to the forecast error variance decomposition  at $12$  month horizon together with $68$\% bootstrap confidence bands.
			\par}
	\end{minipage}
\end{table}

\begin{figure}[h!] 
	\centering
	\caption{Historical decomposition of the real oil price   }
	\includegraphics[width=0.8\textwidth]{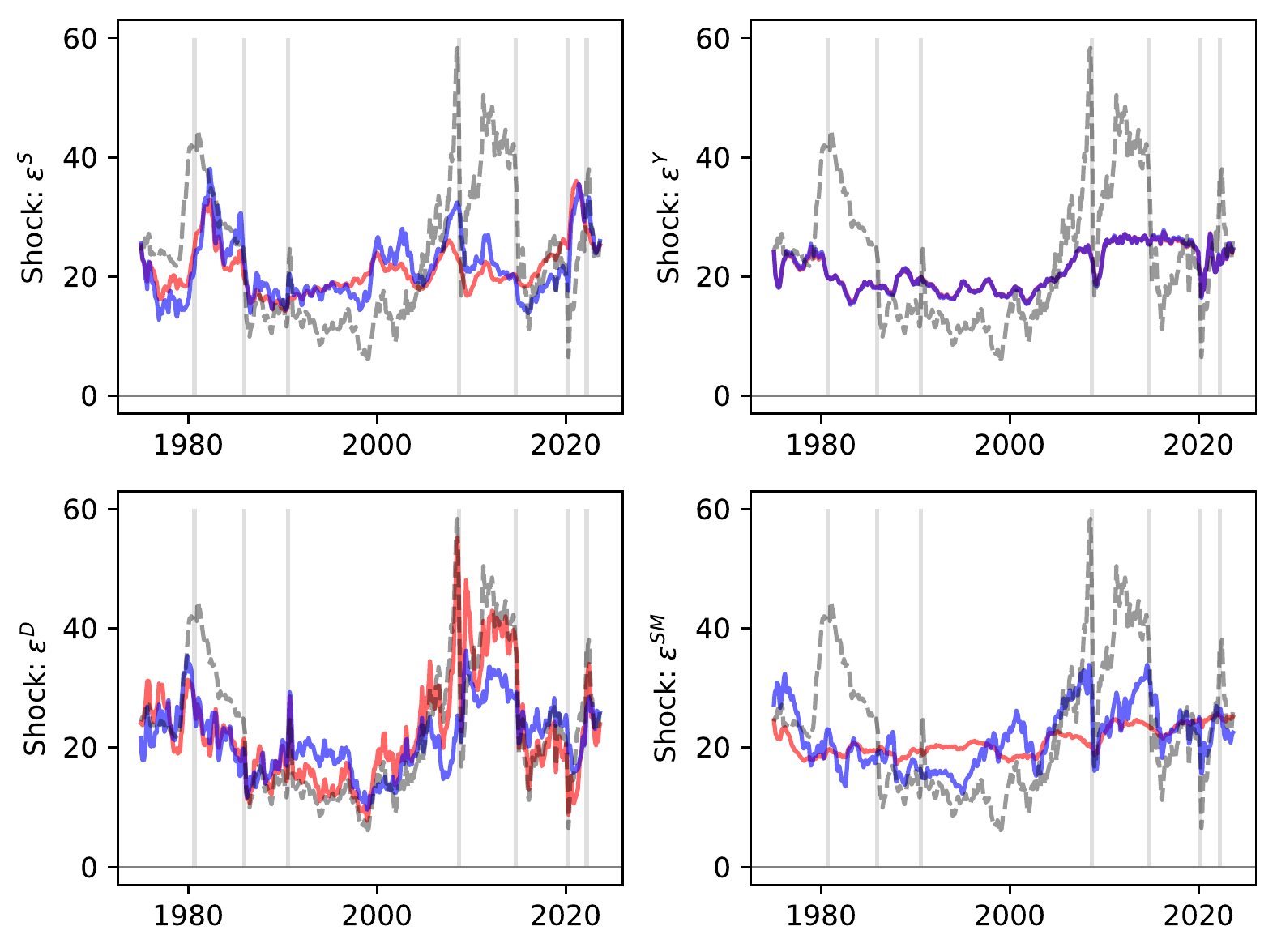}
	\label{fig: ImpactOnrealoilprice}
	\begin{minipage}{1\textwidth} %
		{   \footnotesize  
			\textit{Note:}  Red [blue] shows the decomposition for the recursive [ridge] estimator and grey shows the historical real oil price.
			The vertical bars indicate the following events:  
			Iran Iraq War ($1980$), 
			collapse of OPEC ($1985$), 
			Persian Gulf War ($1990$),  
			the collapse of  Lehman Brothers ($2008$), 
			the oil price  decline in mid $2014$,
			the beginning of the COVID-19 pandemic ($2019$),
			and the invasion of  Ukraine ($2022$). 
			\par}
	\end{minipage} 
\end{figure}

Figure \ref{fig: ImpactOnrealoilprice} shows the historical decomposition of the oil price and sheds light on the importance of supply, demand, and information shocks in different periods.  
In the recursive SVAR, oil-specific demand shocks are the primary driver of the oil price. This pattern is consistent during  events such as the collapse of OPEC in $1985$, the Persian Gulf War in $1990$, the oil price surge in $2007-2008$, the subsequent decline and recovery in oil prices following the collapse of Lehman Brothers in $2008$, the oil price downturn in $2014-2016$, the oil price fluctuations at the start of the COVID-$19$ pandemic, and the recent oil price increase in $2022$ following the Ukraine invasion.
In contrast, the ridge estimator provides a more nuanced picture. 
First, it suggests that supply shocks played a more prominent role during the collapse of OPEC in $1985$, the Persian Gulf War in $1990$, and the upswing of oil prices in $2007-2008$. 
Second, it suggests that information shocks extracted from stock prices contributed largely to the increase in oil prices before $2008$, the decrease in the oil price following the collapse of Lehman Brothers in $2008$, the decrease in the oil price in $2014-2016$, and also to the decrease and recovery of the oil price at the beginning of the COVID-$19$ pandemic.
 
Overall, the analysis suggests an immediate response of the oil price to stock market information shocks. In addition, these information shocks play a significant role in explaining oil price fluctuations. In general, the results highlight the importance of incorporating information from the stock market into the analysis of oil price movements.

\section{Conclusion}
\label{sec: Conclusion}

 	Economically motivated short-run restrictions have been an integral part of identifying SVAR models in numerous applications since \cite{sims1980macroeconomics}.
 	Despite their popularity in applied work, the disadvantageous of restriction based identification methods are well known: incorrect restrictions lead to biased estimates.  
 	Novel identification approaches that rely on stochastic properties of the shocks no longer require short-run restrictions for identification and applications of these approaches oftentimes completely disregard available economically motivated short-run restrictions or only conduct hypothesis tests of restrictions.
 	
 	This study proposes a new approach to combine the rich literature on short-run restrictions with the recent statistical identification literature. By incorporating restrictions via a shrinkage approach alongside non-Gaussian based identification, the estimator combines the strengths of both methodologies. 
 	Simulations show how valid restrictions improve the accuracy of the statistically identified SVAR estimator and that the estimator can detect and reduce the impact of incorrect restrictions as the sample size increases. Therefore, the study underscores the enduring value of over four decades of research into plausible short-run restrictions within the statistical identification framework, where such restrictions are no longer required for identification.

\bibliographystyle{Chicago} 
\bibliography{Bibliography-MM-MC}

\newpage
\appendix

\section{Proof}
\begin{proof}[Proof of Proposition 1:]

	The innovations $e(B)_t$ are equal to $e(B)_t =B^{-1}u_t = B^{-1} B_0 \varepsilon_t$. Define $Q=B^{-1} B_0 $ such that $e_t = Q \varepsilon_t$. If  $e(B)_t$ satisfies the variance and covariance moment conditions, it follows that $Q$ is diagonal.
		W.l.o.g. let  $\tilde{\varepsilon}_{1t}:=[ \varepsilon_{1t},..., \varepsilon_{n_1t}]'$,
	$\tilde{\varepsilon}_{2t}:=[ \varepsilon_{(n_1+1)t},..., \varepsilon_{n_2 t}]'$, 
	and   $\tilde{\varepsilon}_{3t}:=[ \varepsilon_{(n_2+1)t},..., \varepsilon_{n t}]'$.
	
	The following paragraph introduces the  notation used in \cite{mesters2022non}. 
	Define the $r$-th order cumulant tensor $h_r(X) \in S^r( \mathbb{R}^n)$ with entries $h_r(X)_{i_1,...,i_n} = cum(X_{i_1},...,X_{i_n}) $.
	It holds that $h_r(Q \varepsilon_t) = Q \cdot h_r( \varepsilon_t)  $ and the homogeneous polynomial associated with a tensor $T\in S^r( \mathbb{R}^n)$ in variables $x=(x_1,...,x_n)$ is defined as 
	\begin{align}
		f_T(x) = \sum_{i_1=1}^{n}...\sum_{i_r=1}^{n} T_{i_1,...,i_r} x_{i_1}.... x_{i_r}.
	\end{align}

	A tensor $T \in S^r( \mathbb{R}^n)$  is diagonal if all non-diagonal elements are equal to zero, i.e. $T_{i_1,...,i_r}=0 $ if $\exists l,k \in (1,...,r) $  with $l\neq k$ and $i_l \neq i_k$.
	It is easy to verify that mutually mean independent shocks imply that $h_r(\varepsilon_t) $ is diagonal for $r=3$ and mutually independent shocks imply that $h_r(\varepsilon_t) $ is diagonal for $r=4$. 
	If a tensor $T \in S^r( \mathbb{R}^n)$ is diagonal, $f_T(x)$ simplifies to $f_T(x)= \sum_{i=1}^{n} T_{i,...,i} x_i^r$ and $\frac{\partial^2 }{\partial x_j^2} f_T(x) = T_{j,...,j} x_j^{r-2}$.

	\textit{Statement 1: } \\
	Let $T=h_r(\varepsilon_t)$ with $r=3$.
	If $e(B)_t$ satisfies the third order moment conditions, it follows that the tensor $QT$ is diagonal. Let $Q=\begin{bmatrix}
		Q_{11} & Q_{12}\\
		Q_{21} & Q_{22}
	\end{bmatrix} $ with 
	a $n_1 \times n_1$ dimensional matrix $Q_{11}$,
	a $n-n_1 \times n_1$ dimensional matrix $Q_{21}$,
	a $n_1 \times n-n_1$ dimensional matrix $Q_{12}$, and
	a $n-n_1 \times n-n_1$ dimensional matrix $Q_{22}$.

	By Lemma 5.2 in \cite{mesters2022non} it follows that $QT$ is diagonal if and only if $Q_{ij} \frac{\partial^2 }{\partial x_j^2} f_T(Q'x) = D_{ii}Q_{ij}$.

	First, show that $Q$ is block diagonal with $Q_{12}=Q_{21}'=0$.
	Assume that there exists a row $i$ with $q_{ij} \neq 0$ and $q_{ik}\neq 0$ with $j\leq n_1$ and $k>n_1$.
	Therefore, analogous to the proof of Theorem 5.3 in  \cite{mesters2022non} it follows that $T_{j,...,j} x_j^{r-2} = T_{k,...,k} x_k^{r-2}$. For  $k>n_1$ it holds that $T_{k,...,k}=0$ and thus $T_{j,...,j} x_j^{r-2} = 0$ which is a contradiction since $T_{j,...,j} \neq 0 $ for $j\leq n_1$.
	
	Second, show that $Q_{11}$ contains exactly one non-zero element per row. 
	Assume that there exists a row $i$ with $q_{ij} \neq 0$ and $q_{ik}\neq 0$ with $j\leq n_1$ and $k\leq n_1$ and $j \neq k$.
	Therefore,  analogous to the proof of Theorem 5.3 in  \cite{mesters2022non}  it follows that $T_{j,...,j} x_j^{r-2} = T_{k,...,k} x_k^{r-2}$ which only holds if $T_{j,...,j}= T_{k,...,k}=0$ which  is a contradiction since $T_{j,...,j}, T_{k,...,k} \neq 0 $ for $j,k\leq n_1$.
	
	Using $Q=B^{-1} B_0 $  implies that   $\tilde{\varepsilon}_{1t}$  and the corresponding columns of $B_0$ are identified up to sign and permutation.  
	
	\textit{Statement 2: }\\
	Identification of the skewed shocks  $\tilde{\varepsilon}_{1t}$ follows from the first statement, thus, w.l.o.g. let $\tilde{\varepsilon}_{1t}$ be empty.
	Let $T=h_r(\varepsilon_t)$ with $r=4$.
	If $e(B)_t$ satisfies the fourth order moment conditions, it follows that the tensor $QT$ is reflectionally invariant, i.e. the only non-zero entries of $QT$ are those where each index appears an even number of times. Let $Q=\begin{bmatrix}
		Q_{11} & Q_{12}\\
		Q_{21} & Q_{22}
	\end{bmatrix} $ with 
	a $n_2 \times n_2$ dimensional matrix $Q_{11}$,
	a $n-n_2 \times n_2$ dimensional matrix $Q_{21}$,
	a $n_2 \times n-n_2$ dimensional matrix $Q_{12}$, and
	a $n-n_2 \times n-n_2$ dimensional matrix $Q_{22}$.
	
	First, similar to Lemma 5.12 in \cite{mertens2014reconciliation}, it holds  that for the diagonal tensor $T$  with $T_{iiii} \neq 0 $ for $i=1,...,n_2$ and  $T_{iiii} = 0 $ for $i=n_2+1,...,n$   
	\begin{align}
		f_T(x) = f_{AT}(x) 
		\implies
		A =  \begin{bmatrix}
			A_{11} & A_{12}\\
			A_{21} & A_{22}
		\end{bmatrix},
	\end{align}
	where all sub-matrices $A_{ij}$ have the same dimensions as the as the sub-matrices in $Q$  and $A_{11} = \pm I$ and $A_{21}=0$.
	The statement follows with
	\begin{align}
		f_T(x) = \sum_{i=1}^{n} T_{iiii} x_i^4 =  \sum_{i=1}^{n_2} T_{iiii} x_i^4 
	\end{align}
	and
	\begin{align}
		f_{AT}(x) = 	f_{T}(A'x)  = \sum_{i=1}^{n} T_{iiii} (A'x)_i^4 =  \sum_{i=1}^{n_2} T_{iiii} (A'x)_i^4.
	\end{align}
	Therefore, for all $x$, $A$ needs to satisfy $x_i = \pm (A'x)_i$ which implies $a_{ii}= \pm 1$ and $a_{ji}=0$ for $j \neq i$ for all $i=1,...,n_1$, where $a_{kl}$ denotes the element of $A$ in row $k$ and column $l$.

	Second, analogously to the proof of Theorem 5.10 in \cite{mertens2014reconciliation}, a reflectionally invariant tensor $QT$  implies 
	$
	f_{QT}(x) = f_{QT}(Dx)
	$
	for every diagonal matrix $D$ with $\pm 1$ on the diagonal, see Lemma 5.9 in \cite{mertens2014reconciliation}.
	Therefore, it holds that $f_{T}(Q'x) = f_{ T}(Q'Dx)$ and thus $f_{T}( x) = f_{ T}(Q'DQx)$.
	Define $A':= Q'DQ$ and thus, it holds that  $f_{ T}( x) = f_{AT}( x)$ which implies 
	\begin{align}
		\pm I = Q_{11}'D_{11}Q_{11} \quad \text{and} \quad 0 =  Q_{21}'D_{22}Q_{11}.
	\end{align}
	Applying the same argument used in the proof of Theorem 5.10 in yields $Q_{11}$ is a sign permutation matrix and $0 =  Q_{21}$ immediately follows from $0 =  Q_{21}'D_{22}Q_{11}$. 
	Orthogonality of $Q$ implies $Q_{12}=0$ and $Q_{22}$ is orthogonal.
	
	Using $Q=B^{-1} B_0 $  implies that   $\tilde{\varepsilon}_{2t}$  and the corresponding columns of $B_0$ are identified up to sign and permutation.  
	

	\textit{Statement 3: }\\
	 Follows directly from orthogonality and block diagonality of $Q$.

\end{proof}

\section{Appendix - Cross-Validation}
\label{appendix: sec: CV}
This section briefly explains the implementation of the cross-validation.

First, to prevent that  CV estimates  converge  to local minima,  each optimization problem is solved using two different starting values for $B$.
The first starting value is always equal to the unpenalized non-Gaussian estimator using the whole sample.
The second starting value depends on the tuning parameter.
The CV starts with the largest $\lambda$ value and subsequently reduces the tuning parameter to the smallest $\lambda$ value.  
For the largest  tuning parameter, the second starting  value is equal to the estimator obtained using the estimation fold which imposes the restrictions as binding constraints.
Afterwards, the CV solves the optimization problem using  $\lambda$ for both starting values and chooses the solution which leads to the smallest loss in the estimation fold. 
For  each subsequently smaller $\lambda$ lambda value, the second starting  value is equal to the chosen estimator of the last lambda value.

Moreover, the CV involves solving multiple optimization problems and sometimes the optimization does not converge to the desired minima. Specifically, some   solutions correspond  to shocks with extremely large or small variances. For a given estimation, this can easily be detected, however, in the CV an automatic rule to prevent these solutions is required.  
To avoid these solutions in the CV, each estimator in the CV contains an additional variance penalty term. Specifically, the Ridge SVAR-CSUE used in the CV is equal to
\begin{align}
	\label{eq: rcsue}
	\hat{B}_T    := \argmin \limits_{B \in 	\bar{\mathbb{B}}_{\bar{B}} } \text{ }
	g_T(B)'
	W(B)
	g_T(B) + \lambda \sum_{(i,j)\in \mathcal{R}}^{ }  
	v_{ij} B_{ij}^2  +  \frac{1}{n} \sum_{i=1}^{n} (var(e(B)_{it})-1)^2  ,   
\end{align} 
where the last term penalizes deviations of the corresponding innovations from the unit variance assumption, which avoids solutions of the optimization problems corresponding to very large or small variances. The variance regularization term is only used in the CV.

Finally, the loss in the let-out fold at the solution obtain in the estimation fold is calculated using the same moment conditions implied by mean independent shocks. Moreover, each moment condition is weighted by the inverse of the variances of the corresponding moment under the assumption of independent and normal distributed shocks, see \cite{keweloh2023structural}.

 \section{Bootstrap}
 \label{appendix: sec: Bootstrap}
 This section outlines the algorithm used to construct bootstrap confidence bands.
 
 \begin{enumerate}
 	\item Estimate the initial estimator $\hat{A}_i$ for $i=1,...,p$ and $\hat{B}$ with the sample $[y_1,...,y_T]$ and  save the corresponding impulse response functions  denoted by ${irf}_{ij,t}$ for the response of variable $i$ to shock $j$ after $t$ periods. 
 	
 	\item Each step $l=1,...,M$ of the bootstrap algorithm involves i) drawing new reduced form shocks from the reduced form shocks corresponding to the initial estimation in step one, ii) simulating a new sample $[\tilde{y}_1,...,\tilde{y}_T]$ using the new reduced form shocks and $\hat{A}_i$ from the   initial estimation in step one,  iii) estimating the  estimators $\hat{A}_{il}$ for $i=1,...,q$ and $\hat{B}_l$ with the sample $[\tilde{y}_1,...,\tilde{y}_T]$, an iv) estimate an save the corresponding impulse response functions denoted by ${irf}_{ij,t}^l$. 
 	
 	\item  Calculate quantiles $q_{\alpha/2}({irf}_{ij,t}^l)$,
 	$q_{1-\alpha/2}({irf}_{ij,t}^l) $ and $q_{0.5}({irf}_{ij,t}^l)$ over the steps $l=1,...,M$ for $i,j=1,...,n$ and the desired impulse response length $t=1,...,r$. 
 	
 	\item  Construct point-wise $\alpha$\% confidence bands   
 	\begin{align}
 		 [ {irf}_{ij,t} - |q_{\alpha/2}({irf}_{ij,t}^l) - q_{0.5}({irf}_{ij,t}^l) | ,
 		{irf}_{ij,t} + |q_{1-\alpha/2}({irf}_{ij,t}^l) - q_{0.5}({irf}_{ij,t}^l) |  ] .
 	\end{align}
 \end{enumerate}

 \section{Labeling}
 \label{appendix: sec: Labeling}
\begin{figure}[h!] 
 	\centering
 	\caption{Illustration  of labeling  using  $\bar{\mathbb{B}}_{\bar{B}}$ with $\bar{B} \neq B_0$}
 	\includegraphics[width=0.8\textwidth]{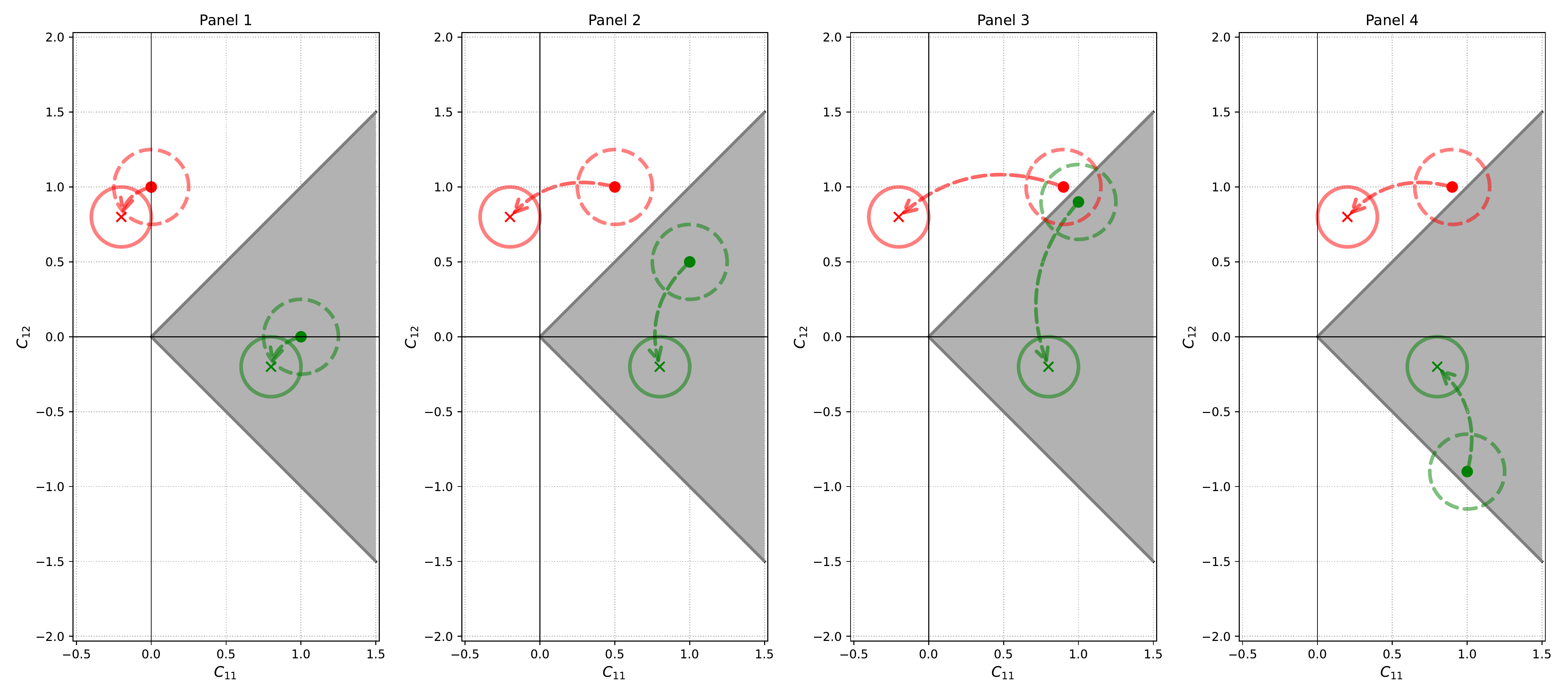} 
 	\includegraphics[width=0.8\textwidth]{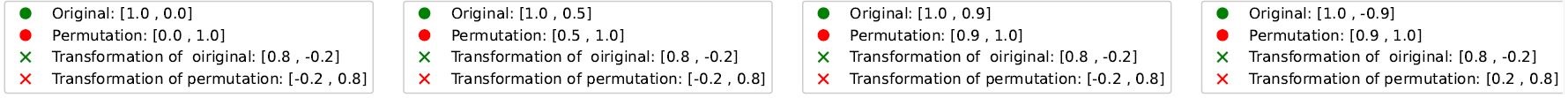} 
 	\label{fig: visualization_set 4} 
 	\begin{minipage}{1\textwidth} %
 		{   \footnotesize  
 			\textit{Note:} 
 			
 					The panels  display  the elements $[c_{11}, c_{12}]$ in the first row of various $B_0$ matrices.
 			The green dots represent the first row of    the following four  matrices: 
 			$B_0 = \begin{bmatrix}
 				1 & 0 \\ 0& 1
 			\end{bmatrix} $ in panel $1$, $ 
 			B_0 = \begin{bmatrix}
 				1 & 0.5 \\ 0& 1
 			\end{bmatrix}  $ in panel $2$, $ 
 			B_0 = \begin{bmatrix}
 				1 & 0.9\\ 0& 1
 			\end{bmatrix} $ in panel $3$, and $ 
 			B_0 = \begin{bmatrix}
 				1 & -0.9 \\ 0& 1
 			\end{bmatrix}   $ in panel $4$. 
 			The red dots display the first row's elements of the sign permutation of the 
 			$B_0$ matrices, which keeps the diagonal elements positive.  
 			Additionally, the figure shows the elements in the first row of the transformed  matrices using $C:= \bar{B}^{-1} B$  with  $\bar{B} =  B_0 +  \begin{bmatrix}
 				0.25 & 0.25 \\ 0 & 0
 			\end{bmatrix}  $   and    $B$ is equal to $B_0$ or equal to the sign permutation of $B_0$, represented by  green and red crosses. 
 			The circles around each point illustrate a set of  matrices close to the corresponding   matrix.
 			The shaded area indicates the set of  matrices contained in $\bar{\mathbb{B}}_{\bar{B}}$. 
 			 
 			\par}
 	\end{minipage}
 \end{figure}

 This section briefly illustrates how using an initial estimator $\bar{B}$ not equal  to $B_0$, but in the proximity of $B_0$  is sufficient to move $B_0$ to the inner area of the set $\bar{\mathbb{B}}_{\bar{B}}$ used to label the shocks. Specifically, Figure \ref{fig: visualization_set 4} is analogously generated to Figure \ref{fig: visualization_set 2}  in the main text, however,   Figure \ref{fig: visualization_set 4} uses  $\bar{B} =  B +  \begin{bmatrix}
 	0.25 & 0.25 \\ 0 & 0
 \end{bmatrix}  $. The figure illustrates how a $\bar{B}$ in the proximity of $B_0$ is sufficient to ensure that estimators  close to $B_0$  are  assigned with the same labels.
 
\section{Appendix - Finite sample performance}
\label{appendix: sec: Finite sample performance} 
This section  illustrates how the cross-validation selects tuning parameters in the the Monte Carlo simulation in Section \ref{sec: Finite Sample Performance}. 
 	Moreover, it  contains  additional simulations,  including SVARs with Gaussian shocks,  a VAR with lags and shocks with a common volatility process, $A$-type restrictions, and  augmented proxy VAR restrictions.

\begin{figure}[h!] 
	\centering
	\caption{Selected tuning parameters in the cross-validation.}
	\includegraphics[width=1\textwidth]{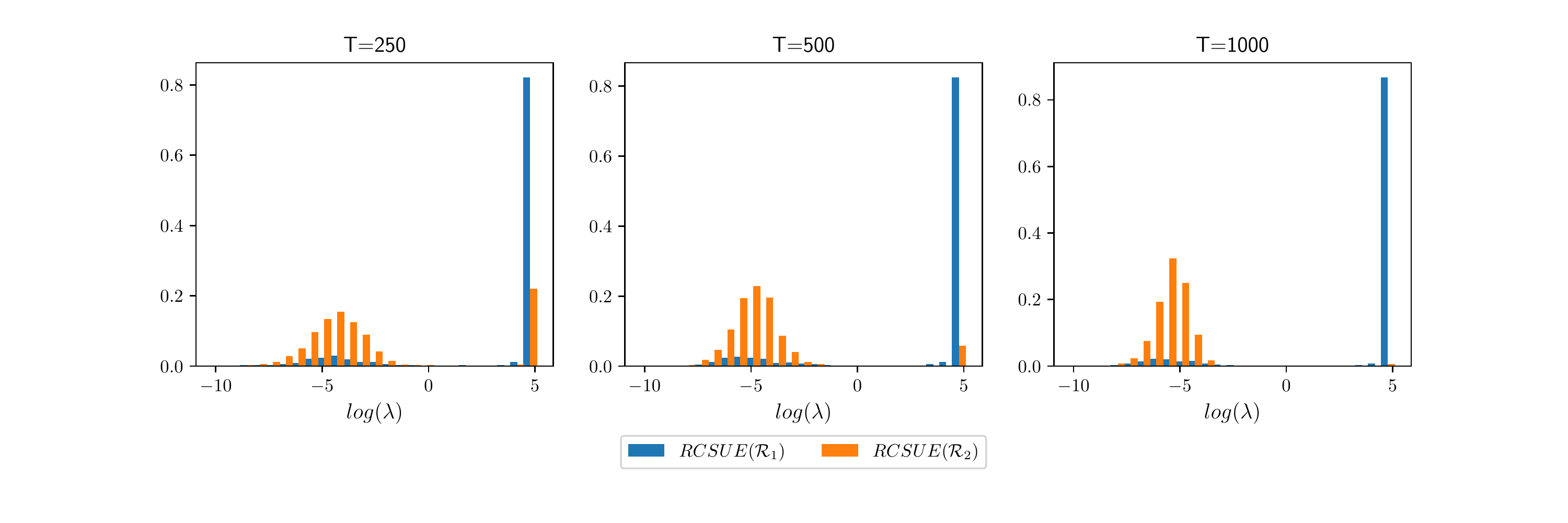}  
	\label{fig: LambdaPlot} 
	\begin{minipage}{1\textwidth} %
		{   \footnotesize  
			\textit{Note:} 	The figure shows how often each tuning parameter was selected by the cross-validation in the Monte Carlo simulation in Section \ref{sec: Finite Sample Performance} for the two ridge estimators considered in the section.
			\par}
	\end{minipage}
\end{figure}

Figure \ref{fig: LambdaPlot} visualizes the results of the cross-validation for the two ridge estimators considered in the Monte Carlo simulation in Section \ref{sec: Finite Sample Performance}. 
In the case of RCSUE($\mathcal{R}_1$), incorporating only correct zero restrictions, the cross-validation consistently favors the largest available tuning parameter, promoting maximal shrinkage towards the specified restrictions.
For the RCSUE($\mathcal{R}_2$) which uses one additional incorrect restriction the cross-validation mainly selects tuning parameters in the center of the available set of tuning parameters. This indicates that the cross-validation  is able to detect that larger tuning parameters lead to dependent shocks in the let-out folds. However, in approximately $20$\% of the simulations using the smallest sample size of $T=250$ observations, the cross-validation selects the largest available tuning parameter, i.e. it is not able to detect and neglect the incorrect restriction. Notably, as the sample size increases, the percentage of simulations with the cross-validation selecting the largest tuning parameter diminishes, dropping below $1$\% for the largest sample size of $T=1000$ observations.

Table \ref{Table: Finite sample performance - Bias MSE - MCRGMMLagsAndHet} and  \ref{Table: Finite sample performance - Coverage - MCRGMMLagsAndHet} 
extend the simulations   considered in the main text with an  VAR $y_t = A_1 y_{t-1} + u_t$ and
\begin{align}
	A_1 = \begin{bmatrix}
		0.5 & 0  & 0 & 0 \\
		0.1 & 0.5  & 0 & 0 \\
		0.1 & 0.1  & 0.5 & 0 \\
		0.1 & 0.1  & 0.1 & 0.5 \\
	\end{bmatrix}.
\end{align}
Moreover, the structural shocks  are no longer independent, but simulated with a common volatility process. Specifically, the simultaneous interaction is given by $u_t = B_0 \tilde{\varepsilon}_t$ with $ \tilde{\varepsilon}_{i,t} = \psi_t \sigma_{i,t} \varepsilon_{i,t} + (1-\psi_t) \varepsilon_{i,t}$ where $\varepsilon_t$ are i.i.d. as defined in the main text, $\psi_t$ is a random variable with a Bernoulli distribution with $Pr(\psi_t=1)=Pr(\psi_t=0)=0.5$, and $\sigma_{i,t}=i$ for $i=1,...,n$. The shocks $\tilde{\varepsilon}_{i,t}$ are normalized to unit variance. The variable $\psi_t$  affects all structural shocks and governs whether the shocks are in a low or high common volatility state.  
The estimators depicted in the table use a two-step estimation approach, where the VAR is estimated in the first step and the second step estimates the simultaneous interaction analogously to the main text using the reduced form shocks from the first step.
The results are similar to results shown in the main text, which indicates the robustness of the proposed method with respect to shocks featuring a common volatility process and to a two-step VAR estimation approach.
\begin{table}
	\caption{Finite sample performance - average and mean squared error - simulation with lags and a common volatility process}
	\label{Table: Finite sample performance - Bias MSE - MCRGMMLagsAndHet}
	\centering
	\renewcommand{\arraystretch}{1}
	\begin{tabular}{c | c c c}
		& $CSUE$ & $RCSUE(\mathcal{R}_1)$ & $RCSUE(\mathcal{R}_2)$ \\
		\hline
		\rotatebox[origin=c]{90}{$T = 250$ } &$\begin{bmatrix}\underset{(0.59)}{9.7} & \underset{(1.36)}{\hphantom{-}0.01} & \underset{(1.16)}{\hphantom{-}0.05} & \underset{(1.23)}{\hphantom{-}0.03} \\ \underset{(1.39)}{4.84} & \underset{(1.22)}{\hphantom{-}9.53} & \underset{(2.78)}{\hphantom{-}0.08} & \underset{(2.55)}{\hphantom{-}0.09} \\ \underset{(1.62)}{4.78} & \underset{(3.28)}{\hphantom{-}4.7} & \underset{(2.2)}{\hphantom{-}9.52} & \underset{(3.8)}{\hphantom{-}4.82} \\ \underset{(1.6)}{4.79} & \underset{(3.15)}{\hphantom{-}4.69} & \underset{(3.56)}{\hphantom{-}4.82} & \underset{(2.46)}{\hphantom{-}9.54} \\ \end{bmatrix}$ &$\begin{bmatrix}\underset{(0.42)}{9.86} & \textcolor{red}{\underset{(0.08)}{0.0}} & \textcolor{red}{\underset{(0.06)}{0.0}} & \textcolor{red}{\underset{(0.06)}{-0.01}} \\ \underset{(0.56)}{4.92} & \underset{(0.67)}{\hphantom{-}9.75} & \textcolor{red}{\underset{(0.23)}{\hphantom{-}0.02}} & \textcolor{red}{\underset{(0.18)}{\hphantom{-}0.03}} \\ \underset{(0.75)}{4.91} & \underset{(1.24)}{\hphantom{-}4.84} & \underset{(1.47)}{\hphantom{-}9.56} & \underset{(3.0)}{\hphantom{-}4.79} \\ \underset{(0.73)}{4.91} & \underset{(1.24)}{\hphantom{-}4.83} & \underset{(2.92)}{\hphantom{-}4.8} & \underset{(1.67)}{\hphantom{-}9.55} \\ \end{bmatrix}$ &$\begin{bmatrix}\underset{(0.42)}{9.83} & \textcolor{red}{\underset{(0.08)}{0.0}} & \textcolor{red}{\underset{(0.1)}{\hphantom{-}0.01}} & \textcolor{red}{\underset{(0.14)}{-0.03}} \\ \underset{(0.58)}{4.9} & \underset{(0.66)}{\hphantom{-}9.69} & \textcolor{red}{\underset{(0.36)}{\hphantom{-}0.02}} & \textcolor{red}{\underset{(0.36)}{-0.04}} \\ \underset{(0.85)}{4.89} & \underset{(1.48)}{\hphantom{-}4.81} & \underset{(1.03)}{\hphantom{-}10.03} & \textcolor{red}{\underset{(13.65)}{\hphantom{-}1.96}} \\ \underset{(0.84)}{4.89} & \underset{(1.49)}{\hphantom{-}4.83} & \underset{(5.44)}{\hphantom{-}6.39} & \underset{(9.47)}{\hphantom{-}7.4} \\ \end{bmatrix}$ \\ [40pt]\rotatebox[origin=c]{90}{$T = 500$ } &$\begin{bmatrix}\underset{(0.23)}{9.87} & \underset{(0.61)}{\hphantom{-}0.03} & \underset{(0.54)}{\hphantom{-}0.02} & \underset{(0.51)}{\hphantom{-}0.03} \\ \underset{(0.61)}{4.9} & \underset{(0.51)}{\hphantom{-}9.8} & \underset{(1.31)}{\hphantom{-}0.03} & \underset{(1.35)}{\hphantom{-}0.07} \\ \underset{(0.75)}{4.9} & \underset{(1.62)}{\hphantom{-}4.87} & \underset{(1.08)}{\hphantom{-}9.78} & \underset{(1.81)}{\hphantom{-}4.94} \\ \underset{(0.71)}{4.89} & \underset{(1.61)}{\hphantom{-}4.84} & \underset{(1.83)}{\hphantom{-}4.9} & \underset{(1.1)}{\hphantom{-}9.82} \\ \end{bmatrix}$ &$\begin{bmatrix}\underset{(0.19)}{9.93} & \textcolor{red}{\underset{(0.03)}{\hphantom{-}0.01}} & \textcolor{red}{\underset{(0.01)}{0.0}} & \textcolor{red}{\underset{(0.01)}{0.0}} \\ \underset{(0.25)}{4.96} & \underset{(0.29)}{\hphantom{-}9.89} & \textcolor{red}{\underset{(0.05)}{\hphantom{-}0.01}} & \textcolor{red}{\underset{(0.07)}{\hphantom{-}0.01}} \\ \underset{(0.31)}{4.98} & \underset{(0.49)}{\hphantom{-}4.94} & \underset{(0.68)}{\hphantom{-}9.8} & \underset{(1.4)}{\hphantom{-}4.89} \\ \underset{(0.29)}{4.97} & \underset{(0.51)}{\hphantom{-}4.93} & \underset{(1.39)}{\hphantom{-}4.91} & \underset{(0.71)}{\hphantom{-}9.79} \\ \end{bmatrix}$ &$\begin{bmatrix}\underset{(0.2)}{9.91} & \textcolor{red}{\underset{(0.02)}{0.0}} & \textcolor{red}{\underset{(0.03)}{0.0}} & \textcolor{red}{\underset{(0.04)}{-0.01}} \\ \underset{(0.25)}{4.94} & \underset{(0.32)}{\hphantom{-}9.85} & \textcolor{red}{\underset{(0.12)}{\hphantom{-}0.01}} & \textcolor{red}{\underset{(0.19)}{-0.05}} \\ \underset{(0.35)}{4.95} & \underset{(0.61)}{\hphantom{-}4.93} & \underset{(0.58)}{\hphantom{-}10.18} & \textcolor{red}{\underset{(8.65)}{\hphantom{-}2.72}} \\ \underset{(0.35)}{4.95} & \underset{(0.66)}{\hphantom{-}4.93} & \underset{(3.54)}{\hphantom{-}6.18} & \underset{(4.82)}{\hphantom{-}8.22} \\ \end{bmatrix}$ \\ [40pt]\rotatebox[origin=c]{90}{$T = 1000$ } &$\begin{bmatrix}\underset{(0.11)}{9.94} & \underset{(0.29)}{\hphantom{-}0.02} & \underset{(0.26)}{\hphantom{-}0.05} & \underset{(0.26)}{\hphantom{-}0.02} \\ \underset{(0.29)}{4.95} & \underset{(0.22)}{\hphantom{-}9.89} & \underset{(0.65)}{\hphantom{-}0.03} & \underset{(0.65)}{\hphantom{-}0.04} \\ \underset{(0.37)}{4.92} & \underset{(0.76)}{\hphantom{-}4.93} & \underset{(0.52)}{\hphantom{-}9.91} & \underset{(0.84)}{\hphantom{-}4.98} \\ \underset{(0.35)}{4.94} & \underset{(0.74)}{\hphantom{-}4.92} & \underset{(0.87)}{\hphantom{-}4.97} & \underset{(0.5)}{\hphantom{-}9.91} \\ \end{bmatrix}$ &$\begin{bmatrix}\underset{(0.09)}{9.97} & \textcolor{red}{\underset{(0.0)}{0.0}} & \textcolor{red}{\underset{(0.0)}{0.0}} & \textcolor{red}{\underset{(0.01)}{0.0}} \\ \underset{(0.11)}{4.99} & \underset{(0.14)}{\hphantom{-}9.93} & \textcolor{red}{\underset{(0.03)}{0.0}} & \textcolor{red}{\underset{(0.03)}{\hphantom{-}0.01}} \\ \underset{(0.13)}{4.99} & \underset{(0.23)}{\hphantom{-}4.96} & \underset{(0.27)}{\hphantom{-}9.9} & \underset{(0.54)}{\hphantom{-}4.97} \\ \underset{(0.13)}{5.0} & \underset{(0.23)}{\hphantom{-}4.97} & \underset{(0.54)}{\hphantom{-}4.95} & \underset{(0.27)}{\hphantom{-}9.9} \\ \end{bmatrix}$ &$\begin{bmatrix}\underset{(0.1)}{9.96} & \textcolor{red}{\underset{(0.01)}{0.0}} & \textcolor{red}{\underset{(0.01)}{0.0}} & \textcolor{red}{\underset{(0.02)}{-0.01}} \\ \underset{(0.12)}{4.98} & \underset{(0.15)}{\hphantom{-}9.9} & \textcolor{red}{\underset{(0.06)}{0.0}} & \textcolor{red}{\underset{(0.07)}{-0.04}} \\ \underset{(0.16)}{4.99} & \underset{(0.3)}{\hphantom{-}4.97} & \underset{(0.31)}{\hphantom{-}10.16} & \textcolor{red}{\underset{(3.83)}{\hphantom{-}3.6}} \\ \underset{(0.16)}{4.99} & \underset{(0.3)}{\hphantom{-}4.98} & \underset{(1.69)}{\hphantom{-}5.76} & \underset{(2.06)}{\hphantom{-}8.91} \\ \end{bmatrix}$ \\ [40pt]

	\end{tabular}
			\begin{minipage}{1\textwidth} %
		{   \footnotesize  
			\textit{Note:} 
			Monte Carlo simulation with $M = 2000$ replications.
			The simulation extends the simultaneous interaction  considered in Section \ref{sec: Finite Sample Performance} with a  VAR $y_t = A_1 y_{t-1} + u_t$ and $u_t = B_0 \tilde{\varepsilon}_t$ with $ \tilde{\varepsilon}_{i,t} = \psi_t \sigma_{i,t} \varepsilon_{i,t} + (1-\psi_t) \varepsilon_{i,t}$, where  $\varepsilon_t$ are i.i.d. as defined in the main text, $\psi_t$ is a random variable with a  Bernoulli distribution with $Pr(\psi_t=1)=Pr(\psi_t=0)=0.5$, and $\sigma_{i,t}=i$ for $i=1,...,n$. 
			 The table
			shows the average, $1/M 
			\sum_{m=1}^{M}
			\hat{b}^{m}_{ij} $ and in parentheses the mean squared error $1/M 
			\sum_{m=1}^{M}
			(\hat{b}^{m}_{ij} - b_ij)^2 $
			of
			each estimated element  $\hat{b}^{m}_{ij}$ in simulation $m$. Penalized elements are highlighted in red.
			\par}
	\end{minipage}
\end{table}
\begin{table}
	\caption{Finite sample performance - coverage and confidence band width - simulation with lags and a common volatility process}
	\label{Table: Finite sample performance - Coverage - MCRGMMLagsAndHet}
	\centering
	\renewcommand{\arraystretch}{1.2}
	\begin{tabular}{c | c c c}
		& $CSUE$ & $RCSUE(\mathcal{R}_1)$ & $RCSUE(\mathcal{R}_2)$ \\
		\hline
		\rotatebox[origin=c]{90}{$T = 250$ } &$\begin{bmatrix}\underset{(1.51)}{63.0} & \underset{(2.48)}{\hphantom{-}70.0} & \underset{(2.34)}{\hphantom{-}70.0} & \underset{(2.25)}{\hphantom{-}68.0} \\ \underset{(2.53)}{70.0} & \underset{(2.12)}{\hphantom{-}62.0} & \underset{(3.33)}{\hphantom{-}69.0} & \underset{(3.28)}{\hphantom{-}69.0} \\ \underset{(2.8)}{70.0} & \underset{(3.68)}{\hphantom{-}70.0} & \underset{(3.01)}{\hphantom{-}67.0} & \underset{(3.79)}{\hphantom{-}69.0} \\ \underset{(2.74)}{70.0} & \underset{(3.64)}{\hphantom{-}70.0} & \underset{(3.81)}{\hphantom{-}69.0} & \underset{(2.99)}{\hphantom{-}68.0} \\ \end{bmatrix}$ &$\begin{bmatrix}\underset{(1.23)}{65.0} & \textcolor{red}{\underset{(0.07)}{\hphantom{-}69.0}} & \textcolor{red}{\underset{(0.07)}{\hphantom{-}69.0}} & \textcolor{red}{\underset{(0.08)}{\hphantom{-}68.0}} \\ \underset{(1.62)}{72.0} & \underset{(1.42)}{\hphantom{-}62.0} & \textcolor{red}{\underset{(0.17)}{\hphantom{-}70.0}} & \textcolor{red}{\underset{(0.17)}{\hphantom{-}72.0}} \\ \underset{(1.87)}{71.0} & \underset{(2.16)}{\hphantom{-}70.0} & \underset{(2.37)}{\hphantom{-}63.0} & \underset{(3.38)}{\hphantom{-}66.0} \\ \underset{(1.85)}{70.0} & \underset{(2.16)}{\hphantom{-}69.0} & \underset{(3.39)}{\hphantom{-}66.0} & \underset{(2.39)}{\hphantom{-}64.0} \\ \end{bmatrix}$ &$\begin{bmatrix}\underset{(1.29)}{63.0} & \textcolor{red}{\underset{(9.13)}{\hphantom{-}70.0}} & \textcolor{red}{\underset{(0.99)}{\hphantom{-}70.0}} & \textcolor{red}{\underset{(6.05)}{\hphantom{-}64.0}} \\ \underset{(2.89)}{71.0} & \underset{(1.82)}{\hphantom{-}62.0} & \textcolor{red}{\underset{(2.67)}{\hphantom{-}69.0}} & \textcolor{red}{\underset{(2.56)}{\hphantom{-}65.0}} \\ \underset{(1.99)}{70.0} & \underset{(2.44)}{\hphantom{-}67.0} & \underset{(2.21)}{\hphantom{-}63.0} & \textcolor{red}{\underset{(2.28)}{\hphantom{-}18.0}} \\ \underset{(3.09)}{69.0} & \underset{(3.03)}{\hphantom{-}66.0} & \underset{(3.82)}{\hphantom{-}35.0} & \underset{(2.33)}{\hphantom{-}13.0} \\ \end{bmatrix}$ \\ [40pt]\rotatebox[origin=c]{90}{$T = 500$ } &$\begin{bmatrix}\underset{(0.99)}{68.0} & \underset{(1.71)}{\hphantom{-}70.0} & \underset{(1.61)}{\hphantom{-}68.0} & \underset{(1.59)}{\hphantom{-}71.0} \\ \underset{(1.72)}{69.0} & \underset{(1.46)}{\hphantom{-}66.0} & \underset{(2.53)}{\hphantom{-}71.0} & \underset{(2.48)}{\hphantom{-}70.0} \\ \underset{(1.92)}{69.0} & \underset{(2.78)}{\hphantom{-}69.0} & \underset{(2.25)}{\hphantom{-}70.0} & \underset{(2.96)}{\hphantom{-}72.0} \\ \underset{(1.89)}{69.0} & \underset{(2.74)}{\hphantom{-}70.0} & \underset{(2.97)}{\hphantom{-}71.0} & \underset{(2.25)}{\hphantom{-}68.0} \\ \end{bmatrix}$ &$\begin{bmatrix}\underset{(0.88)}{68.0} & \textcolor{red}{\underset{(0.04)}{\hphantom{-}69.0}} & \textcolor{red}{\underset{(0.03)}{\hphantom{-}67.0}} & \textcolor{red}{\underset{(0.03)}{\hphantom{-}69.0}} \\ \underset{(1.08)}{72.0} & \underset{(1.02)}{\hphantom{-}67.0} & \textcolor{red}{\underset{(0.08)}{\hphantom{-}72.0}} & \textcolor{red}{\underset{(0.08)}{\hphantom{-}70.0}} \\ \underset{(1.23)}{72.0} & \underset{(1.49)}{\hphantom{-}71.0} & \underset{(1.73)}{\hphantom{-}69.0} & \underset{(2.6)}{\hphantom{-}69.0} \\ \underset{(1.23)}{74.0} & \underset{(1.49)}{\hphantom{-}71.0} & \underset{(2.6)}{\hphantom{-}71.0} & \underset{(1.75)}{\hphantom{-}68.0} \\ \end{bmatrix}$ &$\begin{bmatrix}\underset{(0.88)}{68.0} & \textcolor{red}{\underset{(0.09)}{\hphantom{-}71.0}} & \textcolor{red}{\underset{(0.12)}{\hphantom{-}66.0}} & \textcolor{red}{\underset{(0.13)}{\hphantom{-}67.0}} \\ \underset{(1.07)}{70.0} & \underset{(1.03)}{\hphantom{-}64.0} & \textcolor{red}{\underset{(0.24)}{\hphantom{-}72.0}} & \textcolor{red}{\underset{(0.27)}{\hphantom{-}63.0}} \\ \underset{(1.25)}{71.0} & \underset{(1.56)}{\hphantom{-}67.0} & \underset{(1.39)}{\hphantom{-}60.0} & \textcolor{red}{\underset{(1.54)}{\hphantom{-}22.0}} \\ \underset{(1.25)}{70.0} & \underset{(1.57)}{\hphantom{-}67.0} & \underset{(2.03)}{\hphantom{-}41.0} & \underset{(1.44)}{\hphantom{-}18.0} \\ \end{bmatrix}$ \\ [40pt]\rotatebox[origin=c]{90}{$T = 1000$ } &$\begin{bmatrix}\underset{(0.67)}{67.0} & \underset{(1.18)}{\hphantom{-}70.0} & \underset{(1.11)}{\hphantom{-}70.0} & \underset{(1.08)}{\hphantom{-}70.0} \\ \underset{(1.17)}{69.0} & \underset{(0.98)}{\hphantom{-}68.0} & \underset{(1.79)}{\hphantom{-}71.0} & \underset{(1.78)}{\hphantom{-}70.0} \\ \underset{(1.3)}{69.0} & \underset{(1.98)}{\hphantom{-}73.0} & \underset{(1.56)}{\hphantom{-}69.0} & \underset{(2.09)}{\hphantom{-}70.0} \\ \underset{(1.27)}{69.0} & \underset{(1.96)}{\hphantom{-}73.0} & \underset{(2.08)}{\hphantom{-}71.0} & \underset{(1.58)}{\hphantom{-}69.0} \\ \end{bmatrix}$ &$\begin{bmatrix}\underset{(0.62)}{68.0} & \textcolor{red}{\underset{(0.02)}{\hphantom{-}65.0}} & \textcolor{red}{\underset{(0.02)}{\hphantom{-}67.0}} & \textcolor{red}{\underset{(0.02)}{\hphantom{-}68.0}} \\ \underset{(0.72)}{71.0} & \underset{(0.73)}{\hphantom{-}68.0} & \textcolor{red}{\underset{(0.05)}{\hphantom{-}70.0}} & \textcolor{red}{\underset{(0.05)}{\hphantom{-}71.0}} \\ \underset{(0.82)}{74.0} & \underset{(1.0)}{\hphantom{-}71.0} & \underset{(1.17)}{\hphantom{-}72.0} & \underset{(1.78)}{\hphantom{-}73.0} \\ \underset{(0.81)}{73.0} & \underset{(1.01)}{\hphantom{-}72.0} & \underset{(1.77)}{\hphantom{-}73.0} & \underset{(1.19)}{\hphantom{-}70.0} \\ \end{bmatrix}$ &$\begin{bmatrix}\underset{(0.63)}{66.0} & \textcolor{red}{\underset{(0.06)}{\hphantom{-}71.0}} & \textcolor{red}{\underset{(0.08)}{\hphantom{-}71.0}} & \textcolor{red}{\underset{(0.09)}{\hphantom{-}66.0}} \\ \underset{(0.73)}{69.0} & \underset{(0.74)}{\hphantom{-}64.0} & \textcolor{red}{\underset{(0.16)}{\hphantom{-}73.0}} & \textcolor{red}{\underset{(0.18)}{\hphantom{-}64.0}} \\ \underset{(0.84)}{70.0} & \underset{(1.06)}{\hphantom{-}68.0} & \underset{(1.01)}{\hphantom{-}64.0} & \textcolor{red}{\underset{(1.46)}{\hphantom{-}29.0}} \\ \underset{(0.84)}{70.0} & \underset{(1.07)}{\hphantom{-}69.0} & \underset{(1.57)}{\hphantom{-}50.0} & \underset{(1.05)}{\hphantom{-}25.0} \\ \end{bmatrix}$ \\ [40pt]
		
	\end{tabular}
				\begin{minipage}{1\textwidth} %
		{   \footnotesize  
			\textit{Note:} 
			Monte Carlo simulation with $M = 2000$ replications.
			The simulation extends the simultaneous interaction  considered in Section \ref{sec: Finite Sample Performance} with a  VAR $y_t = A_1 y_{t-1} + u_t$ and $u_t = B_0 \tilde{\varepsilon}_t$ with $ \tilde{\varepsilon}_{i,t} = \psi_t \sigma_{i,t} \varepsilon_{i,t} + (1-\psi_t) \varepsilon_{i,t}$, where  $\varepsilon_t$ are i.i.d. as defined in the main text, $\psi_t$ is a random variable with a  Bernoulli distribution with $Pr(\psi_t=1)=Pr(\psi_t=0)=0.5$, and $\sigma_{i,t}=i$ for $i=1,...,n$. 
			 The table shows the coverage of $68$\% bootstrap confidence bands  and in parentheses the average width of the bands for 
			each estimated element. Penalized elements are highlighted in red.
			\par}
	\end{minipage}
\end{table} 

Table \ref{Table: Finite sample performance - Bias MSE - MCRGMMLagsAndHet - ML} uses the same data-generating process with lags and a common volatility process. The table compares the performance of two statistically identified estimator: i) the CSUE relying on second- to fourth-order moment conditions implied by mean independent shocks and ii) a maximum likelihood estimator denoted by ML assuming independent shocks with a skewed t-distribution, compare to \cite{lanne2017identification}.
The simulation uses larger sample sizes from $T=500$ to $T=5000$ observations to illustrate the difference between the moment based CSUE which only requires mean independent shocks and a non-Gaussian ML estimator assuming independent shocks in the presence of a common volatility process.
The table shows that the CSUE is consistent even though the shocks are not independent. In contrast, the ML estimator is clearly biased. The bias of the ML estimator increases with the impact of the common volatility process, i.e. the bias is stronger in columns corresponding to shocks which are more affected by the common volatility process.
 \begin{table}
 	\caption{Finite sample performance of moment based and maximum likelihood estimation - average and mean squared error - simulation with lags and a common volatility process}
 	\label{Table: Finite sample performance - Bias MSE - MCRGMMLagsAndHet - ML}
 	\centering
 	\renewcommand{\arraystretch}{1}
 	\begin{tabular}{c | c c c}
 		& $CSUE$ & $ML$  \\
 		\hline
 		\rotatebox[origin=c]{90}{$T = 500$ } &
 		$\begin{bmatrix}\underset{(0.24)}{9.86} & \underset{(0.65)}{\hphantom{-}0.01} & \underset{(0.56)}{\hphantom{-}0.03} & \underset{(0.59)}{0.0} \\ \underset{(0.64)}{4.92} & \underset{(0.5)}{\hphantom{-}9.78} & \underset{(1.43)}{\hphantom{-}0.05} & \underset{(1.42)}{\hphantom{-}0.01} \\ \underset{(0.83)}{4.9} & \underset{(1.67)}{\hphantom{-}4.86} & \underset{(1.13)}{\hphantom{-}9.79} & \underset{(1.94)}{\hphantom{-}4.89} \\ \underset{(0.8)}{4.91} & \underset{(1.64)}{\hphantom{-}4.87} & \underset{(1.88)}{\hphantom{-}4.92} & \underset{(1.21)}{\hphantom{-}9.76} \\ \end{bmatrix}$ 
 		
 		&
 	$\begin{bmatrix}\underset{(0.22)}{10.06} & \underset{(0.31)}{\hphantom{-}0.02} & \underset{(0.22)}{\hphantom{-}0.02} & \underset{(0.2)}{\hphantom{-}0.01} \\ \underset{(0.33)}{5.02} & \underset{(0.31)}{\hphantom{-}9.79} & \underset{(0.66)}{\hphantom{-}0.02} & \underset{(0.55)}{-0.01} \\ \underset{(0.35)}{5.01} & \underset{(0.85)}{\hphantom{-}4.9} & \underset{(1.52)}{\hphantom{-}9.06} & \underset{(1.98)}{\hphantom{-}4.23} \\ \underset{(0.32)}{5.02} & \underset{(0.76)}{\hphantom{-}4.91} & \underset{(1.26)}{\hphantom{-}4.58} & \underset{(3.11)}{\hphantom{-}8.52} \\ \end{bmatrix}$ 
 	
 		\\ 
 		[40pt]
 		\rotatebox[origin=c]{90}{$T = 1000$ } &
 	$\begin{bmatrix}\underset{(0.12)}{9.93} & \underset{(0.3)}{\hphantom{-}0.03} & \underset{(0.28)}{\hphantom{-}0.01} & \underset{(0.26)}{0.0} \\ \underset{(0.28)}{4.95} & \underset{(0.22)}{\hphantom{-}9.91} & \underset{(0.66)}{\hphantom{-}0.02} & \underset{(0.63)}{\hphantom{-}0.04} \\ \underset{(0.36)}{4.94} & \underset{(0.78)}{\hphantom{-}4.95} & \underset{(0.52)}{\hphantom{-}9.91} & \underset{(0.85)}{\hphantom{-}4.95} \\ \underset{(0.35)}{4.94} & \underset{(0.74)}{\hphantom{-}4.93} & \underset{(0.88)}{\hphantom{-}4.98} & \underset{(0.52)}{\hphantom{-}9.89} \\ \end{bmatrix}$ 
 	
 		&
 	$\begin{bmatrix}\underset{(0.14)}{10.09} & \underset{(0.15)}{\hphantom{-}0.01} & \underset{(0.11)}{-0.01} & \underset{(0.09)}{\hphantom{-}0.01} \\ \underset{(0.16)}{5.05} & \underset{(0.15)}{\hphantom{-}9.86} & \underset{(0.29)}{0.0} & \underset{(0.25)}{\hphantom{-}0.02} \\ \underset{(0.16)}{5.04} & \underset{(0.39)}{\hphantom{-}4.92} & \underset{(1.1)}{\hphantom{-}9.1} & \underset{(0.93)}{\hphantom{-}4.31} \\ \underset{(0.15)}{5.04} & \underset{(0.36)}{\hphantom{-}4.92} & \underset{(0.66)}{\hphantom{-}4.56} & \underset{(2.21)}{\hphantom{-}8.61} \\ \end{bmatrix}$ 
 	
 		\\ 
 		[40pt]
 		\rotatebox[origin=c]{90}{$T = 5000$ } &
 	$\begin{bmatrix}\underset{(0.02)}{9.98} & \underset{(0.05)}{\hphantom{-}0.01} & \underset{(0.05)}{-0.0} & \underset{(0.05)}{0.0} \\ \underset{(0.05)}{4.99} & \underset{(0.04)}{\hphantom{-}9.98} & \underset{(0.11)}{-0.0} & \underset{(0.12)}{\hphantom{-}0.02} \\ \underset{(0.07)}{4.99} & \underset{(0.14)}{\hphantom{-}4.99} & \underset{(0.09)}{\hphantom{-}9.98} & \underset{(0.15)}{\hphantom{-}4.99} \\ \underset{(0.07)}{4.99} & \underset{(0.14)}{\hphantom{-}4.99} & \underset{(0.15)}{\hphantom{-}5.0} & \underset{(0.1)}{\hphantom{-}9.98} \\ \end{bmatrix}$ 
 	
 		&
 	$\begin{bmatrix}\underset{(0.04)}{10.12} & \underset{(0.03)}{0.0} & \underset{(0.02)}{-0.0} & \underset{(0.02)}{-0.0} \\ \underset{(0.04)}{5.06} & \underset{(0.04)}{\hphantom{-}9.9} & \underset{(0.05)}{-0.01} & \underset{(0.05)}{\hphantom{-}0.01} \\ \underset{(0.03)}{5.06} & \underset{(0.08)}{\hphantom{-}4.96} & \underset{(0.83)}{\hphantom{-}9.12} & \underset{(0.57)}{\hphantom{-}4.3} \\ \underset{(0.03)}{5.06} & \underset{(0.07)}{\hphantom{-}4.95} & \underset{(0.27)}{\hphantom{-}4.57} & \underset{(1.97)}{\hphantom{-}8.62} \\ \end{bmatrix}$ 
 	
 		\\ 
 		[40pt]
 	\end{tabular}
 	\begin{minipage}{1\textwidth} %
 		{   \footnotesize  
 			\textit{Note:} 
 			Monte Carlo simulation with $M = 2000$ replications.
 		The simulation extends the simultaneous interaction considered in Section \ref{sec: Finite Sample Performance} with a  VAR $y_t = A_1 y_{t-1} + u_t$ and $u_t = B_0 \tilde{\varepsilon}_t$ with $ \tilde{\varepsilon}_{i,t} = \psi_t \sigma_{i,t} \varepsilon_{i,t} + (1-\psi_t) \varepsilon_{i,t}$, where  $\varepsilon_t$ are i.i.d. as defined in the main text, $\psi_t$ is a random variable with a  Bernoulli distribution with $Pr(\psi_t=1)=Pr(\psi_t=0)=0.5$, and $\sigma_{i,t}=i$ for $i=1,...,n$. 
 		The table
 		shows the average, $1/M 
 		\sum_{m=1}^{M}
 		\hat{b}^{m}_{ij} $ and in parentheses the mean squared error $1/M 
 		\sum_{m=1}^{M}
 		(\hat{b}^{m}_{ij} - b_ij)^2 $
 		of
 		each estimated element  $\hat{b}^{m}_{ij}$ in simulation $m$. 
 		The ML estimator assumes independent shocks with a skewed t-distribution, compare to \cite{lanne2017identification}.
 			\par}
 	\end{minipage}
 \end{table}

Table \ref{Table: Finite sample performance - average and mean squared error - simulation with smaller a misspecification} shows a simulation anlogously to the main text, however, with a smaller degree of misspecification for the $RCSUE(\mathcal{R}_2)$ estimator. Specifically, the SVAR is equal to
\begin{align} 
\begin{bmatrix}
u_{1t} \\
u_{2t} \\
u_{3t} \\
u_{4t} \\
\end{bmatrix} =
\begin{bmatrix}
10 & 0  & 0 & 0 \\
5 & 10 & 0 & 0\\
5 & 5 & 10 & 1\\
5 & 5 & 5 & 10
\end{bmatrix}
\begin{bmatrix}
\varepsilon_{1t} \\
\varepsilon_{2t} \\
\varepsilon_{3t} \\
\varepsilon_{4t} \\
\end{bmatrix} .
\end{align}  
In contrast to the simulation in the main text, the $b_{34}$ element of the data-generating process is equal to one  instead of five. The $\mathcal{R}_2$ penalty shrinks the $b_{34}$ element to zero and consequently, the degree of misspecification is smaller in this simulation compared to the main text.
The table shows that the incorrect restriction leads to a bias. However, in the smallest considered sample the MSE all elements, including the incorrectly penalized $b_{34}$ element,  of the ridge estimator is smaller compared to the MSE of the unrestricted CSUE. Moreover, summing up the MSE of all elements for each estimator shows that the total MSE of the ridge estimator is smaller compared to the total MSE of the unrestricted CSUE for all sample sizes. Consequently, the simulation shows that smaller misspecifications are beneficial for the MSE of the ridge estimator. However, the simulations also indicate that smaller misspecifications are more difficult to detect and neglect. Specifically, the simulation with the largest sample size of $T=5000$ observations still shows a considerable bias for the $b_{34}$ element of the ridge estimator.
  \begin{table}
 	\caption{Finite sample performance - average and mean squared error - simulation with smaller a misspecification}
 	\label{Table: Finite sample performance - average and mean squared error - simulation with smaller a misspecification}
 	\centering
 	\renewcommand{\arraystretch}{1}
 	\begin{tabular}{c | c c c}
 		& $CSUE$ & $RCSUE(\mathcal{R}_2)$  \\
 		\hline
 		\rotatebox[origin=c]{90}{$T = 500$ } &
$\begin{bmatrix}\underset{(0.23)}{9.88} & \underset{(0.6)}{0.0} & \underset{(0.63)}{\hphantom{-}0.02} & \underset{(0.58)}{-0.01} \\ \underset{(0.64)}{4.97} & \underset{(0.35)}{\hphantom{-}9.88} & \underset{(0.73)}{\hphantom{-}0.01} & \underset{(0.79)}{-0.0} \\ \underset{(0.87)}{4.93} & \underset{(0.8)}{\hphantom{-}4.95} & \underset{(0.51)}{\hphantom{-}9.88} & \underset{(0.97)}{\hphantom{-}0.98} \\ \underset{(1.03)}{4.94} & \underset{(1.04)}{\hphantom{-}4.93} & \underset{(0.97)}{\hphantom{-}4.95} & \underset{(0.64)}{\hphantom{-}9.87} \\ \end{bmatrix}$

 		&
$\begin{bmatrix}\underset{(0.18)}{9.93} & \textcolor{red}{\underset{(0.02)}{0.0}} & \textcolor{red}{\underset{(0.02)}{0.0}} & \textcolor{red}{\underset{(0.02)}{-0.01}} \\ \underset{(0.21)}{4.98} & \underset{(0.19)}{\hphantom{-}9.93} & \textcolor{red}{\underset{(0.02)}{0.0}} & \textcolor{red}{\underset{(0.02)}{-0.02}} \\ \underset{(0.24)}{4.96} & \underset{(0.22)}{\hphantom{-}4.97} & \underset{(0.2)}{\hphantom{-}9.93} & \textcolor{red}{\underset{(0.88)}{\hphantom{-}0.12}} \\ \underset{(0.28)}{4.96} & \underset{(0.3)}{\hphantom{-}4.96} & \underset{(0.67)}{\hphantom{-}5.61} & \underset{(0.5)}{\hphantom{-}9.47} \\ \end{bmatrix}$

 		\\ 
 		[40pt]
 		\rotatebox[origin=c]{90}{$T = 1000$ } &
$\begin{bmatrix}\underset{(0.11)}{9.94} & \underset{(0.29)}{-0.01} & \underset{(0.28)}{\hphantom{-}0.01} & \underset{(0.28)}{0.0} \\ \underset{(0.3)}{4.99} & \underset{(0.17)}{\hphantom{-}9.93} & \underset{(0.36)}{0.0} & \underset{(0.36)}{0.0} \\ \underset{(0.37)}{4.97} & \underset{(0.38)}{\hphantom{-}4.96} & \underset{(0.22)}{\hphantom{-}9.95} & \underset{(0.42)}{\hphantom{-}1.01} \\ \underset{(0.45)}{4.98} & \underset{(0.47)}{\hphantom{-}4.97} & \underset{(0.45)}{\hphantom{-}4.98} & \underset{(0.28)}{\hphantom{-}9.95} \\ \end{bmatrix}$

 		&
$\begin{bmatrix}\underset{(0.09)}{9.95} & \textcolor{red}{\underset{(0.01)}{-0.0}} & \textcolor{red}{\underset{(0.01)}{0.0}} & \textcolor{red}{\underset{(0.01)}{-0.01}} \\ \underset{(0.09)}{4.98} & \underset{(0.09)}{\hphantom{-}9.95} & \textcolor{red}{\underset{(0.01)}{0.0}} & \textcolor{red}{\underset{(0.01)}{-0.02}} \\ \underset{(0.1)}{4.97} & \underset{(0.1)}{\hphantom{-}4.96} & \underset{(0.09)}{\hphantom{-}9.96} & \textcolor{red}{\underset{(0.8)}{\hphantom{-}0.16}} \\ \underset{(0.12)}{4.98} & \underset{(0.13)}{\hphantom{-}4.98} & \underset{(0.52)}{\hphantom{-}5.6} & \underset{(0.34)}{\hphantom{-}9.52} \\ \end{bmatrix}$

 		\\ 
 		[40pt]
 		\rotatebox[origin=c]{90}{$T = 5000$ } &
$\begin{bmatrix}\underset{(0.02)}{9.99} & \underset{(0.05)}{0.0} & \underset{(0.05)}{-0.01} & \underset{(0.05)}{0.0} \\ \underset{(0.06)}{4.99} & \underset{(0.03)}{\hphantom{-}10.0} & \underset{(0.07)}{-0.0} & \underset{(0.07)}{\hphantom{-}0.01} \\ \underset{(0.07)}{5.01} & \underset{(0.07)}{\hphantom{-}4.99} & \underset{(0.04)}{\hphantom{-}9.99} & \underset{(0.08)}{\hphantom{-}1.02} \\ \underset{(0.09)}{5.0} & \underset{(0.09)}{\hphantom{-}5.0} & \underset{(0.09)}{\hphantom{-}4.99} & \underset{(0.06)}{\hphantom{-}10.01} \\ \end{bmatrix}$

 		&
$\begin{bmatrix}\underset{(0.02)}{9.99} & \textcolor{red}{\underset{(0.0)}{0.0}} & \textcolor{red}{\underset{(0.0)}{-0.0}} & \textcolor{red}{\underset{(0.0)}{-0.01}} \\ \underset{(0.02)}{4.99} & \underset{(0.02)}{\hphantom{-}10.0} & \textcolor{red}{\underset{(0.0)}{-0.0}} & \textcolor{red}{\underset{(0.0)}{-0.02}} \\ \underset{(0.02)}{5.0} & \underset{(0.02)}{\hphantom{-}5.0} & \underset{(0.02)}{\hphantom{-}9.98} & \textcolor{red}{\underset{(0.3)}{\hphantom{-}0.55}} \\ \underset{(0.02)}{5.01} & \underset{(0.02)}{\hphantom{-}5.01} & \underset{(0.17)}{\hphantom{-}5.31} & \underset{(0.1)}{\hphantom{-}9.77} \\ \end{bmatrix}$

 		\\ 
 		[40pt]
 	\end{tabular}
 	\begin{minipage}{1\textwidth} %
 		{   \footnotesize  
 			\textit{Note:} 
 			Monte Carlo simulation with $M = 2000$ replications.
 		The simulation is analogous to the simultaneous considered in Section \ref{sec: Finite Sample Performance}, however, it alters the simultaneous interaction and uses     $B_0 = \begin{bmatrix}
10 & 0  & 0 & 0 \\
5 & 10 & 0 & 0\\
5 & 5 & 10 & 1\\
5 & 5 & 5 & 10
\end{bmatrix}$.
 		The table
 		shows the average, $1/M 
 		\sum_{m=1}^{M}
 		\hat{b}^{m}_{ij} $ and in parentheses the mean squared error $1/M 
 		\sum_{m=1}^{M}
 		(\hat{b}^{m}_{ij} - b_ij)^2 $
 		of
 		each estimated element  $\hat{b}^{m}_{ij}$ in simulation $m$. 
 			\par}
 	\end{minipage}
 \end{table}

Table \ref{Table: Finite sample performance - Bias MSE - MC_RGMM_1Gaus} to Table \ref{Table: Finite sample performance - Bias MSE - MC_RGMM_4Gaus} show several simulations with Gaussian shocks.

Table \ref{Table: Finite sample performance - Bias MSE - MC_RGMM_1Gaus} shows a simulation where the second shock is Gaussian. With only one Gaussian shock, the SVAR remains fully identified, the weights of the RCSUE can still be estimated properly, and the cross-validation can detect tuning parameters which lead to dependent shocks. Consequently, the performance of all estimators is similar to the results in the main text.

Table \ref{Table: Finite sample performance - Bias MSE - MC_RGMM_2Gaus} explores a scenario where the second and third shocks are Gaussian. In this case, only the columns of $B_0$ corresponding to non-Gaussian shocks are identified.  Consequently, the second and third columns of the CSUE estimator are not identified. 
Moreover, mixtures of the second and third shock are also Gaussian and the cross-validation is not able to detect these mixtures. Therefore, the data is no longer able to provide evidence against restrictions on the second and third column. By construction, if the data is unable to provide evidence against a given restriction due to the Gaussianity of the shocks, the adaptive RCSUE weights from Equation (\ref{eq: adaptive weights}) of these restrictions converge to infinity. Consequently, both RCSUE estimators  increase shrinkage towards the restrictions in the second and third column. Since all restrictions in the second and third column are correct, the overall performance of both ridge estimators is similar to the performance in the main text.

Table \ref{Table: Finite sample performance - Bias MSE - MC_RGMM_3Gaus} shows a simulation where the second, third, and fourth shocks are Gaussian. Therefore, only the first column of $B_0$ is identified and the RCSUE can not provide evidence against incorrect restrictions on elements in the second, third, and fourth column. The RCSUE($\mathcal{R}_1$) only uses correct restrictions and its performance is similar to the performance in the main text, despite the three Gaussian shocks.
However, the RCSUE($\mathcal{R}_2$) uses an incorrect restriction in the fourth column and since the data is unable to provide evidence against this restriction, the RCSUE($\mathcal{R}_2$) is unable to stop shrinkage towards the incorrect restriction.

In Table \ref{Table: Finite sample performance - Bias MSE - MC_RGMM_4Gaus}, a simulation is conducted with the fourth shock being Gaussian. With only one Gaussian shock, the SVAR retains full identification, resulting in performance comparable to the main text results for all estimators.

\begin{table}
	\caption{Finite sample performance - average and mean squared error - simulation with $\varepsilon_{2t}$ Gaussian}
	\label{Table: Finite sample performance - Bias MSE - MC_RGMM_1Gaus}
	\centering
	\renewcommand{\arraystretch}{1}
	\begin{tabular}{c | c c c}
		& $CSUE$ & $RCSUE(\mathcal{R}_1)$ & $RCSUE(\mathcal{R}_2)$ \\
		\hline
	\rotatebox[origin=c]{90}{$T = 250$ } &$\begin{bmatrix}\underset{(0.59)}{9.73} & \underset{(2.04)}{\hphantom{-}0.02} & \underset{(1.35)}{\hphantom{-}0.01} & \underset{(1.43)}{\hphantom{-}0.05} \\ \underset{(2.58)}{4.92} & \underset{(0.99)}{\hphantom{-}9.6} & \underset{(2.72)}{\hphantom{-}0.11} & \underset{(2.87)}{\hphantom{-}0.02} \\ \underset{(2.38)}{4.86} & \underset{(3.35)}{\hphantom{-}4.78} & \underset{(1.65)}{\hphantom{-}9.77} & \underset{(2.25)}{\hphantom{-}4.88} \\ \underset{(2.36)}{4.84} & \underset{(3.38)}{\hphantom{-}4.83} & \underset{(2.31)}{\hphantom{-}4.91} & \underset{(1.76)}{\hphantom{-}9.71} \\ \end{bmatrix}$ &$\begin{bmatrix}\underset{(0.37)}{9.92} & \textcolor{red}{\underset{(0.03)}{-0.0}} & \textcolor{red}{\underset{(0.08)}{0.0}} & \textcolor{red}{\underset{(0.05)}{\hphantom{-}0.01}} \\ \underset{(0.57)}{4.99} & \underset{(0.22)}{\hphantom{-}9.92} & \textcolor{red}{\underset{(0.14)}{0.0}} & \textcolor{red}{\underset{(0.13)}{\hphantom{-}0.01}} \\ \underset{(0.66)}{4.96} & \underset{(0.65)}{\hphantom{-}5.0} & \underset{(0.77)}{\hphantom{-}9.84} & \underset{(1.28)}{\hphantom{-}4.93} \\ \underset{(0.63)}{4.96} & \underset{(0.64)}{\hphantom{-}5.0} & \underset{(1.36)}{\hphantom{-}4.92} & \underset{(0.71)}{\hphantom{-}9.83} \\ \end{bmatrix}$ &$\begin{bmatrix}\underset{(0.4)}{9.89} & \textcolor{red}{\underset{(0.06)}{0.0}} & \textcolor{red}{\underset{(0.16)}{-0.0}} & \textcolor{red}{\underset{(0.15)}{-0.03}} \\ \underset{(0.63)}{4.96} & \underset{(0.25)}{\hphantom{-}9.89} & \textcolor{red}{\underset{(0.32)}{-0.0}} & \textcolor{red}{\underset{(0.37)}{-0.12}} \\ \underset{(0.82)}{4.95} & \underset{(0.86)}{\hphantom{-}5.0} & \underset{(0.82)}{\hphantom{-}10.32} & \textcolor{red}{\underset{(10.1)}{\hphantom{-}2.46}} \\ \underset{(0.82)}{4.96} & \underset{(0.91)}{\hphantom{-}5.06} & \underset{(4.14)}{\hphantom{-}6.39} & \underset{(5.11)}{\hphantom{-}8.19} \\ \end{bmatrix}$ \\ [40pt]\rotatebox[origin=c]{90}{$T = 500$ } &$\begin{bmatrix}\underset{(0.24)}{9.87} & \underset{(1.06)}{\hphantom{-}0.02} & \underset{(0.59)}{\hphantom{-}0.02} & \underset{(0.65)}{-0.03} \\ \underset{(1.41)}{4.91} & \underset{(0.42)}{\hphantom{-}9.79} & \underset{(1.47)}{\hphantom{-}0.02} & \underset{(1.53)}{-0.0} \\ \underset{(1.13)}{4.92} & \underset{(1.77)}{\hphantom{-}4.91} & \underset{(0.83)}{\hphantom{-}9.87} & \underset{(1.1)}{\hphantom{-}4.93} \\ \underset{(1.17)}{4.94} & \underset{(1.83)}{\hphantom{-}4.92} & \underset{(1.12)}{\hphantom{-}4.93} & \underset{(0.83)}{\hphantom{-}9.85} \\ \end{bmatrix}$ &$\begin{bmatrix}\underset{(0.18)}{9.96} & \textcolor{red}{\underset{(0.01)}{0.0}} & \textcolor{red}{\underset{(0.01)}{0.0}} & \textcolor{red}{\underset{(0.02)}{-0.0}} \\ \underset{(0.25)}{4.97} & \underset{(0.1)}{\hphantom{-}9.96} & \textcolor{red}{\underset{(0.06)}{-0.0}} & \textcolor{red}{\underset{(0.05)}{\hphantom{-}0.01}} \\ \underset{(0.27)}{4.97} & \underset{(0.3)}{\hphantom{-}4.99} & \underset{(0.33)}{\hphantom{-}9.91} & \underset{(0.55)}{\hphantom{-}4.97} \\ \underset{(0.28)}{4.98} & \underset{(0.28)}{\hphantom{-}5.0} & \underset{(0.58)}{\hphantom{-}4.95} & \underset{(0.31)}{\hphantom{-}9.93} \\ \end{bmatrix}$ &$\begin{bmatrix}\underset{(0.19)}{9.94} & \textcolor{red}{\underset{(0.02)}{-0.0}} & \textcolor{red}{\underset{(0.05)}{0.0}} & \textcolor{red}{\underset{(0.06)}{-0.03}} \\ \underset{(0.29)}{4.96} & \underset{(0.11)}{\hphantom{-}9.93} & \textcolor{red}{\underset{(0.14)}{-0.01}} & \textcolor{red}{\underset{(0.19)}{-0.09}} \\ \underset{(0.35)}{4.97} & \underset{(0.39)}{\hphantom{-}5.01} & \underset{(0.45)}{\hphantom{-}10.25} & \textcolor{red}{\underset{(4.78)}{\hphantom{-}3.37}} \\ \underset{(0.36)}{4.99} & \underset{(0.4)}{\hphantom{-}5.05} & \underset{(2.07)}{\hphantom{-}5.92} & \underset{(2.25)}{\hphantom{-}8.88} \\ \end{bmatrix}$ \\ [40pt]\rotatebox[origin=c]{90}{$T = 1000$ } &$\begin{bmatrix}\underset{(0.1)}{9.93} & \underset{(0.48)}{0.0} & \underset{(0.27)}{-0.02} & \underset{(0.28)}{-0.01} \\ \underset{(0.62)}{4.98} & \underset{(0.18)}{\hphantom{-}9.9} & \underset{(0.68)}{0.0} & \underset{(0.7)}{-0.01} \\ \underset{(0.5)}{4.99} & \underset{(0.8)}{\hphantom{-}4.96} & \underset{(0.38)}{\hphantom{-}9.95} & \underset{(0.52)}{\hphantom{-}4.94} \\ \underset{(0.51)}{4.99} & \underset{(0.79)}{\hphantom{-}4.96} & \underset{(0.52)}{\hphantom{-}4.99} & \underset{(0.38)}{\hphantom{-}9.92} \\ \end{bmatrix}$ &$\begin{bmatrix}\underset{(0.08)}{9.97} & \textcolor{red}{\underset{(0.01)}{0.0}} & \textcolor{red}{\underset{(0.01)}{0.0}} & \textcolor{red}{\underset{(0.01)}{0.0}} \\ \underset{(0.11)}{5.0} & \underset{(0.05)}{\hphantom{-}9.98} & \textcolor{red}{\underset{(0.02)}{-0.0}} & \textcolor{red}{\underset{(0.02)}{-0.0}} \\ \underset{(0.11)}{4.99} & \underset{(0.12)}{\hphantom{-}5.0} & \underset{(0.14)}{\hphantom{-}9.97} & \underset{(0.23)}{\hphantom{-}4.97} \\ \underset{(0.12)}{5.0} & \underset{(0.12)}{\hphantom{-}5.01} & \underset{(0.23)}{\hphantom{-}5.0} & \underset{(0.14)}{\hphantom{-}9.95} \\ \end{bmatrix}$ &$\begin{bmatrix}\underset{(0.09)}{9.97} & \textcolor{red}{\underset{(0.01)}{0.0}} & \textcolor{red}{\underset{(0.02)}{-0.0}} & \textcolor{red}{\underset{(0.02)}{-0.02}} \\ \underset{(0.14)}{4.99} & \underset{(0.05)}{\hphantom{-}9.97} & \textcolor{red}{\underset{(0.04)}{-0.01}} & \textcolor{red}{\underset{(0.07)}{-0.06}} \\ \underset{(0.15)}{5.0} & \underset{(0.16)}{\hphantom{-}5.02} & \underset{(0.2)}{\hphantom{-}10.18} & \textcolor{red}{\underset{(1.53)}{\hphantom{-}4.08}} \\ \underset{(0.16)}{5.01} & \underset{(0.17)}{\hphantom{-}5.04} & \underset{(0.72)}{\hphantom{-}5.55} & \underset{(0.71)}{\hphantom{-}9.39} \\ \end{bmatrix}$ \\ [40pt]
	
	\end{tabular}
				\begin{minipage}{1\textwidth} %
		{   \footnotesize  
			\textit{Note:} 
			Monte Carlo simulation with $M = 2000$ replications.
				The simulation is analogous to the simultaneous in     Section \ref{sec: Finite Sample Performance}, however, the shock $\varepsilon_{2t}$ is now drawn from a standard normal distribution.
			The table
			shows the average, $1/M 
			\sum_{m=1}^{M}
			\hat{b}^{m}_{ij} $ and in parentheses the mean squared error $1/M 
			\sum_{m=1}^{M}
			(\hat{b}^{m}_{ij} - b_ij)^2 $
			of
			each estimated element  $\hat{b}^{m}_{ij}$ in simulation $m$. Penalized elements are highlighted in red.
			\par}
	\end{minipage}
\end{table}

\begin{table}
	\caption{Finite sample performance - average and mean squared error - simulation with $\varepsilon_{2t}$ and $\varepsilon_{3t}$ Gaussian}
	\label{Table: Finite sample performance - Bias MSE - MC_RGMM_2Gaus}
	\centering
	\renewcommand{\arraystretch}{1}
	\begin{tabular}{c | c c c}
		& $CSUE$ & $RCSUE(\mathcal{R}_1)$ & $RCSUE(\mathcal{R}_2)$ \\
		\hline
	 \rotatebox[origin=c]{90}{$T = 250$ } &
	 $\begin{bmatrix}\underset{(0.67)}{9.65} & / & / & \underset{(1.48)}{\hphantom{-}0.06} \\ \underset{(2.66)}{4.83} & / & / & \underset{(3.02)}{\hphantom{-}0.08} \\ \underset{(3.68)}{4.82} & / & / & \underset{(3.78)}{\hphantom{-}4.88} \\ \underset{(2.74)}{4.8} & / & / & \underset{(2.17)}{\hphantom{-}9.68} \\ \end{bmatrix}$

	 &$\begin{bmatrix}\underset{(0.35)}{9.91} & \textcolor{red}{\underset{(0.0)}{0.0}} & \textcolor{red}{\underset{(0.0)}{0.0}} & \textcolor{red}{\underset{(0.05)}{\hphantom{-}0.01}} \\ \underset{(0.5)}{4.96} & \underset{(0.2)}{\hphantom{-}9.92} & \textcolor{red}{\underset{(0.03)}{-0.0}} & \textcolor{red}{\underset{(0.14)}{\hphantom{-}0.01}} \\ \underset{(0.7)}{4.99} & \underset{(0.6)}{\hphantom{-}4.99} & \underset{(0.68)}{\hphantom{-}9.85} & \underset{(2.19)}{\hphantom{-}4.91} \\ \underset{(0.65)}{4.99} & \underset{(0.59)}{\hphantom{-}4.97} & \underset{(1.78)}{\hphantom{-}4.96} & \underset{(0.95)}{\hphantom{-}9.81} \\ \end{bmatrix}$ &$\begin{bmatrix}\underset{(0.39)}{9.87} & \textcolor{red}{\underset{(0.0)}{0.0}} & \textcolor{red}{\underset{(0.0)}{-0.0}} & \textcolor{red}{\underset{(0.11)}{-0.01}} \\ \underset{(0.58)}{4.92} & \underset{(0.22)}{\hphantom{-}9.89} & \textcolor{red}{\underset{(0.03)}{-0.0}} & \textcolor{red}{\underset{(0.34)}{-0.06}} \\ \underset{(0.91)}{4.93} & \underset{(0.75)}{\hphantom{-}4.95} & \underset{(1.0)}{\hphantom{-}10.64} & \textcolor{red}{\underset{(13.4)}{\hphantom{-}1.94}} \\ \underset{(0.94)}{4.94} & \underset{(0.86)}{\hphantom{-}4.97} & \underset{(7.4)}{\hphantom{-}7.12} & \underset{(6.69)}{\hphantom{-}7.82} \\ \end{bmatrix}$ \\ [40pt]
	 
	 \rotatebox[origin=c]{90}{$T = 500$ } &$\begin{bmatrix}\underset{(0.25)}{9.84} & / & / & \underset{(0.65)}{-0.0} \\ \underset{(1.42)}{4.84} & / & / & \underset{(1.57)}{0.0} \\ \underset{(1.94)}{4.85} & / & / & \underset{(2.07)}{\hphantom{-}4.9} \\ \underset{(1.34)}{4.87} & /& / & \underset{(1.13)}{\hphantom{-}9.81} \\ \end{bmatrix}$

	 &$\begin{bmatrix}\underset{(0.16)}{9.97} & \textcolor{red}{\underset{(0.0)}{0.0}} & \textcolor{red}{\underset{(0.0)}{-0.0}} & \textcolor{red}{\underset{(0.02)}{-0.0}} \\ \underset{(0.2)}{4.98} & \underset{(0.1)}{\hphantom{-}9.97} & \textcolor{red}{\underset{(0.01)}{-0.0}} & \textcolor{red}{\underset{(0.06)}{-0.01}} \\ \underset{(0.27)}{5.02} & \underset{(0.25)}{\hphantom{-}5.01} & \underset{(0.26)}{\hphantom{-}9.94} & \underset{(0.91)}{\hphantom{-}4.93} \\ \underset{(0.24)}{5.03} & \underset{(0.23)}{\hphantom{-}5.0} & \underset{(0.69)}{\hphantom{-}5.0} & \underset{(0.4)}{\hphantom{-}9.89} \\ \end{bmatrix}$ &$\begin{bmatrix}\underset{(0.2)}{9.95} & \textcolor{red}{\underset{(0.0)}{0.0}} & \textcolor{red}{\underset{(0.0)}{-0.0}} & \textcolor{red}{\underset{(0.06)}{-0.02}} \\ \underset{(0.28)}{4.96} & \underset{(0.11)}{\hphantom{-}9.95} & \textcolor{red}{\underset{(0.0)}{-0.0}} & \textcolor{red}{\underset{(0.2)}{-0.08}} \\ \underset{(0.42)}{4.99} & \underset{(0.34)}{\hphantom{-}4.99} & \underset{(0.6)}{\hphantom{-}10.5} & \textcolor{red}{\underset{(8.11)}{\hphantom{-}2.8}} \\ \underset{(0.42)}{5.01} & \underset{(0.39)}{\hphantom{-}5.01} & \underset{(4.22)}{\hphantom{-}6.5} & \underset{(3.81)}{\hphantom{-}8.5} \\ \end{bmatrix}$ \\ [40pt]
	 
	 \rotatebox[origin=c]{90}{$T = 1000$ } &$\begin{bmatrix}\underset{(0.11)}{9.93} & / & / & \underset{(0.29)}{\hphantom{-}0.01} \\ \underset{(0.72)}{4.96} & / & / & \underset{(0.73)}{\hphantom{-}0.01} \\ \underset{(0.85)}{4.96} &/& / & \underset{(0.87)}{\hphantom{-}4.95} \\ \underset{(0.6)}{4.95} & /& / & \underset{(0.46)}{\hphantom{-}9.91} \\ \end{bmatrix}$ 
	 
	 &$\begin{bmatrix}\underset{(0.07)}{9.98} & \textcolor{red}{\underset{(0.0)}{0.0}} & \textcolor{red}{\underset{(0.0)}{0.0}} & \textcolor{red}{\underset{(0.01)}{0.0}} \\ \underset{(0.08)}{5.0} & \underset{(0.04)}{\hphantom{-}9.99} & \textcolor{red}{\underset{(0.0)}{0.0}} & \textcolor{red}{\underset{(0.02)}{-0.0}} \\ \underset{(0.09)}{5.0} & \underset{(0.09)}{\hphantom{-}5.0} & \underset{(0.08)}{\hphantom{-}9.98} & \underset{(0.26)}{\hphantom{-}4.97} \\ \underset{(0.09)}{5.01} & \underset{(0.1)}{\hphantom{-}5.0} & \underset{(0.21)}{\hphantom{-}5.0} & \underset{(0.12)}{\hphantom{-}9.96} \\ \end{bmatrix}$ &$\begin{bmatrix}\underset{(0.1)}{9.99} & \textcolor{red}{\underset{(0.0)}{0.0}} & \textcolor{red}{\underset{(0.0)}{0.0}} & \textcolor{red}{\underset{(0.02)}{-0.01}} \\ \underset{(0.13)}{5.0} & \underset{(0.05)}{\hphantom{-}9.99} & \textcolor{red}{\underset{(0.0)}{0.0}} & \textcolor{red}{\underset{(0.07)}{-0.05}} \\ \underset{(0.18)}{5.01} & \underset{(0.16)}{\hphantom{-}5.01} & \underset{(0.3)}{\hphantom{-}10.35} & \textcolor{red}{\underset{(3.05)}{\hphantom{-}3.73}} \\ \underset{(0.17)}{5.02} & \underset{(0.17)}{\hphantom{-}5.04} & \underset{(1.67)}{\hphantom{-}5.88} & \underset{(1.34)}{\hphantom{-}9.18} \\ \end{bmatrix}$ \\ [40pt]
	
	\end{tabular}
				\begin{minipage}{1\textwidth} %
		{   \footnotesize  
			\textit{Note:} 
			Monte Carlo simulation with $M = 2000$ replications.
			The simulation is analogous to the simultaneous in     Section \ref{sec: Finite Sample Performance}, however, the shocks $\varepsilon_{2t}$ and $\varepsilon_{3t}$ are now drawn from a standard normal distribution.
			The table
			shows the average, $1/M 
			\sum_{m=1}^{M}
			\hat{b}^{m}_{ij} $ and in parentheses the mean squared error $1/M 
			\sum_{m=1}^{M}
			(\hat{b}^{m}_{ij} - b_ij)^2 $
			of
			each estimated element  $\hat{b}^{m}_{ij}$ in simulation $m$. Penalized elements are highlighted in red. The CSUE only shows identified elements.
			\par}
	\end{minipage}
\end{table}

\begin{table}
	\caption{Finite sample performance - average and mean squared error - simulation with $\varepsilon_{2t}$, $\varepsilon_{3t}$, and $\varepsilon_{4t}$ Gaussian}
	\label{Table: Finite sample performance - Bias MSE - MC_RGMM_3Gaus}
	\centering
	\renewcommand{\arraystretch}{1}
	\begin{tabular}{c | c c c}
		& $CSUE$ & $RCSUE(\mathcal{R}_1)$ & $RCSUE(\mathcal{R}_2)$ \\
		\hline
	\rotatebox[origin=c]{90}{$T = 250$ } 
	&$\begin{bmatrix}\underset{(0.73)}{9.61} &/ & / & / \\ \underset{(2.77)}{4.84} & / & / & / \\ \underset{(4.03)}{4.83} & / & / & / \\ \underset{(4.03)}{4.82} & / & / & / \\ \end{bmatrix}$ 
	
	&$\begin{bmatrix}\underset{(0.35)}{9.92} & \textcolor{red}{\underset{(0.0)}{0.0}} & \textcolor{red}{\underset{(0.02)}{-0.0}} & \textcolor{red}{\underset{(0.01)}{0.0}} \\ \underset{(0.46)}{4.99} & \underset{(0.2)}{\hphantom{-}9.96} & \textcolor{red}{\underset{(0.03)}{0.0}} & \textcolor{red}{\underset{(0.01)}{0.0}} \\ \underset{(0.62)}{5.02} & \underset{(0.58)}{\hphantom{-}5.03} & \underset{(0.89)}{\hphantom{-}9.79} & \underset{(2.5)}{\hphantom{-}4.96} \\ \underset{(0.62)}{5.03} & \underset{(0.56)}{\hphantom{-}5.04} & \underset{(2.54)}{\hphantom{-}4.89} & \underset{(0.84)}{\hphantom{-}9.83} \\ \end{bmatrix}$ &$\begin{bmatrix}\underset{(0.37)}{9.85} & \textcolor{red}{\underset{(0.01)}{0.0}} & \textcolor{red}{\underset{(0.03)}{0.0}} & \textcolor{red}{\underset{(0.02)}{-0.0}} \\ \underset{(0.5)}{4.91} & \underset{(0.21)}{\hphantom{-}9.93} & \textcolor{red}{\underset{(0.04)}{-0.01}} & \textcolor{red}{\underset{(0.03)}{-0.0}} \\ \underset{(0.69)}{4.91} & \underset{(0.68)}{\hphantom{-}4.96} & \underset{(1.39)}{\hphantom{-}11.04} & \textcolor{red}{\underset{(24.59)}{\hphantom{-}0.07}} \\ \underset{(0.7)}{4.92} & \underset{(0.66)}{\hphantom{-}4.97} & \underset{(16.59)}{\hphantom{-}8.74} & \underset{(13.23)}{\hphantom{-}6.65} \\ \end{bmatrix}$ \\ [40pt]
	
	\rotatebox[origin=c]{90}{$T = 500$ } &$\begin{bmatrix}\underset{(0.29)}{9.81} & / & / & /\\ \underset{(1.6)}{4.95} & / & / &/ \\ \underset{(2.11)}{4.9} & /& /& /\\ \underset{(2.09)}{4.9} & / & /& / \\ \end{bmatrix}$ 
	
	&$\begin{bmatrix}\underset{(0.15)}{9.96} & \textcolor{red}{\underset{(0.0)}{0.0}} & \textcolor{red}{\underset{(0.01)}{0.0}} & \textcolor{red}{\underset{(0.0)}{0.0}} \\ \underset{(0.19)}{5.0} & \underset{(0.09)}{\hphantom{-}9.96} & \textcolor{red}{\underset{(0.0)}{0.0}} & \textcolor{red}{\underset{(0.0)}{0.0}} \\ \underset{(0.21)}{5.01} & \underset{(0.2)}{\hphantom{-}5.02} & \underset{(0.27)}{\hphantom{-}9.92} & \underset{(0.77)}{\hphantom{-}5.0} \\ \underset{(0.21)}{5.01} & \underset{(0.2)}{\hphantom{-}5.02} & \underset{(0.77)}{\hphantom{-}4.96} & \underset{(0.26)}{\hphantom{-}9.95} \\ \end{bmatrix}$ &$\begin{bmatrix}\underset{(0.22)}{9.9} & \textcolor{red}{\underset{(0.0)}{0.0}} & \textcolor{red}{\underset{(0.0)}{0.0}} & \textcolor{red}{\underset{(0.0)}{0.0}} \\ \underset{(0.28)}{4.93} & \underset{(0.11)}{\hphantom{-}9.93} & \textcolor{red}{\underset{(0.0)}{-0.0}} & \textcolor{red}{\underset{(0.0)}{0.0}} \\ \underset{(0.39)}{4.91} & \underset{(0.42)}{\hphantom{-}4.95} & \underset{(1.46)}{\hphantom{-}11.1} & \textcolor{red}{\underset{(24.92)}{\hphantom{-}0.02}} \\ \underset{(0.38)}{4.9} & \underset{(0.42)}{\hphantom{-}4.95} & \underset{(15.33)}{\hphantom{-}8.87} & \underset{(11.2)}{\hphantom{-}6.68} \\ \end{bmatrix}$ \\ [40pt]
	
	\rotatebox[origin=c]{90}{$T = 1000$ } &$\begin{bmatrix}\underset{(0.12)}{9.91} & / & / & / \\ \underset{(0.73)}{4.93}& / & / & /  \\ \underset{(1.06)}{4.96} & / & / & /  \\ \underset{(1.09)}{4.92} & / & / & / \\ \end{bmatrix}$ 
	
	&$\begin{bmatrix}\underset{(0.06)}{9.98} & \textcolor{red}{\underset{(0.0)}{0.0}} & \textcolor{red}{\underset{(0.0)}{0.0}} & \textcolor{red}{\underset{(0.0)}{0.0}} \\ \underset{(0.07)}{5.01} & \underset{(0.04)}{\hphantom{-}9.98} & \textcolor{red}{\underset{(0.0)}{0.0}} & \textcolor{red}{\underset{(0.0)}{0.0}} \\ \underset{(0.07)}{5.0} & \underset{(0.07)}{\hphantom{-}5.01} & \underset{(0.06)}{\hphantom{-}9.97} & \underset{(0.16)}{\hphantom{-}5.0} \\ \underset{(0.07)}{5.01} & \underset{(0.07)}{\hphantom{-}5.01} & \underset{(0.15)}{\hphantom{-}4.98} & \underset{(0.06)}{\hphantom{-}9.98} \\ \end{bmatrix}$ &$\begin{bmatrix}\underset{(0.19)}{9.92} & \textcolor{red}{\underset{(0.0)}{0.0}} & \textcolor{red}{\underset{(0.0)}{0.0}} & \textcolor{red}{\underset{(0.0)}{0.0}} \\ \underset{(0.23)}{4.94} & \underset{(0.07)}{\hphantom{-}9.95} & \textcolor{red}{\underset{(0.0)}{0.0}} & \textcolor{red}{\underset{(0.0)}{0.0}} \\ \underset{(0.33)}{4.91} & \underset{(0.24)}{\hphantom{-}4.92} & \underset{(1.26)}{\hphantom{-}11.06} & \textcolor{red}{\underset{(25.0)}{0.0}} \\ \underset{(0.33)}{4.91} & \underset{(0.23)}{\hphantom{-}4.92} & \underset{(14.76)}{\hphantom{-}8.82} & \underset{(11.07)}{\hphantom{-}6.69} \\ \end{bmatrix}$ \\ [40pt]
	
	\end{tabular}
				\begin{minipage}{1\textwidth} %
		{   \footnotesize  
			\textit{Note:} 
			Monte Carlo simulation with $M = 2000$ replications.
					The simulation is analogous to the simultaneous in     Section \ref{sec: Finite Sample Performance}, however, the shocks $\varepsilon_{2t}$, $\varepsilon_{3t}$, and $\varepsilon_{4t}$ are now drawn from a standard normal distribution.
			The table
			shows the average, $1/M 
			\sum_{m=1}^{M}
			\hat{b}^{m}_{ij} $ and in parentheses the mean squared error $1/M 
			\sum_{m=1}^{M}
			(\hat{b}^{m}_{ij} - b_ij)^2 $
			of
			each estimated element  $\hat{b}^{m}_{ij}$ in simulation $m$. Penalized elements are highlighted in red. The CSUE only shows identified elements.
			\par}
	\end{minipage}
\end{table}

\begin{table}
	\caption{Finite sample performance - average and mean squared error - simulation with $\varepsilon_{4t}$ Gaussian}
	\label{Table: Finite sample performance - Bias MSE - MC_RGMM_4Gaus}
	\centering
	\renewcommand{\arraystretch}{1}
	\begin{tabular}{c | c c c}
		& $CSUE$ & $RCSUE(\mathcal{R}_1)$ & $RCSUE(\mathcal{R}_2)$ \\
		\hline
	 \rotatebox[origin=c]{90}{$T = 250$ } &$\begin{bmatrix}\underset{(0.69)}{9.68} & \underset{(1.52)}{\hphantom{-}0.05} & \underset{(1.36)}{\hphantom{-}0.01} & \underset{(2.16)}{\hphantom{-}0.12} \\ \underset{(1.72)}{4.86} & \underset{(1.0)}{\hphantom{-}9.69} & \underset{(1.8)}{-0.0} & \underset{(2.76)}{\hphantom{-}0.2} \\ \underset{(2.47)}{4.81} & \underset{(2.52)}{\hphantom{-}4.85} & \underset{(1.78)}{\hphantom{-}9.69} & \underset{(3.39)}{\hphantom{-}4.95} \\ \underset{(3.39)}{4.77} & \underset{(3.49)}{\hphantom{-}4.76} & \underset{(3.29)}{\hphantom{-}4.84} & \underset{(2.05)}{\hphantom{-}9.71} \\ \end{bmatrix}$ &$\begin{bmatrix}\underset{(0.4)}{9.9} & \textcolor{red}{\underset{(0.08)}{\hphantom{-}0.01}} & \textcolor{red}{\underset{(0.11)}{\hphantom{-}0.01}} & \textcolor{red}{\underset{(0.04)}{-0.0}} \\ \underset{(0.53)}{5.0} & \underset{(0.42)}{\hphantom{-}9.89} & \textcolor{red}{\underset{(0.06)}{0.0}} & \textcolor{red}{\underset{(0.06)}{\hphantom{-}0.01}} \\ \underset{(0.74)}{4.97} & \underset{(0.66)}{\hphantom{-}4.98} & \underset{(0.99)}{\hphantom{-}9.8} & \underset{(1.85)}{\hphantom{-}4.93} \\ \underset{(0.76)}{4.98} & \underset{(0.76)}{\hphantom{-}4.93} & \underset{(2.3)}{\hphantom{-}4.89} & \underset{(0.74)}{\hphantom{-}9.82} \\ \end{bmatrix}$ &$\begin{bmatrix}\underset{(0.42)}{9.85} & \textcolor{red}{\underset{(0.17)}{\hphantom{-}0.02}} & \textcolor{red}{\underset{(0.22)}{\hphantom{-}0.01}} & \textcolor{red}{\underset{(0.12)}{-0.02}} \\ \underset{(0.62)}{4.95} & \underset{(0.47)}{\hphantom{-}9.85} & \textcolor{red}{\underset{(0.23)}{-0.02}} & \textcolor{red}{\underset{(0.18)}{-0.03}} \\ \underset{(0.95)}{4.92} & \underset{(0.9)}{\hphantom{-}4.96} & \underset{(1.15)}{\hphantom{-}10.66} & \textcolor{red}{\underset{(17.41)}{\hphantom{-}1.23}} \\ \underset{(0.96)}{4.92} & \underset{(1.0)}{\hphantom{-}4.92} & \underset{(8.84)}{\hphantom{-}7.52} & \underset{(7.9)}{\hphantom{-}7.45} \\ \end{bmatrix}$ \\ [40pt]\rotatebox[origin=c]{90}{$T = 500$ } &$\begin{bmatrix}\underset{(0.24)}{9.87} & \underset{(0.66)}{\hphantom{-}0.04} & \underset{(0.71)}{\hphantom{-}0.01} & \underset{(1.07)}{\hphantom{-}0.01} \\ \underset{(0.71)}{4.9} & \underset{(0.39)}{\hphantom{-}9.89} & \underset{(0.91)}{\hphantom{-}0.04} & \underset{(1.36)}{\hphantom{-}0.01} \\ \underset{(1.21)}{4.93} & \underset{(1.2)}{\hphantom{-}4.92} & \underset{(0.84)}{\hphantom{-}9.85} & \underset{(1.71)}{\hphantom{-}4.91} \\ \underset{(1.69)}{4.94} & \underset{(1.63)}{\hphantom{-}4.93} & \underset{(1.67)}{\hphantom{-}4.93} & \underset{(0.98)}{\hphantom{-}9.8} \\ \end{bmatrix}$ &$\begin{bmatrix}\underset{(0.18)}{9.97} & \textcolor{red}{\underset{(0.03)}{\hphantom{-}0.01}} & \textcolor{red}{\underset{(0.03)}{0.0}} & \textcolor{red}{\underset{(0.01)}{0.0}} \\ \underset{(0.22)}{4.99} & \underset{(0.19)}{\hphantom{-}9.97} & \textcolor{red}{\underset{(0.04)}{\hphantom{-}0.01}} & \textcolor{red}{\underset{(0.02)}{0.0}} \\ \underset{(0.3)}{5.0} & \underset{(0.29)}{\hphantom{-}4.98} & \underset{(0.46)}{\hphantom{-}9.9} & \underset{(0.86)}{\hphantom{-}4.96} \\ \underset{(0.34)}{5.01} & \underset{(0.31)}{\hphantom{-}4.98} & \underset{(1.1)}{\hphantom{-}4.94} & \underset{(0.31)}{\hphantom{-}9.91} \\ \end{bmatrix}$ &$\begin{bmatrix}\underset{(0.2)}{9.94} & \textcolor{red}{\underset{(0.09)}{\hphantom{-}0.01}} & \textcolor{red}{\underset{(0.1)}{-0.01}} & \textcolor{red}{\underset{(0.05)}{-0.02}} \\ \underset{(0.28)}{4.95} & \underset{(0.22)}{\hphantom{-}9.94} & \textcolor{red}{\underset{(0.13)}{0.0}} & \textcolor{red}{\underset{(0.1)}{-0.06}} \\ \underset{(0.44)}{4.98} & \underset{(0.44)}{\hphantom{-}4.97} & \underset{(0.79)}{\hphantom{-}10.6} & \textcolor{red}{\underset{(12.17)}{\hphantom{-}2.05}} \\ \underset{(0.46)}{4.99} & \underset{(0.49)}{\hphantom{-}4.99} & \underset{(6.42)}{\hphantom{-}7.07} & \underset{(5.33)}{\hphantom{-}8.05} \\ \end{bmatrix}$ \\ [40pt]\rotatebox[origin=c]{90}{$T = 1000$ } &$\begin{bmatrix}\underset{(0.11)}{9.94} & \underset{(0.27)}{-0.01} & \underset{(0.27)}{-0.01} & \underset{(0.46)}{\hphantom{-}0.02} \\ \underset{(0.29)}{4.99} & \underset{(0.18)}{\hphantom{-}9.94} & \underset{(0.37)}{-0.0} & \underset{(0.57)}{\hphantom{-}0.06} \\ \underset{(0.52)}{4.98} & \underset{(0.54)}{\hphantom{-}4.93} & \underset{(0.4)}{\hphantom{-}9.93} & \underset{(0.81)}{\hphantom{-}5.01} \\ \underset{(0.76)}{4.97} & \underset{(0.77)}{\hphantom{-}4.91} & \underset{(0.84)}{\hphantom{-}4.95} & \underset{(0.42)}{\hphantom{-}9.95} \\ \end{bmatrix}$ &$\begin{bmatrix}\underset{(0.08)}{9.98} & \textcolor{red}{\underset{(0.01)}{-0.0}} & \textcolor{red}{\underset{(0.01)}{-0.0}} & \textcolor{red}{\underset{(0.01)}{-0.0}} \\ \underset{(0.09)}{5.0} & \underset{(0.09)}{\hphantom{-}9.98} & \textcolor{red}{\underset{(0.01)}{-0.0}} & \textcolor{red}{\underset{(0.0)}{0.0}} \\ \underset{(0.12)}{5.01} & \underset{(0.12)}{\hphantom{-}4.99} & \underset{(0.19)}{\hphantom{-}9.96} & \underset{(0.34)}{\hphantom{-}5.0} \\ \underset{(0.14)}{5.01} & \underset{(0.14)}{\hphantom{-}4.98} & \underset{(0.45)}{\hphantom{-}4.97} & \underset{(0.13)}{\hphantom{-}9.97} \\ \end{bmatrix}$ &$\begin{bmatrix}\underset{(0.1)}{9.96} & \textcolor{red}{\underset{(0.03)}{-0.01}} & \textcolor{red}{\underset{(0.03)}{-0.02}} & \textcolor{red}{\underset{(0.02)}{-0.02}} \\ \underset{(0.13)}{4.99} & \underset{(0.12)}{\hphantom{-}9.96} & \textcolor{red}{\underset{(0.05)}{-0.01}} & \textcolor{red}{\underset{(0.03)}{-0.03}} \\ \underset{(0.21)}{5.0} & \underset{(0.21)}{\hphantom{-}4.98} & \underset{(0.47)}{\hphantom{-}10.45} & \textcolor{red}{\underset{(4.82)}{\hphantom{-}3.3}} \\ \underset{(0.24)}{5.0} & \underset{(0.24)}{\hphantom{-}4.98} & \underset{(2.93)}{\hphantom{-}6.28} & \underset{(1.99)}{\hphantom{-}8.93} \\ \end{bmatrix}$ \\ [40pt]

	\end{tabular}
				\begin{minipage}{1\textwidth} %
		{   \footnotesize  
			\textit{Note:} 
			Monte Carlo simulation with $M = 2000$ replications.
			The simulation is analogous to the simultaneous in     Section \ref{sec: Finite Sample Performance}, however, the shock  $\varepsilon_{4t}$ is now drawn from a standard normal distribution.
			The table
			shows the average, $1/M 
			\sum_{m=1}^{M}
			\hat{b}^{m}_{ij} $ and in parentheses the mean squared error $1/M 
			\sum_{m=1}^{M}
			(\hat{b}^{m}_{ij} - b_ij)^2 $
			of
			each estimated element  $\hat{b}^{m}_{ij}$ in simulation $m$. Penalized elements are highlighted in red.
			\par}
	\end{minipage}
\end{table}

Table \ref{Table: Finite sample performance - Bias MSE - MCRGMMA} demonstrates the application of the ridge estimator in imposing $A$-type restrictions. The simulation employs the same SVAR model as in the main text, but the two ridge estimators now enforce restrictions on $A$, which is the inverse of $B$. Specifically, $A_0 u_t = \varepsilon_t$ and $A_0 = B_0^{-1}$ with
\begin{align}
	A_0 = \begin{bmatrix}
		0.1 & 0 & 0 &  0 \\
		-0.05 & 0.1 & 0 &  0 \\
		-0.02 & -0.03 & 0.13 &  -0.07 \\
		-0.02 &  -0.03 & -0.07  &  0.13 \\
	\end{bmatrix}.
\end{align}
In this simulation, the RCSUE is equal to 
 \begin{align} 
 	\label{eq: rcsue A}
	\hat{B}_T    := \argmin \limits_{B \in 	\bar{\mathbb{B}}_{\bar{B}} } \text{ }
	g_T(B)'
	W(B)
	g_T(B) + \lambda \sum_{(i,j)\in \mathcal{R}}^{ }  
	v_{ij} A_{ij}^2  ,   
\end{align} 
where $A=B^{-1}$ and $v_{ij} = \frac{1}{\hat{A}_{ij}^2}$ where $\hat{A}$ is the first step estimator based on the unpenalized CSUE of $A_0$.
The $RCSUE(\mathcal{R}_1)$ imposes the correct zero restrictions on the $A$ matrix with 
$\mathcal{R}_1 = \{(1,2) ,(1,3),(1,4),(2,3),(2,4) \}$, 
 while the $RCSUE(\mathcal{R}_2)$ imposes the one additional incorrect   restrictions on the $A$ matrix with 
 $\mathcal{R}_2 = \{(1,2) ,(1,3),(1,4),(2,3),(2,4),(3,4) \}$.
 The results are similar to the results reported in Section \ref{sec: Finite Sample Performance}, illustrating how the ridge estimator can effectively handle $A$-type restrictions.
\begin{table}
	\caption{Finite sample performance - average and mean squared error - simulation with $A$-type restrictions}
	\label{Table: Finite sample performance - Bias MSE - MCRGMMA}
	\centering
	\renewcommand{\arraystretch}{1}
	\begin{tabular}{c | c c c}
		& $CSUE$ & $RCSUE(\mathcal{R}_1)$ & $RCSUE(\mathcal{R}_2)$ \\
		\hline
		 \rotatebox[origin=c]{90}{$T = 250$ } &$\begin{bmatrix}\underset{(0.61)}{9.73} & \underset{(1.32)}{-0.04} & \underset{(1.45)}{\hphantom{-}0.03} & \underset{(1.28)}{\hphantom{-}0.04} \\ \underset{(1.39)}{4.93} & \underset{(0.83)}{\hphantom{-}9.75} & \underset{(1.82)}{\hphantom{-}0.04} & \underset{(1.83)}{\hphantom{-}0.09} \\ \underset{(2.21)}{4.9} & \underset{(2.1)}{\hphantom{-}4.84} & \underset{(1.53)}{\hphantom{-}9.77} & \underset{(2.2)}{\hphantom{-}4.89} \\ \underset{(2.17)}{4.9} & \underset{(2.17)}{\hphantom{-}4.82} & \underset{(2.29)}{\hphantom{-}4.91} & \underset{(1.43)}{\hphantom{-}9.75} \\ \end{bmatrix}$ &$\begin{bmatrix}\underset{(0.41)}{9.88} & \textcolor{red}{\underset{(0.08)}{-0.0}} & \textcolor{red}{\underset{(0.09)}{0.0}} & \textcolor{red}{\underset{(0.09)}{\hphantom{-}0.01}} \\ \underset{(0.49)}{4.96} & \underset{(0.41)}{\hphantom{-}9.9} & \textcolor{red}{\underset{(0.1)}{0.0}} & \textcolor{red}{\underset{(0.1)}{\hphantom{-}0.02}} \\ \underset{(0.75)}{4.97} & \underset{(0.63)}{\hphantom{-}4.95} & \underset{(0.74)}{\hphantom{-}9.83} & \underset{(1.57)}{\hphantom{-}4.9} \\ \underset{(0.76)}{4.97} & \underset{(0.66)}{\hphantom{-}4.96} & \underset{(1.49)}{\hphantom{-}4.92} & \underset{(0.83)}{\hphantom{-}9.8} \\ \end{bmatrix}$ &$\begin{bmatrix}\underset{(0.55)}{9.86} & \textcolor{red}{\underset{(0.18)}{-0.02}} & \textcolor{red}{\underset{(0.21)}{-0.0}} & \textcolor{red}{\underset{(0.16)}{\hphantom{-}0.01}} \\ \underset{(0.66)}{4.98} & \underset{(0.63)}{\hphantom{-}9.87} & \textcolor{red}{\underset{(0.28)}{0.0}} & \textcolor{red}{\underset{(0.22)}{\hphantom{-}0.02}} \\ \underset{(1.05)}{4.98} & \underset{(1.12)}{\hphantom{-}4.93} & \underset{(1.13)}{\hphantom{-}10.36} & \textcolor{red}{\underset{(10.47)}{\hphantom{-}2.48}} \\ \underset{(1.53)}{4.93} & \underset{(1.07)}{\hphantom{-}4.93} & \underset{(5.09)}{\hphantom{-}6.42} & \underset{(5.7)}{\hphantom{-}8.46} \\ \end{bmatrix}$ \\ [40pt]\rotatebox[origin=c]{90}{$T = 500$ } &$\begin{bmatrix}\underset{(0.23)}{9.87} & \underset{(0.62)}{\hphantom{-}0.05} & \underset{(0.66)}{\hphantom{-}0.04} & \underset{(0.64)}{\hphantom{-}0.03} \\ \underset{(0.67)}{4.91} & \underset{(0.34)}{\hphantom{-}9.9} & \underset{(0.79)}{\hphantom{-}0.04} & \underset{(0.8)}{\hphantom{-}0.02} \\ \underset{(1.1)}{4.89} & \underset{(0.98)}{\hphantom{-}4.94} & \underset{(0.65)}{\hphantom{-}9.89} & \underset{(1.1)}{\hphantom{-}4.96} \\ \underset{(1.09)}{4.89} & \underset{(0.97)}{\hphantom{-}4.95} & \underset{(0.99)}{\hphantom{-}4.98} & \underset{(0.68)}{\hphantom{-}9.9} \\ \end{bmatrix}$ &$\begin{bmatrix}\underset{(0.18)}{9.94} & \textcolor{red}{\underset{(0.02)}{0.0}} & \textcolor{red}{\underset{(0.01)}{-0.0}} & \textcolor{red}{\underset{(0.02)}{0.0}} \\ \underset{(0.21)}{4.98} & \underset{(0.19)}{\hphantom{-}9.95} & \textcolor{red}{\underset{(0.03)}{-0.0}} & \textcolor{red}{\underset{(0.03)}{-0.0}} \\ \underset{(0.31)}{4.98} & \underset{(0.28)}{\hphantom{-}4.97} & \underset{(0.33)}{\hphantom{-}9.9} & \underset{(0.63)}{\hphantom{-}4.97} \\ \underset{(0.3)}{4.98} & \underset{(0.28)}{\hphantom{-}4.97} & \underset{(0.61)}{\hphantom{-}4.96} & \underset{(0.34)}{\hphantom{-}9.92} \\ \end{bmatrix}$ &$\begin{bmatrix}\underset{(0.2)}{9.94} & \textcolor{red}{\underset{(0.06)}{0.0}} & \textcolor{red}{\underset{(0.07)}{0.0}} & \textcolor{red}{\underset{(0.06)}{\hphantom{-}0.01}} \\ \underset{(0.26)}{4.97} & \underset{(0.21)}{\hphantom{-}9.95} & \textcolor{red}{\underset{(0.08)}{-0.01}} & \textcolor{red}{\underset{(0.07)}{-0.0}} \\ \underset{(0.44)}{4.98} & \underset{(0.37)}{\hphantom{-}4.99} & \underset{(0.43)}{\hphantom{-}10.29} & \textcolor{red}{\underset{(4.87)}{\hphantom{-}3.42}} \\ \underset{(0.47)}{4.97} & \underset{(0.4)}{\hphantom{-}4.96} & \underset{(2.37)}{\hphantom{-}6.01} & \underset{(2.08)}{\hphantom{-}9.07} \\ \end{bmatrix}$ \\ [40pt]\rotatebox[origin=c]{90}{$T = 1000$ } &$\begin{bmatrix}\underset{(0.1)}{9.96} & \underset{(0.28)}{\hphantom{-}0.02} & \underset{(0.28)}{\hphantom{-}0.02} & \underset{(0.28)}{\hphantom{-}0.01} \\ \underset{(0.32)}{4.99} & \underset{(0.16)}{\hphantom{-}9.96} & \underset{(0.35)}{0.0} & \underset{(0.35)}{-0.0} \\ \underset{(0.48)}{4.97} & \underset{(0.48)}{\hphantom{-}5.02} & \underset{(0.28)}{\hphantom{-}9.95} & \underset{(0.48)}{\hphantom{-}4.97} \\ \underset{(0.48)}{4.97} & \underset{(0.47)}{\hphantom{-}5.02} & \underset{(0.44)}{\hphantom{-}4.99} & \underset{(0.3)}{\hphantom{-}9.94} \\ \end{bmatrix}$ &$\begin{bmatrix}\underset{(0.09)}{9.98} & \textcolor{red}{\underset{(0.01)}{0.0}} & \textcolor{red}{\underset{(0.01)}{0.0}} & \textcolor{red}{\underset{(0.01)}{0.0}} \\ \underset{(0.1)}{5.0} & \underset{(0.09)}{\hphantom{-}9.97} & \textcolor{red}{\underset{(0.01)}{-0.0}} & \textcolor{red}{\underset{(0.01)}{-0.0}} \\ \underset{(0.12)}{5.0} & \underset{(0.12)}{\hphantom{-}5.01} & \underset{(0.14)}{\hphantom{-}9.96} & \underset{(0.25)}{\hphantom{-}4.97} \\ \underset{(0.12)}{5.0} & \underset{(0.13)}{\hphantom{-}5.02} & \underset{(0.25)}{\hphantom{-}4.98} & \underset{(0.14)}{\hphantom{-}9.95} \\ \end{bmatrix}$ &$\begin{bmatrix}\underset{(0.1)}{9.98} & \textcolor{red}{\underset{(0.02)}{0.0}} & \textcolor{red}{\underset{(0.02)}{0.0}} & \textcolor{red}{\underset{(0.02)}{\hphantom{-}0.01}} \\ \underset{(0.12)}{5.01} & \underset{(0.1)}{\hphantom{-}9.98} & \textcolor{red}{\underset{(0.02)}{-0.0}} & \textcolor{red}{\underset{(0.02)}{-0.0}} \\ \underset{(0.18)}{5.0} & \underset{(0.17)}{\hphantom{-}5.02} & \underset{(0.2)}{\hphantom{-}10.2} & \textcolor{red}{\underset{(1.47)}{\hphantom{-}4.13}} \\ \underset{(0.18)}{4.99} & \underset{(0.17)}{\hphantom{-}5.02} & \underset{(0.78)}{\hphantom{-}5.59} & \underset{(0.57)}{\hphantom{-}9.53} \\ \end{bmatrix}$ \\ [40pt]

	\end{tabular}
				\begin{minipage}{1\textwidth} %
		{   \footnotesize  
			\textit{Note:} 
			Monte Carlo simulation with $M = 2000$ replications.
		The simulation is analogous to the simultaneous in     Section \ref{sec: Finite Sample Performance}, however, the ridge estimators impose $A$-type restrictions as described in Equation (\ref{eq: rcsue A}).
			The table
			shows the average, $1/M 
			\sum_{m=1}^{M}
			\hat{b}^{m}_{ij} $ and in parentheses the mean squared error $1/M 
			\sum_{m=1}^{M}
			(\hat{b}^{m}_{ij} - b_ij)^2 $
			of
			each estimated element  $\hat{b}^{m}_{ij}$ in simulation $m$. Penalized elements are highlighted in red.
			\par}
	\end{minipage}
\end{table}

Table \ref{Table: Finite sample performance - Bias MSE - MCRGMMProxyExog} 
simulates a system where $u_{4t}$ can be seen as a linear proxy variable affected by the target shock $\varepsilon_{1t}$ and a noise term  $\varepsilon_{4t}$ with $u_{4t}=5 \varepsilon_{1t}  + 10 \varepsilon_{4t}$ such that 
\begin{align}
	\label{eq: MC proxy} 
	\begin{bmatrix}
		u_{1t} \\
		u_{2t} \\
		u_{3t} \\
		u_{4t} \\
	\end{bmatrix} =
	\begin{bmatrix}
		10 & 0  & 0 & 0 \\
		5 & 10 & 0 & 0\\
		5 & 5 & 10 & 0\\
		5 & 0 & 0 & 10
	\end{bmatrix}
	\begin{bmatrix}
		\varepsilon_{1t} \\
		\varepsilon_{2t} \\
		\varepsilon_{3t} \\
		\varepsilon_{4t} \\
	\end{bmatrix}.
\end{align}  
The proxy restrictions  $ \mathcal{R} =\{(1,4) ,(2,4),(3,4),(4,2),(4,3) \}$ impose a penalty $B$ to ensure that the proxy is exogenous and that the noise term does not affect the other variables. Imposing the proxy penalty leads to a smaller bias and to a two too three times smaller MSE compared to the unpenalized CSUE and thus,   illustrates that the RCSUE can also be used to impose restrictions implied by proxy variables.
\begin{table}
	\caption{Finite sample performance - average and mean squared error - simulation with proxy variable restrictions}
	\label{Table: Finite sample performance - Bias MSE - MCRGMMProxyExog}
	\centering
	\renewcommand{\arraystretch}{1}
	\begin{tabular}{c | c c  }
		& $CSUE$ & $RCSUE(\mathcal{R}_1)$  \\
		\hline
		  \rotatebox[origin=c]{90}{$T = 250$ } &$\begin{bmatrix}\underset{(0.54)}{9.76} & \underset{(1.27)}{\hphantom{-}0.07} & \underset{(1.26)}{\hphantom{-}0.1} & \underset{(1.37)}{\hphantom{-}0.01} \\ \underset{(1.4)}{4.85} & \underset{(0.75)}{\hphantom{-}9.8} & \underset{(1.64)}{\hphantom{-}0.06} & \underset{(1.75)}{\hphantom{-}0.1} \\ \underset{(1.76)}{4.82} & \underset{(1.62)}{\hphantom{-}4.93} & \underset{(1.01)}{\hphantom{-}9.79} & \underset{(2.06)}{\hphantom{-}0.08} \\ \underset{(1.92)}{9.76} & \underset{(2.59)}{\hphantom{-}0.01} & \underset{(2.63)}{\hphantom{-}0.07} & \underset{(1.8)}{\hphantom{-}9.75} \\ \end{bmatrix}$ &
		  $\begin{bmatrix}\underset{(0.37)}{9.85} & \textcolor{black}{\underset{(0.37)}{\hphantom{-}0.01}} & \textcolor{black}{\underset{(0.36)}{\hphantom{-}0.02}} & \textcolor{red}{\underset{(0.05)}{-0.01}} \\ \underset{(0.73)}{4.94} & \underset{(0.5)}{\hphantom{-}9.86} & \textcolor{black}{\underset{(1.28)}{\hphantom{-}0.02}} & \textcolor{red}{\underset{(0.06)}{\hphantom{-}0.01}} \\ \underset{(0.92)}{4.93} & \underset{(1.34)}{\hphantom{-}4.95} & \underset{(0.77)}{\hphantom{-}9.84} & \textcolor{red}{\underset{(0.08)}{0.0}} 
		  	\\ \underset{(0.68)}{9.93} & \textcolor{red}{\underset{(0.11)}{\hphantom{-}0.01}} & \textcolor{red}{\underset{(0.15)}{\hphantom{-}0.01} }& \underset{(0.5)}{\hphantom{-}9.86} \\ \end{bmatrix}$ 
		  
		  \\ [40pt]\rotatebox[origin=c]{90}{$T = 500$ } &$\begin{bmatrix}\underset{(0.24)}{9.87} & \underset{(0.6)}{-0.0} & \underset{(0.56)}{\hphantom{-}0.03} & \underset{(0.61)}{\hphantom{-}0.01} \\ \underset{(0.67)}{4.93} & \underset{(0.37)}{\hphantom{-}9.85} & \underset{(0.85)}{\hphantom{-}0.04} & \underset{(0.91)}{\hphantom{-}0.03} \\ \underset{(0.77)}{4.89} & \underset{(0.92)}{\hphantom{-}4.91} & \underset{(0.49)}{\hphantom{-}9.88} & \underset{(1.11)}{\hphantom{-}0.04} \\ \underset{(0.81)}{9.87} & \underset{(1.24)}{-0.0} & \underset{(1.21)}{\hphantom{-}0.02} & \underset{(0.76)}{\hphantom{-}9.89} \\ \end{bmatrix}$ &
		  $\begin{bmatrix}\underset{(0.17)}{9.91} & \textcolor{black}{\underset{(0.15)}{-0.02}} & \textcolor{black}{\underset{(0.15)}{0.0}} & \textcolor{red}{\underset{(0.02)}{-0.0}} \\ \underset{(0.31)}{4.97} & \underset{(0.24)}{\hphantom{-}9.89} & \textcolor{black}{\underset{(0.64)}{\hphantom{-}0.02}} & \textcolor{red}{\underset{(0.02)}{0.0}} \\ \underset{(0.37)}{4.96} & \underset{(0.7)}{\hphantom{-}4.94} & \underset{(0.36)}{\hphantom{-}9.91} & \textcolor{red}{\underset{(0.03)}{0.0}} 
		  	\\ \underset{(0.33)}{9.95} & \textcolor{red}{\underset{(0.03)}{-0.0}} & \textcolor{red}{\underset{(0.03)}{\hphantom{-}0.01}} & \underset{(0.23)}{\hphantom{-}9.94} \\ \end{bmatrix}$ 
		  
		  \\ [40pt]\rotatebox[origin=c]{90}{$T = 1000$ } &$\begin{bmatrix}\underset{(0.1)}{9.95} & \underset{(0.27)}{\hphantom{-}0.01} & \underset{(0.28)}{\hphantom{-}0.02} & \underset{(0.26)}{\hphantom{-}0.02} \\ \underset{(0.29)}{4.97} & \underset{(0.16)}{\hphantom{-}9.94} & \underset{(0.38)}{\hphantom{-}0.03} & \underset{(0.37)}{-0.0} \\ \underset{(0.36)}{4.96} & \underset{(0.38)}{\hphantom{-}4.97} & \underset{(0.22)}{\hphantom{-}9.97} & \underset{(0.43)}{-0.01} \\ \underset{(0.35)}{9.95} & \underset{(0.59)}{\hphantom{-}0.02} & \underset{(0.58)}{\hphantom{-}0.03} & \underset{(0.33)}{\hphantom{-}9.97} \\ \end{bmatrix}$ &
		  $\begin{bmatrix}\underset{(0.07)}{9.97} & \textcolor{black}{\underset{(0.07)}{\hphantom{-}0.01}} & \textcolor{black}{\underset{(0.07)}{-0.0}} & \textcolor{red}{\underset{(0.0)}{0.0}} \\ \underset{(0.14)}{4.98} & \underset{(0.11)}{\hphantom{-}9.96} & \textcolor{black}{\underset{(0.24)}{\hphantom{-}0.02}} & \textcolor{red}{\underset{(0.01)}{-0.0}} \\ \underset{(0.17)}{4.99} & \underset{(0.25)}{\hphantom{-}4.98} & \underset{(0.15)}{\hphantom{-}9.98} & \textcolor{red}{\underset{(0.01)}{-0.0}} 
		  	\\ \underset{(0.14)}{10.0} & \textcolor{red}{\underset{(0.01)}{\hphantom{-}0.01}} & \textcolor{red}{\underset{(0.01)}{0.0}} & \underset{(0.1)}{\hphantom{-}9.98} \\ \end{bmatrix}$ \\ [40pt]

	\end{tabular}
				\begin{minipage}{1\textwidth} %
		{   \footnotesize  
			\textit{Note:} 
			Monte Carlo simulation with $M = 2000$ replications.
				The simulation is analogous to the simultaneous in      Section \ref{sec: Finite Sample Performance}, however, the simultaneous interaction motivated by a proxy variable and is given by Equation (\ref{eq: MC proxy}). The ridge penalty is equal to  $ \mathcal{R} =\{(1,4) ,(2,4),(3,4),(4,2),(4,3) \}$, which corresponds to an exogenous proxy variable.
			The table
			shows the average, $1/M 
			\sum_{m=1}^{M}
			\hat{b}^{m}_{ij} $ and in parentheses the mean squared error $1/M 
			\sum_{m=1}^{M}
			(\hat{b}^{m}_{ij} - b_ij)^2 $
			of
			each estimated element  $\hat{b}^{m}_{ij}$ in simulation $m$. Penalized elements are highlighted in red.
			\par}
	\end{minipage}
\end{table}

\section{Appendix - Application} 
\label{appendix: sec: Application} 
This section contains supplementary material  for the application in Section \ref{sec: Application}.

\begin{figure}[h!] 
 	\centering
 	\caption{Data}
 	\includegraphics[width=0.8\textwidth]{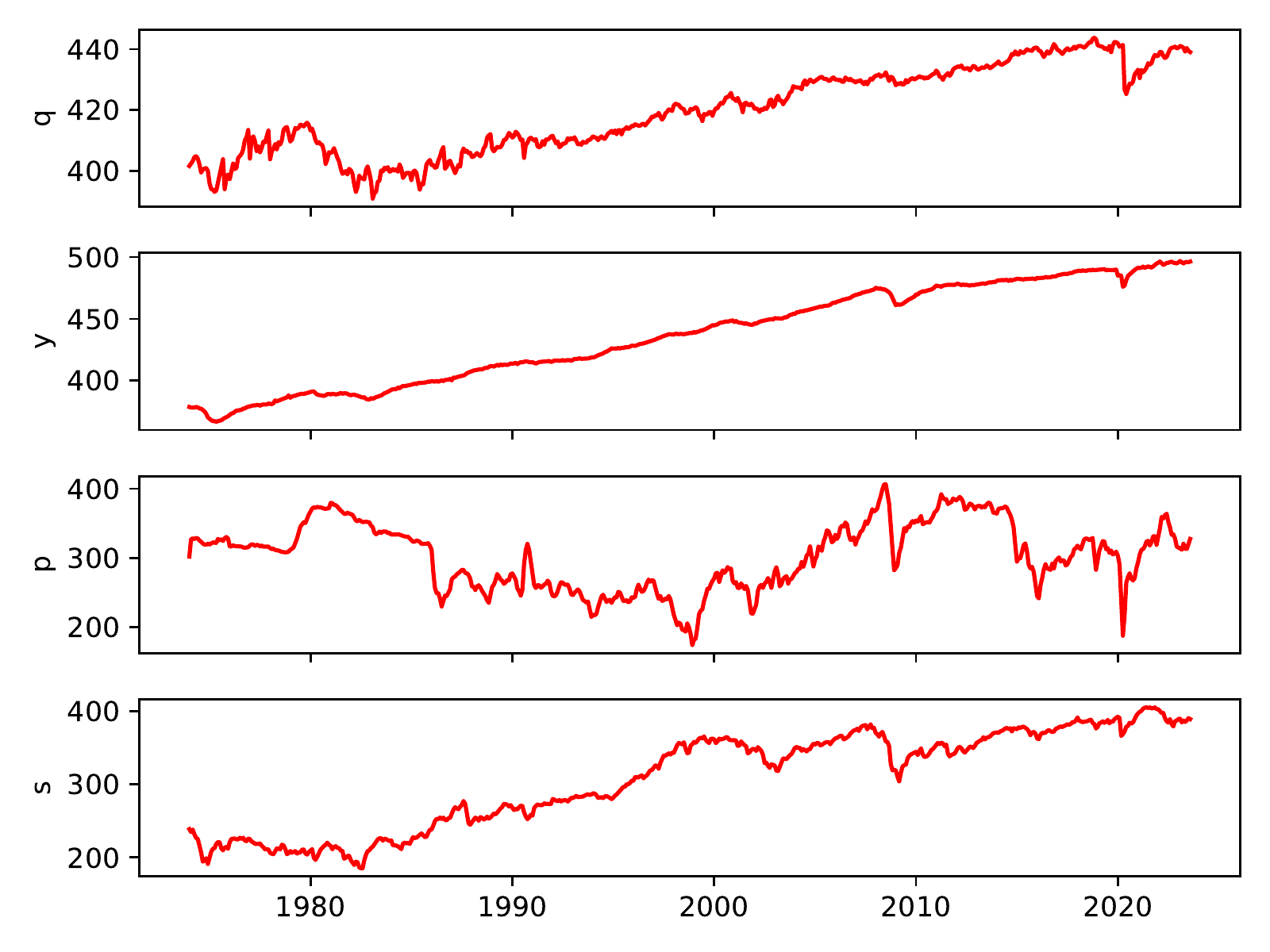}  
 	\label{fig: data } 
 	\begin{minipage}{1\textwidth} %
 		{   \footnotesize  
 			\textit{Note:} Log   of world crude oil production	$q_t$, log   of  global industrial production  $y_t$, log   of the real oil price $p_t$,   and log of a monthly U.S. stock price index $s_t$. Global oil production is given by the global crude oil including lease condensate production obtained from the U.S. EIA. Global industrial production is given by the  monthly industrial production index in the OECD and six major other
 			countries obtained from \cite{baumeister2019structural}. The real oil price is equal to the refiner's acquisition cost of imported crude oil  from the U.S. EIA deflated by the U.S. CPI. Real stock prices correspond to the aggregate   U.S. stock index constructed by the OECD deflated by the U.S. CPI.
 			\par}
 	\end{minipage}
 \end{figure}

Figure \ref{fig: data }  shows the data and contains the data sources.
 
   Table \ref{Table: Non-Gaussianity    rec} to \ref{Table: Non-Gaussianity    unrestricted} show the skewness, kurtosis, and the p-value of the Jarque-Bera test of the estimated structural shocks from the recursive estimator, the estimated structural shocks from the ridge estimator, and the estimated structural
  	 shocks from the unrestricted CSUE. The tables show clear evidence of non-Gaussianity for at least three out of four structural shocks. Moreover, in all models the skewness of all shocks, except one,  exceeds a value of one in absolute terms. For comparison, a standard normal random variable with the same sample size will   exceed a sample skewness of one with probability less than $0.00$\%. Therefore, the results indicate sufficiently skewed shocks to ensure identification based on mutually mean independent shocks as stated in Proposition \ref{proposition: 1}.
\begin{table}[h]
	\caption{Skewness, kurtosis, and the p-value of the Jarque-Bera test of the estimated structural shocks with $16$\% and $84$\%  bootstrap quantiles - recursive estimator}
	\label{Table: Non-Gaussianity    rec}
	\begin{center} 
		\begin{tabular}{ c|  c c c c         }
			&$\varepsilon_{S,t}$  &$\varepsilon_{Y,t}$  &$\varepsilon_{D,t}$ & $\varepsilon_{SM,t}$    
			\\ \hline
 			Skewness &
			$ \underset{(-0.68 / -1.43)}{-1.248 }   $ & 
			$\underset{(-0.26/ -2.08)}{-1.573} $    &
			$ \underset{(-0.05 / -0.56)}{-0.342} $   &
			$ \underset{(-0.53 / -1.11)}{-1.035} $ 
			\\
			Kurtosis &    
			$ \underset{(7.48 / 10.07)}{10.178}   $ & 
			$\underset{(9.74 / 19.31)}{18.684} $    &
			$ \underset{( 3.6 / 5.02)}{5.195} $   &
			$ \underset{(5.08 /7.95)}{7.852} $ 
			\\
			JB-Test &  $0 $   &  $0  $   &  $0 $ & $ 0 $
		\end{tabular} 
	\end{center}
\end{table} 
  \begin{table}[h]
  	\caption{Skewness, kurtosis, and the p-value of the Jarque-Bera test of the estimated structural shocks  with $16$\% and $84$\%  bootstrap quantiles - ridge estimator}
  	\label{Table: Non-Gaussianity    ridge}
  	\begin{center} 
  		\begin{tabular}{ c|  c c c c         }
  			&$\varepsilon_{S,t}$  &$\varepsilon_{Y,t}$  &$\varepsilon_{D,t}$ & $\varepsilon_{SM,t}$    
  			\\ \hline
  			Skewness &
  		$ \underset{(-0.78 / -1.52)}{-1.336 }   $ & 
  		$\underset{(-0.30/ -2.07)}{-1.566} $    &
  		$ \underset{(-0.01 / -0.27)}{-0.164} $   &
  		$ \underset{(-0.72 / -1.77)}{-1.492} $ 
  		\\
  		Kurtosis &    
  		$ \underset{(7.5 / 10.34)}{ 9.938 }   $ & 
  		$\underset{(9.64 / 19.5)}{18.483} $    &
  		$ \underset{( 3.13 / 3.97)}{3.464} $   &
  		$ \underset{(5.97 /13.00)}{11.496} $ 
  			\\
  			JB-Test &  $0 $   &  $0  $   &  $0.02 $ & $ 0 $
  		\end{tabular} 
  	\end{center}
  \end{table} 	
   \begin{table}[h]
 	\caption{Skewness, kurtosis, and the p-value of the Jarque-Bera test of the estimated structural shocks  with $16$\% and $84$\%  bootstrap quantiles - unrestricted estimator}
 	\label{Table: Non-Gaussianity    unrestricted}
 	\begin{center} 
 		\begin{tabular}{ c|  c c c c         }
 			&$\varepsilon_{S,t}$  &$\varepsilon_{Y,t}$  &$\varepsilon_{D,t}$ & $\varepsilon_{SM,t}$    
 			\\ \hline
 			Skewness &
 			$ \underset{(-0.79 / -1.54)}{-1.325}   $ & 
 			$\underset{(-0.25 / -2.15)}{-1.633} $    &
 			$ \underset{(-0.00 / -0.29)}{-0.197} $   &
 			$ \underset{(-0.69 / -1.77)}{-1.487} $ 
 			\\
 			Kurtosis &    
 			$ \underset{(7.56 / 10.66)}{9.828}   $ & 
 			$\underset{(9.23 / 20.31)}{ 19.08} $    &
 			$ \underset{(3.22 / 4.02)}{3.483} $   &
 			$ \underset{(5.87 /13.21)}{11.724} $ 
 			\\
 			JB-Test &  $0 $   &  $0  $   &  $0.01 $ & $ 0 $
 		\end{tabular} 
 	\end{center}
 \end{table}

Table \ref{Table: dependency} displays the symmetric-fourth order co-moments of the estimated shocks. For independent shocks, the moments should approach one asymptotically. For dependent shocks, specifically shocks with a common volatility process, the moments should deviate from one.
Using the recursive estimator, the moment $E[\varepsilon_{D,t}^2 \varepsilon_{SM,t}^2]$ clearly deviates from one. The ridge estimator and the unrestricted estimator show a similar result and indicate a dependency of the volatility process of the oil-specific demand and stock market information shock. Additionally, both estimators show some evidence for a common volatility process of the oil supply and oil-specific demand shocks. 
Overall, the results indicate that the shocks are not independent but affected by a common volatility process.
 \begin{table}[h]
 	\caption{Symmetric fourth-order moments with $16$\% and $84$\%  bootstrap quantiles }
 	\label{Table: dependency}
 	\begin{center} 
 		\begin{tabular}{ c|  c c c c   c c       }
 			&$E[\varepsilon_{S,t}^2 \varepsilon_{Y,t}^2]$  &
 			$E[\varepsilon_{S,t}^2 \varepsilon_{D,t}^2]$  &
 			$E[\varepsilon_{S,t}^2 \varepsilon_{SM,t}^2]$ & 
 			$E[\varepsilon_{Y,t}^2 \varepsilon_{D,t}^2]$    &
 			$E[\varepsilon_{Y,t}^2 \varepsilon_{SM,t}^2]$ & 
 			$E[\varepsilon_{D,t}^2 \varepsilon_{SM,t}^2]$  
 			\\ \hline
 		 \shortstack{Recursive \\ estimator} &
 			$ \underset{(0.86 / 1.41)}{1.188 }   $ & 
 			$\underset{(0.93/ 1.95)}{1.705} $    &
 			$ \underset{(0.95 / 1.53)}{1.298} $   &
 			$ \underset{(1.02 / 1.39)}{1.257} $ &
 			$ \underset{(0.97 / 1.44)}{1.319} $ &
 			$ \underset{(1.55 / 4.04)}{3.529} $ 
 			\\ 
 				 \shortstack{Ridge \\ estimator} &
 			$ \underset{(0.92 / 1.63)}{1.399 }   $ & 
			$\underset{(1.01/ 2.82)}{2.873} $    &
			$ \underset{(0.90 / 1.34)}{1.278} $   &
			$ \underset{(0.84 / 1.24)}{1.069} $ &
			$ \underset{(0.92 / 1.51)}{1.366} $ &
			$ \underset{(1.03 / 1.53)}{1.428} $ 
 				\\ 
 			\shortstack{Unrestricted \\ estimator} &
 			$ \underset{(0.85 / 1.67)}{1.445 }   $ & 
			$\underset{(0.93/ 2.83)}{2.931} $    &
			$ \underset{(0.78 / 1.25)}{1.108} $   &
			$ \underset{(0.81 / 1.15)}{0.967} $ &
			$ \underset{(0.81 / 1.61)}{1.247} $ &
			$ \underset{(0.98 / 1.45)}{1.364} $ 
 		\end{tabular} 
 	\end{center}
 \end{table}

\begin{figure}[h!] 
 	\centering
 	\caption{Impulse responses. Red: Unrestricted estimator. Blue: Ridge estimator.} 
 	\includegraphics[width=0.80\textwidth]{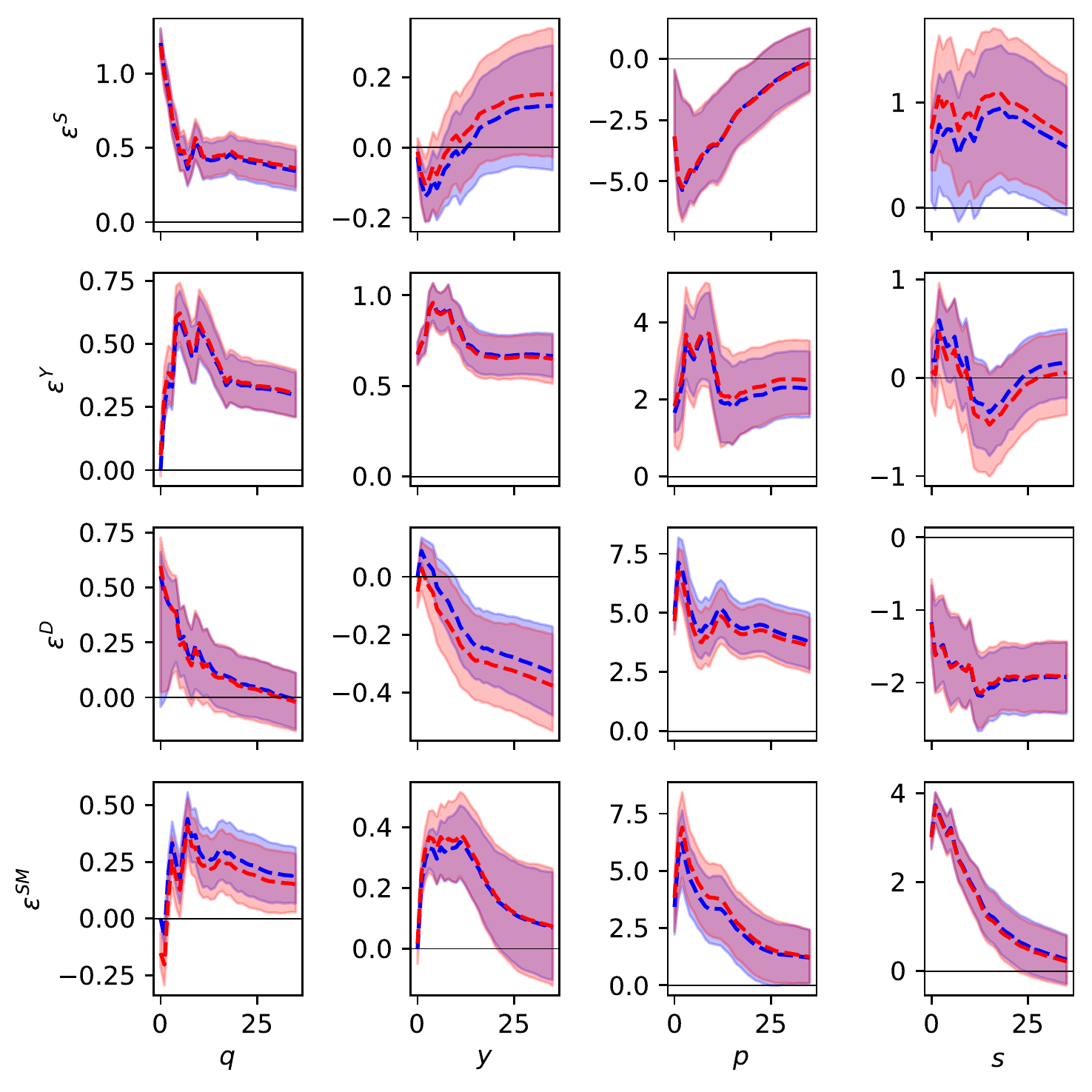}
 	\label{fig: IRF CSUE} 
 	\begin{minipage}{1\textwidth} %
 		{   \footnotesize  
 			\textit{Note:} Impulse responses to one standard deviation shocks with $68$\% bootstrap confidence bands.
 			The ridge estimator in blue is equal to the RCSUE analyzed in Section \ref{sec: Application}.
 			The unrestricted estimator in red is equal to the CSUE relying only on non-Gaussianity and (mean-)independent shocks without a restriction penalty.
 			\par}
 	\end{minipage}
 \end{figure}

Figure \ref{fig: IRF CSUE}  compares the impulse responses of ridge estimator considered in the main text to the  unrestricted CSUE. The results are overall similar and are in line with the simulations in Section \ref{sec: Finite Sample Performance} the confidence bands of the ridge estimator are smaller. 
 
Equation (\ref{eq: recursive A}) displays the $A$-type transformation of the recursive estimator with $68$\% bootstrap confidence bands.
 Equation (\ref{eq: Ridge A}) displays analogous results for the ridge estimator, and Equation (\ref{eq: Unrestricted A}) displays the results for the unrestricted estimator.
 Comparing the response of the unrestricted and the ridge estimator illustrates how including the restrictions using the penalty term leads to smaller confidence bands. 
 \begin{gather}
 	\nonumber
 	\textit{Recursive estimator} 
 	\\
 	\label{eq: recursive A}
 	\begin{vmatrix} 
 		q_t 
 		&=&  
 		& &  \underset{(-0.0 / -0.0)}{-0.0} 	y_t 
 		&+&  \underset{(-0.0 / -0.0)}{-0.0} 	p_t 
 		&+&  \underset{(-0.0 / -0.0)}{-0.0}     s_t   
 		&+\varepsilon^S_t
 		\\
 		y_t 
 		&=&   \underset{(-0.03 / 0.01)}{-0.01}  q_t
 		& & 
 		&+& \underset{(-0.0 / -0.0)}{-0.0}  p_t 
 		&+& \underset{(-0.0 / -0.0)}{-0.0}   s_t 
 		&+\varepsilon^Y_t 
 		\\
 		q_t 
 		&=&
 		& &  \underset{(1.5 / 4.06)}{2.353}  y_t 
 		&+& \underset{(-2.0 / -0.6)}{-1.039}  p_t 
 		&+&  \underset{(-0.0 / -0.0)}{-0.0} s_t
 		&+\varepsilon^D_t
 		\\
 		s_t 
 		&=&   \underset{(-0.15 / 0.05)}{-0.049}   q_t 
 		&+& \underset{(-0.27 / 0.29)}{0.0} y_t 
 		&+&   \underset{(0.01 / 0.1)}{0.05} p_t 
 		& & 
 		&\text{ }+\varepsilon^{SM}_t  
 	\end{vmatrix}
 \end{gather} 
 \begin{gather}
 	\nonumber
 	\textit{Ridge estimator} 
 	\\
 	\label{eq: Ridge A}
 	\begin{vmatrix} 
 		q_t 
 		&=&  
 		& & \underset{(-0.28 / -0.03)}{-0.191}	y_t 
 		&+& \underset{(-0.01 / 0.11)}{0.088}	p_t 
 		&+& \underset{(-0.15 / -0.01)}{-0.1}    s_t   
 		&+\varepsilon^S_t
 		\\
 		y_t 
 		&=&  \underset{(-0.04 / 0.0)}{-0.018}  q_t
 		& & 
 		&+& \underset{(0.0 / 0.0)}{0.002} p_t 
 		&+& \underset{(-0.0 / -0.0)}{-0.002}  s_t 
 		&+\varepsilon^Y_t 
 		\\
 		q_t 
 		&=&
 		& &  \underset{(0.18 / 1.19)}{0.702} y_t 
 		&+& \underset{(-0.42 / -0.23)}{-0.325}   p_t 
 		&+&  \underset{(0.04 / 0.54)}{0.367} s_t
 		&+\varepsilon^D_t
 		\\
 		s_t 
 		&=&   \underset{(-0.3 / 0.01)}{-0.138}  q_t 
 		&+& \underset{(0.39 / 1.29)}{0.81} y_t 
 		&+&  \underset{(-0.34 / -0.12)}{-0.222} p_t 
 		& & 
 		&\text{ }+\varepsilon^{SM}_t  
 	\end{vmatrix}
 \end{gather} 
 \begin{gather}
 	\nonumber
 	\textit{Unrestricted estimator} 
 	\\
 	\label{eq: Unrestricted A}
 	\begin{vmatrix} 
 		q_t 
 		&=&  
 		& & \underset{(-0.35 / 0.06)}{-0.13} 	y_t 
 		&+& \underset{(-0.01 / 0.12)}{0.087} 	p_t 
 		&+& \underset{(-0.22 / -0.07)}{-0.16}    s_t   
 		&+\varepsilon^S_t
 		\\
 		y_t 
 		&=&  \underset{(-0.06 / 0.0)}{-0.029} q_t
 		& & 
 		&+& \underset{(-0.02 / 0.01)}{-0.005}  p_t 
 		&+& \underset{(-0.01 / 0.03)}{0.01}   s_t 
 		&+\varepsilon^Y_t 
 		\\
 		q_t 
 		&=&
 		& &  \underset{(0.07 / 3.61)}{0.864} y_t 
 		&+& \underset{(-1.67 / -0.24)}{-0.301}  p_t 
 		&+&   \underset{(0.14 / 1.86)}{0.326} s_t
 		&+\varepsilon^D_t
 		\\
 		s_t 
 		&=&  \underset{(-0.2 / 0.15)}{-0.011} q_t 
 		&+&  \underset{(0.16 / 1.54)}{0.767} y_t 
 		&+& \underset{(-0.36 / -0.11)}{-0.245} p_t 
 		& & 
 		&\text{ }+\varepsilon^{SM}_t  
 	\end{vmatrix}
 \end{gather}

\begin{figure}[h!] 
	\centering
	\caption{Impulse responses. Red: Ridge ($A$-type) estimator. Blue: Ridge estimator.} 
	\includegraphics[width=0.80\textwidth]{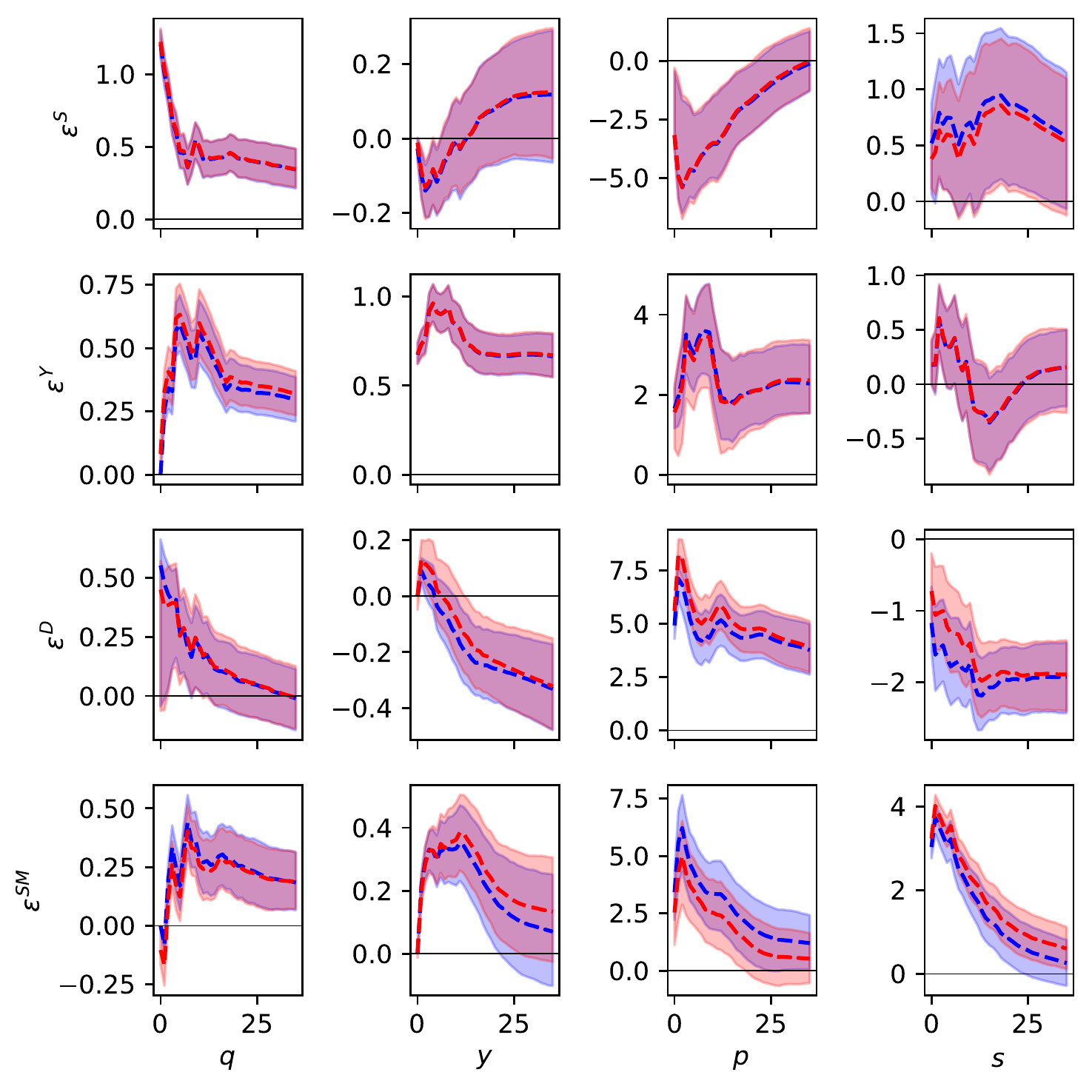}
	\label{fig: IRF Arest2} 
	\begin{minipage}{1\textwidth} %
		{   \footnotesize  
			\textit{Note:} Impulse responses to one standard deviation shocks with $68$\% bootstrap confidence bands.   
				The ridge estimator in blue is equal to the RCSUE analyzed in Section \ref{sec: Application}.
				The ridge estimator depicted in red is equal to the RCSUE imposing $A$-type penalties and shrinking towards the following $A$-type model:  
					i) $q_t = a_{13} p_t + \varepsilon^S_t $, 
					ii) $y_t = a_{23} p_t + \varepsilon^Y_t$, 
					iii) $q_t = a_{32} y_t +  a_{33} p_t +  a_{33} s_t+ \varepsilon^D_t$, 
					and iv) $s_t = a_{41} q_t +  a_{42} y_t +  a_{43} p_t+ \varepsilon^{SM}_t $
				
			\par}
	\end{minipage}
\end{figure}

  Figure \ref{fig: IRF Arest2}  compares the impulse responses of ridge estimator considered in the main text to 
the impulse responses based on a ridge estimator using  $A$-type restrictions  such that
\begin{align}
	q_t &= a_{13} p_t + \varepsilon^S_t \\
	y_t &= a_{23} p_t + \varepsilon^Y_t \\
	q_t &= a_{32} y_t +  a_{33} p_t +  a_{33} s_t+ \varepsilon^D_t\\
	s_t &= a_{41} q_t +  a_{42} y_t +  a_{43} p_t+ \varepsilon^{SM}_t ,
\end{align}
which represent an oil supply equation, an economic activity equation, an oil demand equation, and a stock market equation, compare \cite{baumeister2019structural} and \cite{braun2021importance}.
The  $A$-type penalty leads to similar results compared to the main text. Specifically, both estimators find a simultaneous response of the oil price to the information shock $\varepsilon^{SM}_t$.

\end{document}